\documentclass{emulateapj}
\usepackage{amsmath} 
\usepackage{apjfonts} 
\usepackage{amssymb} 
\usepackage{xspace}
\usepackage[bookmarks=false]{hyperref} 
\usepackage{multirow}
\usepackage{graphicx}
\usepackage{epstopdf}
\usepackage{epsf}
\usepackage{subfigure}
\usepackage{CJK}
\usepackage[usenames,dvipsnames]{xcolor}

\def\gsim{\gtrsim}
\def\lsim{\lesssim}    

\begin{document}
\begin{CJK*}{UTF8}{gkai}
\shorttitle{Automated Lensing Learner}
\shortauthors{Avestruz, Li, Zhu, et al.}
\submitted{The Astrophysical Journal} 
\slugcomment{The Astrophysical Journal, submitted}

\title{Automated Lensing Learner: Automated Strong Lensing Identification with a Computer Vision Technique }

\author{Camille Avestruz\altaffilmark{1-2}\thanks{E-mail:
    avestruz@uchicago.edu}, Nan Li\altaffilmark{2-5}, Hanjue
  Zhu (朱涵珏) \altaffilmark{6}, Matthew Lightman\altaffilmark{7}, Thomas
  E.~Collett\altaffilmark{8}, Wentao Luo\altaffilmark{9}}

\affil{
$^1${Enrico Fermi Institute, The University of Chicago, Chicago, IL 60637 U.S.A.}\\
$^2${Kavli Institute for Cosmological Physics, The University of Chicago, Chicago, IL 60637 U.S.A.}\\
$^3${Department of Astronomy \& Astrophysics, The University of Chicago, Chicago, IL 60637 U.S.A.};\\
$^4${High Energy Physics Division, Argonne National Laboratory,
    Lemont, IL 60439};\\
$^5${School of Physics and Astronomy, University of Nottingham, University Park, Nottingham, NG7 2RD, UK; nan.li@nottingham.ac.uk};\\
$^6${The University of Chicago, Chicago, IL 60637 U.S.A.};\\
$^7${JPMorgan Chase, Chicago, IL 60603 U.S.A.};\\
$^8${Institute of Cosmology and Gravitation, University of Portsmouth, Burnaby Rd, Portsmouth, PO1 3FX, UK};\\
$^9${Center for Astronomy and Astrophysics, Shanghai Jiaotong University,
800 Dongchuan Road, Shanghai 200240, China};\\
  \href{mailto:avestruz@uchicago.edu}{avestruz@uchicago.edu}\\
}

  
\begin{abstract} 
Forthcoming surveys such as the Large Synoptic Survey Telescope (LSST)
and Euclid necessitate automatic and efficient identification methods
of strong lensing systems. We present a strong lensing identification
approach that utilizes a feature extraction method from computer
vision, the Histogram of Oriented Gradients (HOG), to capture edge
patterns of arcs.  We train a supervised classifier model on the HOG
of mock strong galaxy-galaxy lens images similar to observations from
the Hubble Space Telescope (HST) and LSST. We assess model performance
with the area under the curve (AUC) of a Receiver Operating
Characteristic (ROC) curve. Models trained on 10,000 lens and non-lens
containing images images exhibit an AUC of 0.975 for an HST-like
sample, 0.625 for one exposure of LSST, and 0.809 for 10-year mock
LSST observations. Performance appears to continually improve with the
training set size. Models trained on fewer images perform better in
absence of the lens galaxy light. However, with larger training data
sets, information from the lens galaxy actually improves model
performance, indicating that HOG captures much of the morphological
complexity of the arc finding problem. We test our classifier on data
from the Sloan Lens ACS Survey and find that small scale image
features reduces the efficiency of our trained model. However, these
preliminary tests indicate that some parameterizations of HOG can
compensate for differences between observed mock data. {One example
  best-case parameterization results in an AUC of 0.6 in the F814
  filter image with other parameterization results equivalent to
  random performance}.
\end{abstract}

\section{Introduction}

Gravitational lensing occurs when intermediate fluctuations in the
matter density field deflect light from background sources
\citep[see][for a review]{kneibandnatarajan11}.  Strong gravitational
lensing can manifest as visible giant arcs magnifying high redshift
galaxies \citep{lyndsandpetrosian86,gladdersetal03}, multiply imaged
quasars \citep{walshetal79}, multiply imaged galaxies
\citep{sharonetal05}, and arclets \citep{bezecourtetal98}.  Lensing
signatures probe the underlying dark matter distribution of the lens
\citep{warrenanddye03}, and high redshift galaxy formation
\citep{allametal07}. Strong lenses also provide a geometric test of
cosmology via comparison of predicted arc abundances with observed
abundances \citep{kochanek96,chae03,linder04}, via time-delay
between signals from multiply imaged quasars
\citep{suyuetal14,suyuetal16,bonvinetal17}, and via
distance ratios in lenses with sources at multiple redshifts
\citep{julloetal10,collettetal12,collettandauger14}.

The application of strong gravitational lensing to constrain the mass
distribution of strong lenses, such as early-type galaxies (ETGs),
necessitates large samples of galaxy-galaxy strong lensing systems.
\citet{miraldaescudeandlehar92} first suggested that massive
ellipticals would likely be frequent strong lensing sources in optical
surveys.  These systems contain a background source galaxy that the
lens galaxy deflects into a partial or full arc shaped Einstein ring.
The strong lensing signature directly probes the underlying matter.
The identification of such systems in upcoming surveys is the first
step in constraining the mass-to-light ratio for a large number of
objects in this mass range.

Over the last decade, infrastructure for both large scale visual and
automated image classification emerged. By nature, the human eye is
one of the best discriminators for image classification.  Visual arc
identification has been effective through the use of citizen science
platforms.  {\em SpaceWarps} is an example of citizen science based
image classification of strong lensing systems in Canada France Hawaii
Telescope Science (CFHTLS) telescope observations
\citep{marshalletal16,moreetal16}.  These platforms are quite
successful for a dataset like CFHTLS; here, 3,000 candidate images
were identified in eight months, resulting in 89 final candidates.
However, future datasets like Euclid \citep{oguriandmarshall10} and
LSST expect to find hundreds of thousands galaxy-scale strong lenses
\citep{collett15}.  The volume of upcoming data challenges the
scalability of a pure citizen science approach.

Recent efforts have focused on the development of automated methods
with performance comparable to or better than humans. {\em SpaceWarps}
is a part of the Zooniverse Project \citep{marshalletal16}, which
also includes {\em Galaxy Zoo}, the citizen science based image
classification of galaxy types \citep{lintottetal08}.  Galaxy
classification is an early example in which machine
learning algorithms successfully trained models to classify astronomical
images with comparable performance to humans \citep{dielemanetal15}.

Earlier efforts on automated, or ``robot'', identification of strong
lensing systems have two distinct generalized steps.  The first
enhances and extracts characteristic features, and the second uses
some form of pattern recognition in the features to classify lens and
non-lens containing systems.  Among others, selected features might
include shape parameterization
\citep[see][e.g.]{,alard06,kuboanddellantonio08,xuetal16,lee17},
colors in multi-band imaging \citep{maturietal14,gavazzietal14},
light profiles \citep{braultandgavazzi15}, and characteristics of
potential lens galaxies \citep{marshalletal09}.  Pattern recognition
often incorporates cutoffs in selected parameter space distributions
\citep{lenzenetal04,josephetal14,paraficzetal16}.  A number of
these have been publicly distributed, with specific end applications.
For example, {\em ArcFinder} is one such code that finds arcs in
groups or cluster scale lens \citep{seidelandbartelmann07}, and {\em
  RingFinder} is an analogous tool in searching for multiply imaged
quasars \citep{gavazzietal14}.  Codes like these have been
complementary to human identification \citep{moreetal16}.

Pattern recognition methods have transitioned to using ``machine
learning'' algorithms in place of manual cutoffs.  With machine
learning, we can train a model to separate a dataset into a known set
of classes, such as ``lensing systems'' and ``non-lensing
systems''. However, many classic machine learning algorithms do not
work well in image space, i.e. directly using the raw pixel values of
the image, but first require a process of {\em feature engineering}.
{\em Feature engineering} uses domain expertise to extract variables
that are more directly related to the classification task at hand.  An
example would be the application of an edge detector.  The optimal
weights and cutoffs for these derived variables that are used to
determine the class label of an image are found automatically by the
algorithm.

Some more recent works have made use of a subset of machine
learning algorithms called neural networks to classify images,
either from derived image
parameters or directly in image space.  In \citet{estradaetal07},
authors used derived shape parameters to train neural nets to identify
arc candidates.  \citet{agnelloetal15} used neural networks trained
on data from multiband magnitudes, and \citet{bometal16} used
extracted morphological parameters.

The most advanced use of neural networks for strong lensing
classification operate directly in image space.
\citep{petrilloetal17} used mock Kilo Degree Survey (KiDS) data to train
convolutional neural networks, with a training set size of six million
images. \citet{lanusseetal17} used state-of-the-art deep residual
neural networks to also work directly in the image space with minimal
image pre-processing.  A major strength of the \citet{lanusseetal17}
implementation is that in comparison to deep convolutional neural
networks, deep residual neural networks have been found to be easier
to train and perform better on simulated data
\citep{metcalfetal18}.

A lens-finding challenge conducted in 2017 compared results of various
lens-finding methods from several teams \citep{metcalfetal18}.  The
challenge included both ground- and space-based data, respectively
using mock and real KiDS data, and mock Euclid data. Using the area
under the curve (AUC) of the Receiver Operating Characteristic curve
(see Section~\ref{sec:ROC} for more details) as a metric, the
performance for ground-based-like data was as follows.  Neural
networks performed best with a top AUC=0.98
(e.g. \citet{lanusseetal17,jacobsetal17}), human inspection with an
AUC=0.89, and the method we present in this paper at AUC=0.84.  AUC
values closer to 1 indicate better identification of lenses. It is
worth noting, however, that the AUC worsened for all methods when
evaluated on true ground-based data alone.

To date, the most significant challenge has been in translating
lens-finding algorithms to perform well on observed data.
\citet{jacobsetal17} provides the first example of a trained neural
network successfully applied to data from the Canada-France-Hawaii
Telescope Legacy Survey.  The success in application to observations
stems from the creation of training sets that incorporate real survey
galaxies and real survey backgrounds.

While there has been a recent surge in the use of deep neural networks
applied to image classification problems in astronomy, it is not
always easy to scale these techniques to large data sets, nor are the
necessary computational resources and hardware, such as graphical
processing units (GPUs), easily accessible to the entire scientific
community.  We present a first application of the Histogram of
Oriented Gradients (HOG) as a feature extraction method and classify
strong lensing systems with a basic Logistic Regression algorithm
(LR).  HOG is a fast feature extraction procedure that quantifies
edges in images, commonly used to identify humans in security
software.  The authors know of only one other use of HOG in astronomy,
in the recent context of spectral line observations to characterize
atomic and molecular gas \citep{soleretal18}.  LR is a linear
classifier, making its scalability relatively straightforward with
existing open source tools.  This paper also serves as both a
presentation of the mock data set.  We test the methods on mock HST
and LSST data, which will also be made publicly available.  Results
are reproducible on personal computers, as both the pipeline and the
data will be publicly distributed.

We show results of the pipeline on mock galaxy-galaxy lens systems
observed by HST and LSST as respective examples of classifier
performance on optical space- and ground-based observations.  We train
and test our pipeline on subsamples from 10,000 mock observed strong
lensing systems and 10,000 non-strong lensing systems, each centered
on a potential lens galaxy.  We additionally assess model performance
on observed data from The Sloan Lens ACS Survey (SLACS)
\citep{boltonetal08}.  We discuss limitations of our methods in the
context of what simulations are able to capture.

Our paper is organized as follows. In Section~\ref{sec:methods} we
briefly describe the methods to generate the mock HST and LSST data
and our overall image processing and classification pipeline. We
present our results in Section~\ref{sec:results}, and our summary and
discussions in Section~\ref{sec:conclusions}.

\section{Methodology}
\label{sec:methods}
\subsection{Mock Images}
To train and test our model, we create mock observations using two
different codes.  We generate mock {\em Hubble Space
  Telescope}\footnote{https://www.spacetelescope.org/} (HST) with {\em
  PICS} (Pipeline for Images of Cosmological Strong lensing)
\citep{lietal16}.  From PICS, we also have simulated strong lenses
without PSF and noise, on which we then apply {\em LensPop}
  \citep{collett15} to the {\em PICS} generated images to perform
  mock {\em Large Synoptic Survey
    Telescope}\footnote{https://www.lsst.org} (LSST) observations.
  Note, both codes are used because the mock observing module in {\em
    PICS} does not include LSST.  We specifically use {\em LensPop} to
  implement the mock observing for LSST based on the simulated images
  created by PICS.

There are 10,000 lens containing mock observations and 10,000
non-lens containing mock observations for running our parameter search. 
We keep a hold-out set of 1,000 lens containing mock observations
and 1,000 non-lens containing mock observations on which we evaluate
the final trained model.

Mock observations of lensing systems include the lens galaxy, lensed
images of the source galaxy, and galaxies along the line of sight.
Mock observations of non-lensing systems include all but the images of
a lensed source galaxy.  We convolve each of the $2\times 10,000$
train/test images and $2\times 1,000$ hold-out images to produce three
separate sets of mock ``observations''.  These each have equal numbers
of lens and non-lens containing systems: (1) HST-like observations,
(2) best single exposure LSST-like observations, and (3) LSST-like
observations over the span of ten years.  We respectively label these
observations as HST, LSST-best, and LSST10.

\subsubsection{Modeling the Mass Distribution of Lens Galaxies}\label{sec:methods:massmodel}

To produce simulated lensed galaxies, we first model the mass of lens
galaxies as a Singular Isothermal Ellipsoid (SIE). This model is
both analytically tractable and is consistent with models of
individual lenses and lens statistics on the length scales relevant
for strong lensing
\citep[e.g.]{koopmansetal06,gavazzietal07,dyeetal08,liandchen09}.

The normalized convergence map of the SIE model is defined as:
\begin{equation}\label{eqn:sie}
\kappa = \frac{\theta_{\rm E}}{2}\frac{1}{\sqrt{x_1^2/q+x_2^2 q}},
\end{equation}
where $\theta_{\rm E}$ is the Einstein Radius and $q$ is the axis ratio.
The Einstein radius can be calculated from the redshift of the
lens, the redshift of the source, and the velocity dispersion of the
lens galaxy as follows,
\begin{equation}
\theta_{\rm E} = 4\pi\left(\frac{\sigma_v}{c}\right)^2\frac{D_{ls}}{D_{s}}, 
\end{equation}
Here, $c$ is the speed of light, $\sigma_v$ is the velocity dispersion
of the lens galaxy, $D_{ls}$ and $D_{s}$ are respectively the angular
diameter distances from the source plane to the lens plane and from
the source plane to the observer.

To rotate the lenses with random orientation angle $\phi$, we adopt
the transformation below:
\begin{eqnarray}
\label{eqn:rotation}
\begin{bmatrix}
    x_{1}\prime      \\
    x_{2}\prime 
\end{bmatrix}
= 
\begin{bmatrix}
    \cos \phi  & -\sin \phi      \\
    \sin \phi  & ~\cos \phi      
\end{bmatrix} 
\begin{bmatrix}
    x_1      \\
    x_2      
\end{bmatrix}.
\end{eqnarray}

From Equations~\ref{eqn:sie}~-~\ref{eqn:rotation}, there are five
independent parameters from which we can derive the lensing map:
velocity dispersion $\sigma_v$, axis ratio or ellipticity $q$,
orientation angle $\phi$, redshift of the lens $z_l$, and redshift of
the source $z_s$.  

We choose $\sigma_v$, $q$, and $\phi$ randomly (flat prior) from
typical ranges of observed galaxies: $\sigma_v \in [200, 320]~km/s$,
$q \in [0.5, 1.0]$, and $\phi \in [0, 360]$.  While a flat prior is
not realistic, we use this as a starting point to test our pipeline
(see Section~\ref{sec:conclusions} for more details on future
work). We obtain the redshift of the lens galaxy by matching the
velocity dispersion drawn to generate the simulations to the
corresponding redshift from the catalog of elliptical galaxies in the
COSMOS survey\footnote{http://cosmos.astro.caltech.edu/} from
\citet{zahidetal15}.  This catalog contains both the redshift and
  velocity dispersion measurements from BOSS
  spectra\footnote{http://www.sdss3.org/surveys/boss.php} for
  massive early-type galaxies in COSMOS that are analogs to our lens
  galaxies. This results in lens galaxy redshifts in the range, $z_l
\in [0.2,0.7]$.  Note, the distribution of COSMOS galaxies leads to a
selection of brighter and larger objects at higher redshifts.

\subsubsection{Modeling Images of the Lens Galaxies}
We model the light distributions of the lens galaxies with an
elliptical Sersic profile,
\begin{equation}\label{eqn:sersic} 
I(R) = I_{\rm eff}~{\rm exp} \left\lbrace -b_{n} \left[ \left(
  \frac{R}{R_{\rm eff}}\right)^{1/n} - 1 \right ]
\right\rbrace 
\end{equation}
where, $R = \sqrt[]{x_1^2 /q+x_2^2 q }$, $R_{\rm eff}$ is the
effective radius in units of arcseconds, $I_{\rm eff}$ is the
intensity at the effective radius, $n$ is the index of the Sersic
profile, and $q$ is the ellipticity of the lens galaxy.  We perform a
similar transformation as Equation~\ref{eqn:rotation} to orient the
source galaxies and assume that the distribution of light follows that
of mass.  The ellipticity and orientation are therefore the same as in
the SIE model.

To create images of the lens galaxies, {we also utilize the
  catalog to match the velocity dispersion with an assigned effective
  radius and effective luminosity to the light profile. As described
  in Section~\ref{sec:methods:massmodel}, the catalog consists of both
  COSMOS imaging and BOSS spectroscopy for velocity dispersion
  measurements of massive early type galaxies, providing sufficient
  information to construct a fundamental plane of relations between
  all relevant quantities.  We explicitly use the fundamental plane
  from these observations to relate $\sigma_v$ in the lensing map with
  $R_{eff}$ and $I_{eff}$ in Equation~\ref{eqn:sersic}.}

To create images of the lens galaxies, we use the COSMOS morphological
catalog \citep{zahidetal15} to match the velocity dispersion with an
assigned effective radius, effective luminosity, and index to the
light profile.  This catalog uses SDSS/BOSS selection criteria.  Note,
we explicitly use the fundamental plane from these observations to
relate $\sigma_v$ in the lensing map with $R_{\rm eff}$, $I_{\rm
  eff}$, and $n$ in Equation~\ref{eqn:sersic}.

While the galaxy distributions of the COSMOS data at higher redshifts
are biased towards larger galaxies and the highest surface brightness
galaxies for fixed size, this selection should mimic those in surveys
such as the SLACS sample described in Section~\ref{sec:SLACS}.  We
next assume the light center is on top of the mass center and create
noiseless images of lens galaxies.

We construct galaxies along the line of sight by cutting light-cones
from the Hubble Ultra Deep
Field\footnote{http://www.spacetelescope.org/science/deep\_fields/}.
We create a composite of these images with the lens galaxy image and
the lensed source galaxy, calibrating the magnitude of all three
components.

\subsubsection{Modeling Images of the Source Galaxies}
The background source galaxies come from a set of detailed images of
low-redshift bright galaxies ($z\sim0.45$) that have been extracted
from mosaics produced by the CANDELS team
\citep{groginetal11,koekemoeretal11} and selected from the CANDELS
UDS catalog \citep{galametzetal13}.  For these source images, we use
the F606W band, which is the closest filter available in CANDELS data
to the g-band of LSST.  We rescale these background source galaxies to
artificially redshift the sources to $z_s=2$, and select source
positions near caustics of the lensing system.  The rescaling involves
a correction to both the size and magnitude corresponding to the
change in cosmological distances with $z_s=2$.  The rescaling did not
involve a color adjustment, as our mock data is achromatic.

\begin{figure*}[t]
  \centering 
  \mbox{
    \subfigure{\includegraphics[width=.333\columnwidth,trim=115 50 105 25,
        clip]{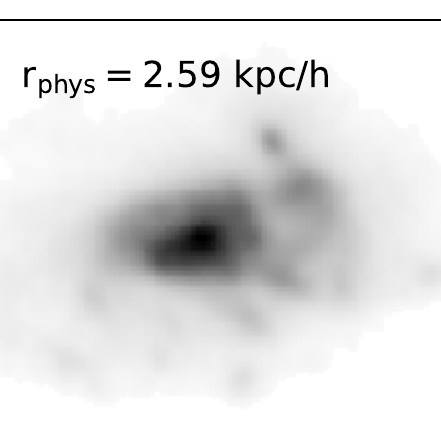}}\hfill
    \subfigure{\includegraphics[width=.333\columnwidth,trim=115 50 105 25,
        clip]{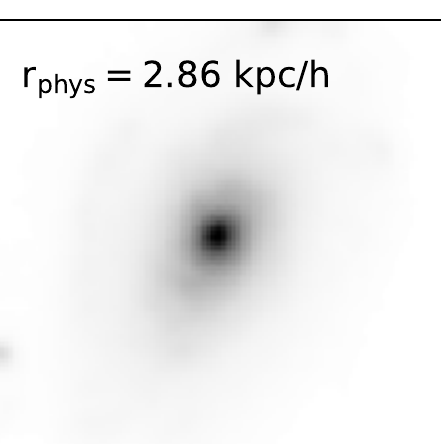}}\hfill
    \subfigure{\includegraphics[width=.333\columnwidth,trim=115 50 105 25,
        clip]{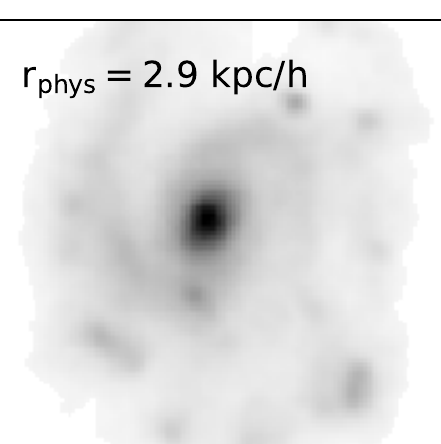}\hfill}
    \subfigure{\includegraphics[width=.333\columnwidth,trim=115 50 105 25,
        clip]{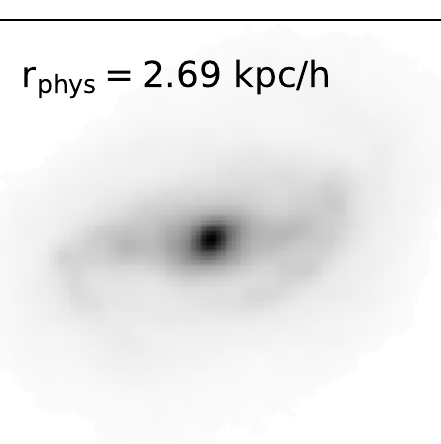}}\hfill
    \subfigure{\includegraphics[width=.333\columnwidth,trim=115 50 105 25,
        clip]{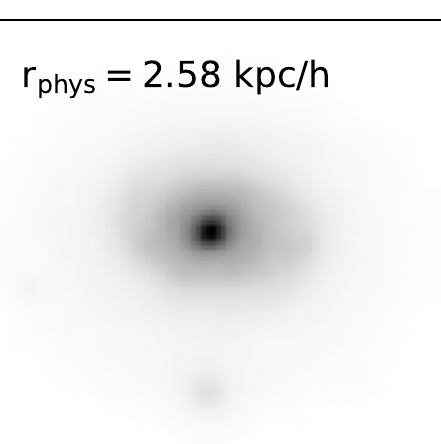}}\hfill
    \subfigure{\includegraphics[width=.333\columnwidth,trim=115 50 105 25,
        clip]{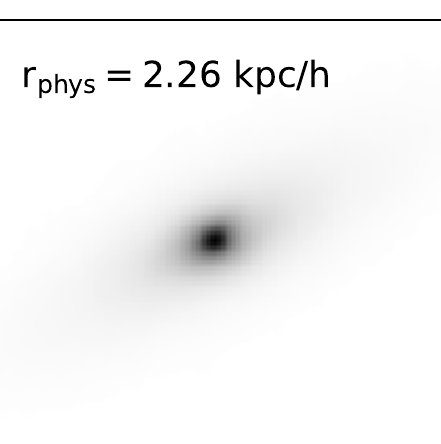}\hfill}
  }\\[-.5cm]
  \mbox{
    \subfigure{\includegraphics[width=.333\columnwidth,trim=115 50 105 25,
        clip]{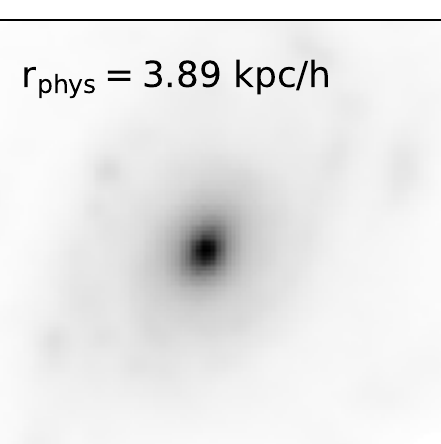}}\hfill
    \subfigure{\includegraphics[width=.333\columnwidth,trim=115 50 105 25,
        clip]{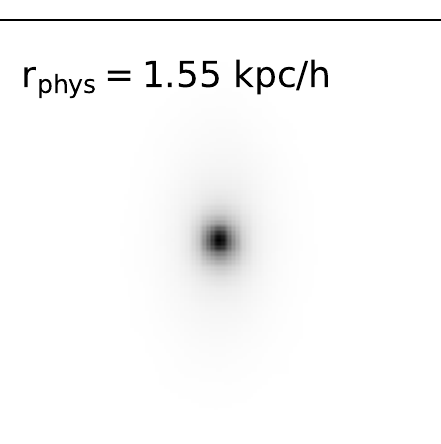}}\hfill
    \subfigure{\includegraphics[width=.333\columnwidth,trim=115 50 105 25,
        clip]{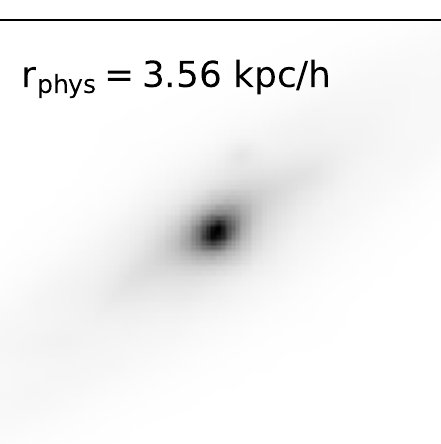}\hfill}
    \subfigure{\includegraphics[width=.333\columnwidth,trim=115 50 105 25,
        clip]{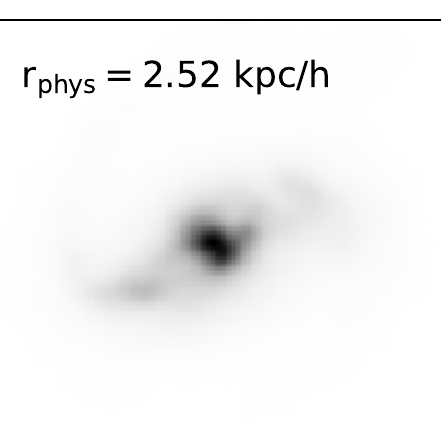}}\hfill
    \subfigure{\includegraphics[width=.333\columnwidth,trim=115 50 105 25,
        clip]{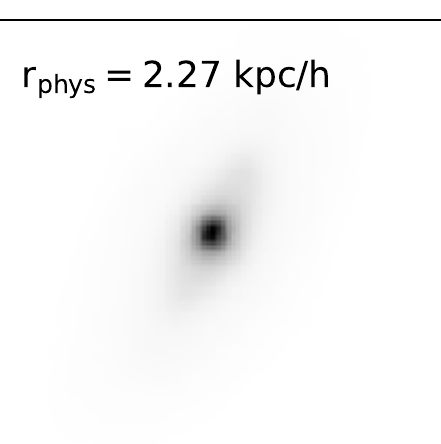}}\hfill
    \subfigure{\includegraphics[width=.333\columnwidth,trim=115 50 105 25,
        clip]{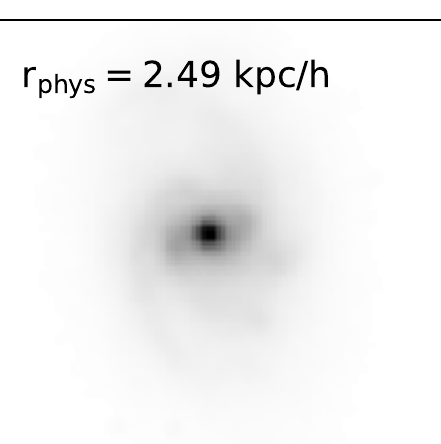}\hfill}
  }\\ [-.5cm]
  \mbox{
    \subfigure{\includegraphics[width=.333\columnwidth,trim=115 50 105 25,
        clip]{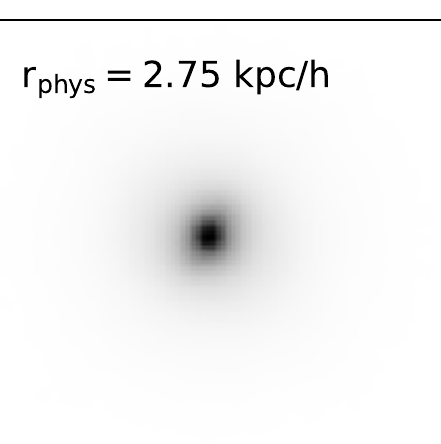}}\hfill
    \subfigure{\includegraphics[width=.333\columnwidth,trim=115 50 105 25,
        clip]{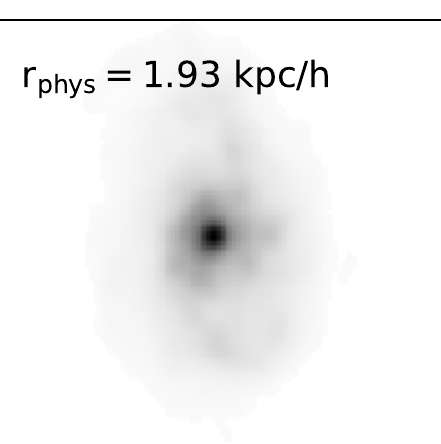}}\hfill
    \subfigure{\includegraphics[width=.333\columnwidth,trim=115 50 105 25,
        clip]{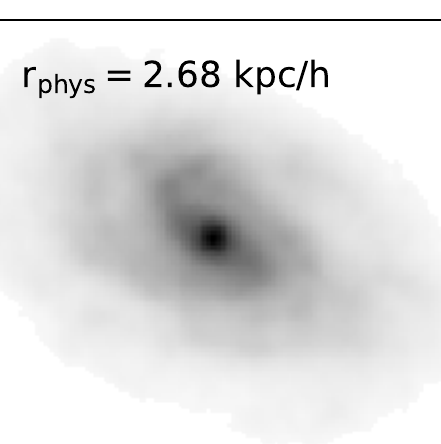}\hfill}
    \subfigure{\includegraphics[width=.333\columnwidth,trim=115 50 105 25,
        clip]{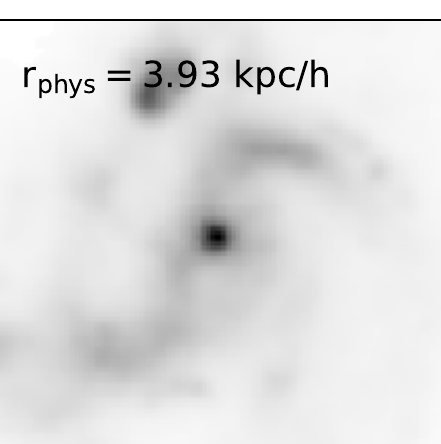}}\hfill
    \subfigure{\includegraphics[width=.333\columnwidth,trim=115 50 105 25,
        clip]{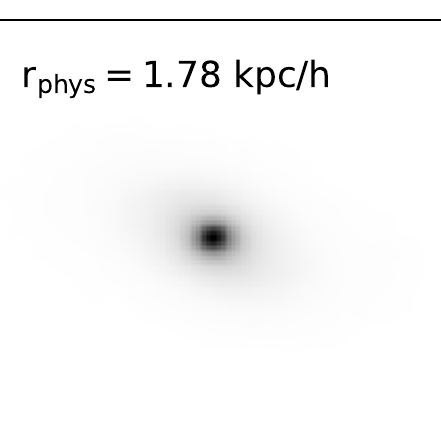}}\hfill
    \subfigure{\includegraphics[width=.333\columnwidth,trim=115 50 105 25,
        clip]{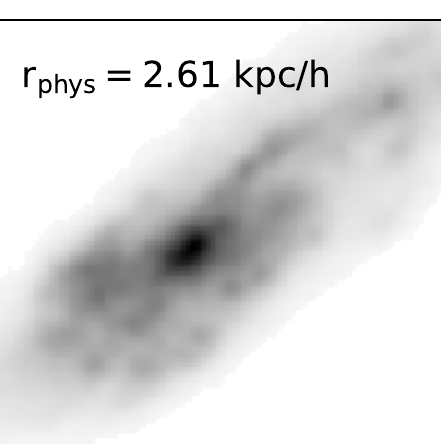}\hfill}
  }\\[-.5cm]
  \mbox{
    \subfigure{\includegraphics[width=.333\columnwidth,trim=115 50 105 25,
        clip]{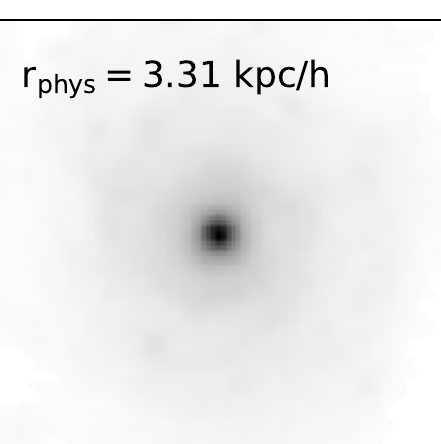}}\hfill
    \subfigure{\includegraphics[width=.333\columnwidth,trim=115 50 105 25,
        clip]{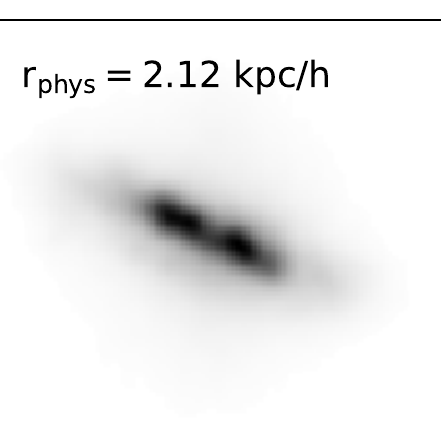}}\hfill
    \subfigure{\includegraphics[width=.333\columnwidth,trim=115 50 105 25,
        clip]{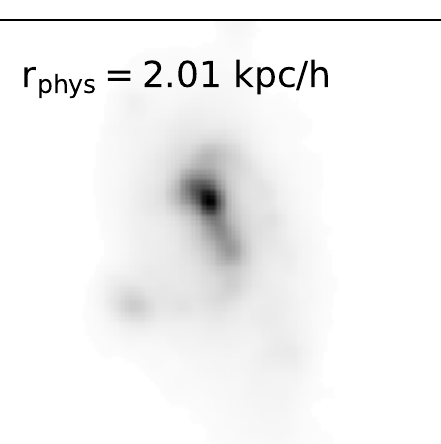}\hfill}
    \subfigure{\includegraphics[width=.333\columnwidth,trim=115 50 105 25,
        clip]{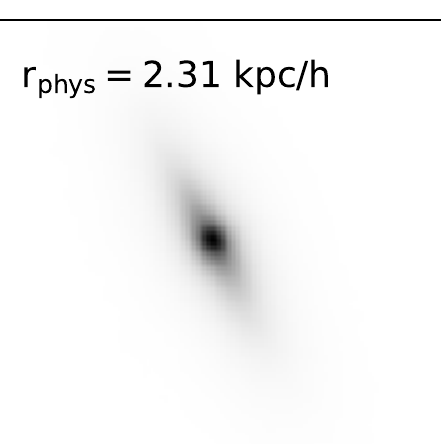}}\hfill
    \subfigure{\includegraphics[width=.333\columnwidth,trim=115 50 105 25,
        clip]{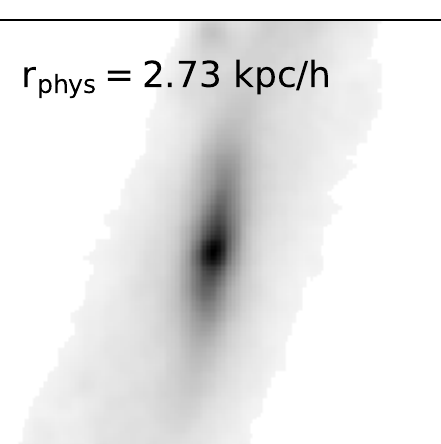}}\hfill
    \subfigure{\includegraphics[width=.333\columnwidth,trim=115 50 105 25,
        clip]{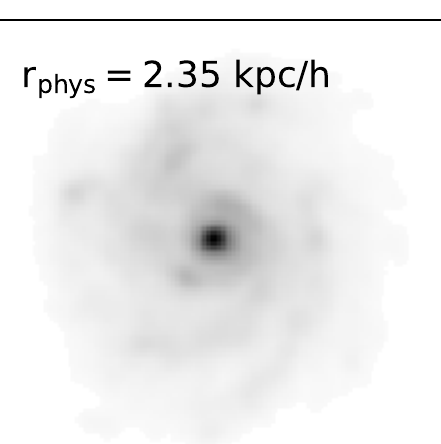}\hfill}
  }\\[-.5cm]
  \mbox{
    \subfigure{\includegraphics[width=.333\columnwidth,trim=115 50 105 25,
        clip]{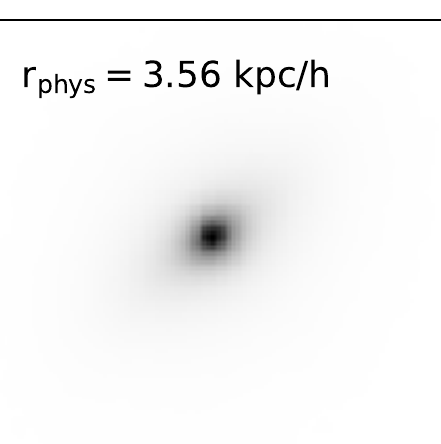}}\hfill
    \subfigure{\includegraphics[width=.333\columnwidth,trim=115 50 105 25,
        clip]{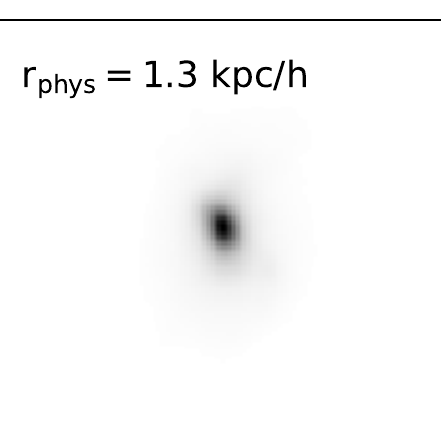}}\hfill
    \subfigure{\includegraphics[width=.333\columnwidth,trim=115 50 105 25,
        clip]{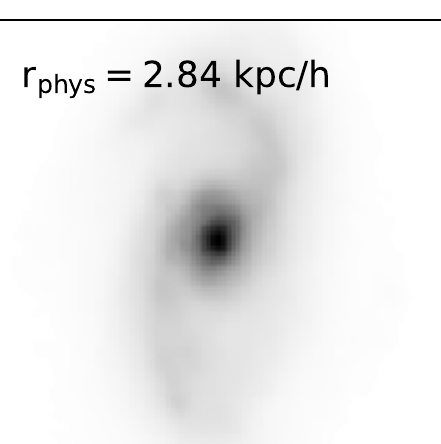}\hfill}
    \subfigure{\includegraphics[width=.333\columnwidth,trim=115 50 105 25,
        clip]{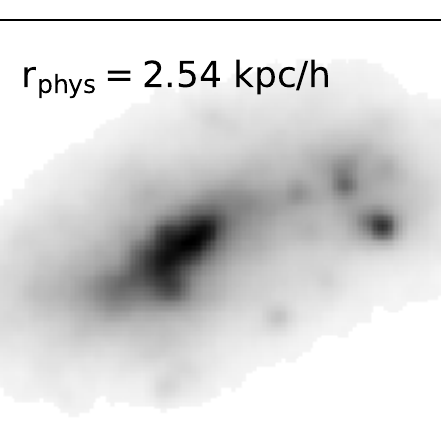}}\hfill
    \subfigure{\includegraphics[width=.333\columnwidth,trim=115 50 105 25,
        clip]{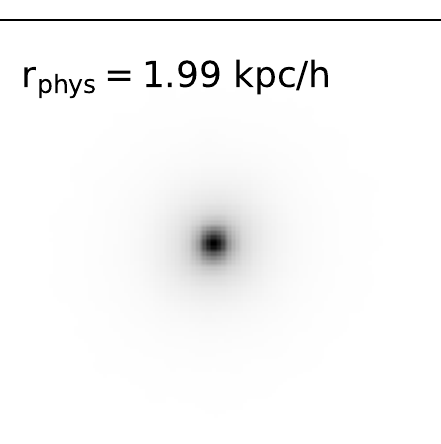}}\hfill
    \subfigure{\includegraphics[width=.333\columnwidth,trim=115 50 105 25,
        clip]{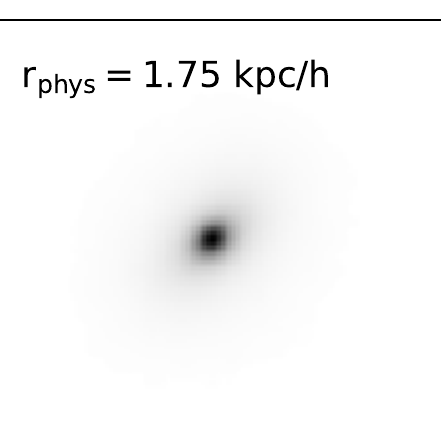}\hfill}
  }\\[-.2cm]
\caption{Images of our source galaxies from the CANDELS catalog.  We
  annotate the physical size of each galaxy when redshifted to $z_s=2$
  in the top left corner of each thumbnail image. At the source
  redshift, each thumbnail is 0.96~arcsec across; the resolution of
  each is 128~pixels$^2$ at 0.0075~arsec/pixel.}\vspace{10pt}
\label{fig:sources}
\end{figure*}

We add the caveat that the selection function of our galaxies is
biased as this sample is not complete.  Here, we select bright large
objects with apparent magnitude $<20$, and effective radius above
3~arcsec.  We define an effective radius as the radius enclosing
pixels with values above 3$\sigma$ of the background noise.  While
this leaves us with only 31 such sources, we leave a full exploration
of the statistical bias of source galaxy distribution in training sets
to future work.  A thumbnail panel of 30 of these images is in
Figure~\ref{fig:sources}, each annotated with the physical size of the
galaxy at $z_s=2.0$.  The physical sizes range from 1.3-3.93~$kpc/h$.

The selected positions are within a $2''\times2''$ square box centered
on the lens galaxy center of mass. We then fix the projected position
of the lens galaxy at the center of the field of view for each image.

Next, we perform ray-tracing simulations using the modeled galaxy
images and parameters.  The lens galaxy, source galaxy, and parameters
of the image simulations (i.e., $300 \times 300~{\rm pixels}^2$ with
0.03~arcsec/pixel for an HST-like image) are inputs to the simulation
that produce ideal lensed images with the appropriate resolution.

\subsubsection{Mock Observing}

The final part is to perform the mock observation from the ideal
images produced in the previous step.  Both our HST- and LSST-like
observations are monochromatic.  The HST-like sample is in the F606W
band and the LSST-like sample is in the g-band.  We note that the use
of a single band limits the potential performance of our models.  We
refer the reader to \citet{metcalfetal18} for an example of our model
trained on multi-band data where we concatenated the feature vector
from each band.

We create a composite image of the lens galaxy, source image, and
galaxies along the line of sight.  While we adjust the source size and
magnitude to correspond to the change $D_s$ when placed at $z_s$, we
do not adjust colors.  The distribution of lens properties is not
realistic with a constant $z_s$, but provides a sufficient start in
covering the feature space to test supervised classification methods.
Incorporating the lens, source image, line of sight galaxies, and
noise are particularly crucial in methods that use edge features.

For HST-like observations, we do the following for each component.
The component that mimics along the line of sight galaxies is a cutout
from the HUDF.  The inclusion of this cutout results in an image where
noise and the point spread function (PSF) for HST is in the field of
view.  The lens galaxies have been convolved with the HST PSF, but
their angular extent is significantly larger than the PSF of
$\sim0.03''$.  Convolution of the lens galaxy component will not
noticeably alter its appearance.  The source images are from a ray
traced CANDELS galaxy observed with HST PSF.  This procedure does not
capture the true clumpiness of these sources.  We then create a
composite image and magnitude calibrate all components to produce the
final HST-like images, which are $300\times300~{\rm pix}^2$ with
$0.03$~arcsec per pixel.

Note, we do not convolve the lensed image with the HST PSF in a final
step after rescaling and ray-tracing.  The original source image is a
real HST galaxy that already has the HST PSF.  We then magnify the
source galaxy via the lensing procedure.  It does not then make sense
to perform an additional final convolution that would erase the
clumpiness potentially captured in HST lensed arcs.  Note, the
observation of a true lensing effect would have a better PSF providing
finer details than in the unlensed HST images that we have used.

For LSST-like observations, we use the {\em LensPop} software
\citep{collett15}.  We resample images to match the detector pixel
scale and convolve the resampled image with a circularly symmetric
Gaussian Point Spread Function discretised at the same pixel scale. To
generate a noisy realization of the image, we assume a Poisson model
based on the sky plus signal, and an additional Gaussian read-out
noise. Parameters for these simulations follow \citet{collett15} and
are based on the LSST observation simulator \citep{connollyetal10}.

To account for variations in seeing and sky-brightness over the course
of the survey, we draw each simulated exposure from a stochastic
distribution of these parameters. We then consider two different
strategies to use the simulated exposures. First, we build one
single-epoch image for each field (hereafter labeled as LSST-best) by
keeping only the best seeing exposure.  Second, we build another
``worst-case'' stacked image by degrading all individual exposures,
ten per filter per year, to match the one with the worst seeing and
co-add all exposures to a single image (hereafter labeled as LSST10).
These two sets of images will allow us to investigate the trade-off
between resolution and signal-to-noise for our automated lens search.

\begin{figure*}[t]
\begin{center}
\includegraphics[width=2\columnwidth]{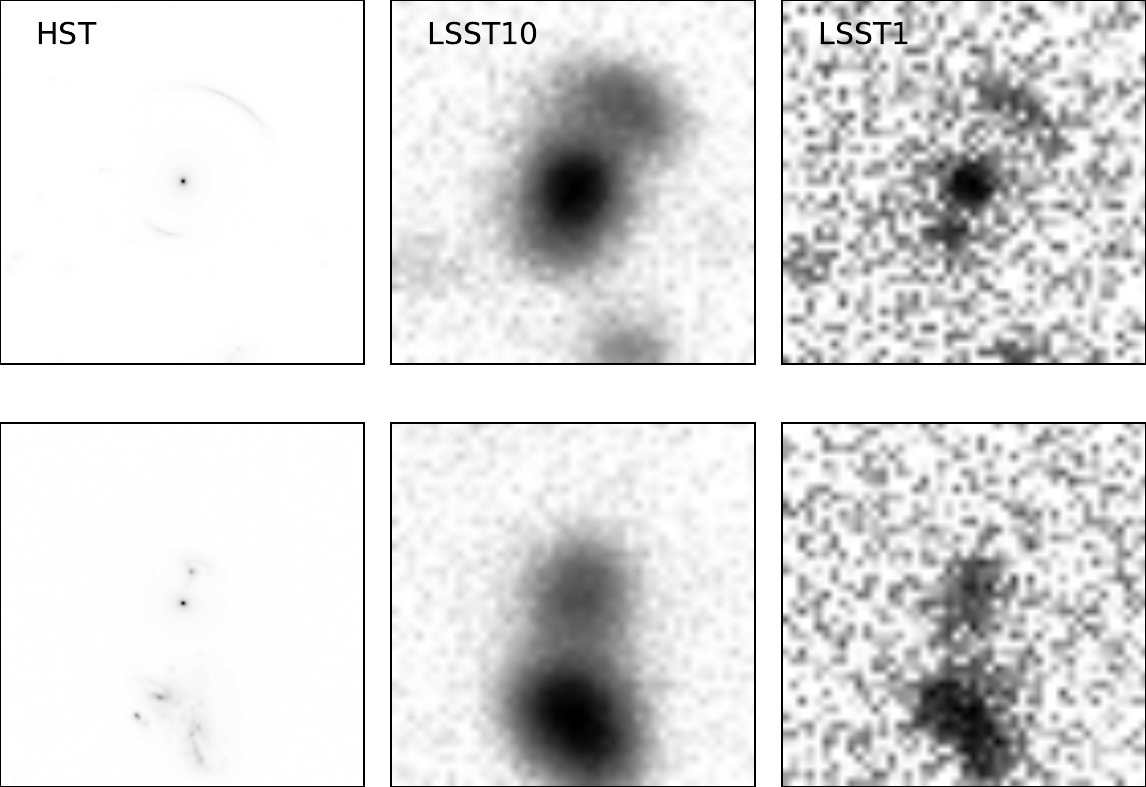}
\caption{Left to right show example mock HST, LSST 10 year, and
  LSST-best images.  The top row corresponds to a lensing system with
  a very visible arc signature, and the bottom row to a lensing system
  that is less obvious.  Example mock HST images have
  $n_\text{pix}\times n_\text{pix}=300\times300$.  Example mock LSST
  images have $n_\text{pix}\times n_\text{pix}=45\times45$.  The
  resolution and noise of a ground based telescope is noticeably
  worse. Visual identification of giant arcs in the LSST images in the
  bottom row is very difficult.}\label{fig:mockimages}
\end{center}
\end{figure*}

Figure~\ref{fig:mockimages} illustrates sample mock observations with
a strong lensing signature from each telescope.  The left-most column
corresponds to a mock HST lensing system with a highly magnified
source galaxy (top) and a less visible image of the source galaxy
(bottom). For the HST-like dataset, many arcs are visually obvious due
to the exquisite spatial resolution and quality of space based
imaging.

The middle column corresponds to the same simulated systems for
LSST10.  The right-most column corresponds to the simulated systems of
LSST-best.  These images have resolution $45\times45~{\rm pixels}^2$
with $0.18$~arcsec/pixel.  LSST10 images visually exhibit the improved
signal to noise ratio, recovering the arc feature, albeit at a much
lower resolution than with the HST-like image or the LSST-best image.
The top images of LSST10 and LSST-best show a visible lensed source
galaxy image.  The bottom images are washed out in the bottom row,
where the magnification of the source galaxy is not as large.  The
ground based noise, PSF, and limited resolution of the LSST-best make
visual giant arc identification difficult, except in systems with the
most magnified source galaxies.

Our mock observations also include non-lens containing images.  The
procedure is similar to mock lensed images but we do not perform
ray-tracing, so these images do not have lensed source galaxies.  

Furthermore, we investigate the influence of light from the lens
galaxies on the performance of our lens identification pipeline.  We
generate another set of each HST, LSST-best, and LSST10 images without the
lens galaxy.  We respectively label these nHST, nLSST-best, and nLSST10.

The final data set is then comprised of $6\times10,000$ lens
containing images, and $6\times 10,000$ non-lens containing images.
We also keep a hold-out set of $6\times 1,000$ lens containing images,
and $6\times 1,000$ non-lens containing images.

\subsection{Strong Arc Lensing Identification Pipeline}

To perform our analysis, we have used tools from {\em Scikit-learn}
\citep{pedregosaetal12} to identify galaxy-galaxy strong lensing
systems through supervised classification.  Supervised classification
is a class of machine learning where the class labels in the training
set are known.  In our case, the labels are ``lens'' and ``non-lens''
containing images.

The first step of our pipeline consists of a feature extraction stage,
where our feature vector is a {\em histogram of oriented gradients}
(HOG) \citep{dalalandtriggs05} that quantifies edges in the image.  We
describe the method and parameter search in Section~\ref{sec:hog}.  We
then use {\em Logistic Regression} (LR), a machine learning algorithm
described in Section~\ref{sec:LR}, to train a classifier model on a
subset of our images.  LR requires a parameter search over the
regression coefficient, $C_{LogReg}$, which we explore in
Section~\ref{sec:regularization}.  We briefly comment that our initial
tests with a {\em Support Vector Machine} (SVM) using radial basis
functions as an alternative
machine learning algorithm yielded negligible performance improvement,
and significantly increased computation cost.  This indicated that the
features of lens and non-lens images are relatively well separated by
hyperplanes in feature space.  For these reasons, we do not include
SVM in our final analysis and comparisons and continue all discussions
with a linear classifier.

Both the feature extrator, HOG, and the linear classifier, LR, contain
parameters, which must be tested and optimized for peak model
performance.  We use {\em GridSearchCV} from {\em Scikit-learn} to
select cross-validated parameters, and discuss this step of our
methodology in Section~\ref{sec:gridsearch}.

The second step of our analysis is to test our trained model on an
independent subset of the images to assess the model performance.
Here, we evaluate the model on each test image, predicting a
likelihood (``score'') between 0 and 1 that image contains a lensing
system.  This ``holdout set'' is not used in any of our parameter
searches to keep our test metric independent of tuning.

\subsubsection{Feature Extractor: Histogram Oriented Gradients}\label{sec:hog}
Originally created for human detection in computer vision, histogram of
oriented gradients (HOG) is a feature extraction method that computes
centered horizontal and vertical gradients. HOG is relatively robust
to noise in the image, and is a fairly fast transform that describes
edges.  Details can be found in \citet{dalalandtriggs05}, but we
describe the procedure here.  The end result of HOG is a one
dimensional histogram computed as follows.

HOG first divides the image into blocks of 50\% overlap.  Each block
contains $m\times m$ cells-per-block that each contain $n_{pix}\times
n_{pix}$ pixels-per-cell.  The computed gradient orientation is
quantized into $N_{orient}$ bins.  Each gradient is computed using a
$[-1,0,1]$ and $[-1,0,1]^T$ filter kernel to provide the x and y
components of the gradient.

The orientation gradient of all pixels within each cell are binned
into the quantized orientations, providing a net gradient description
within that cell.  As an example, for $N_{orient}=3$, our bins are
centered at $\theta=0, 2\pi/3, 4\pi/3$ in radians.  If a cell
only has a gradient in the
$\theta=\pi/2$ direction, it will contribute 75\% of its magnitude to the
$\theta=2\pi/3$ bin, and 25\% of its magnitude to the $\theta=0$ bin.
The bins in all cells are then concatenated to make a larger feature
vector that is $N_{orient}\times N_{cells}$.

The last step is a normalization procedure to control for illumination
effects.  Here, the sub-histograms of each cell within the same block
are normalized with respect to one another before the transformation
returns the final feature vector.  The division of the image and the
quantization of orientations are thus controlled by 3 parameters in HOG:
$N_{orient}$, cells-per-block, and pixels-per-cell.  We discuss how we
select parameters using cross-validation in
Section~\ref{sec:gridsearch}

\subsubsection{Optimized Pipeline Parameters with a Grid Search}\label{sec:gridsearch}

\begin{table}
\caption{Grid Search of Pipeline Parameters}
\begin{center}
\begin{tabular}{ccccccc}
\hline \\ [-0.2cm]
N$_{orient}$ & Pixels/Cell & Cells/Block & N$_{feat}$ & C$_{LogReg}$ & Score &\\ 
\hline \\ [-0.1cm]
\multicolumn{6}{c}{(a) HST-like data} \\ [0.01cm]
\hline \\ [-0.2cm]
9 & (8, 8) & (4, 4) & 166464 & 10 & $0.8764\pm0.0064$ &\\ [0.01cm]
9 & (16, 16) & (4, 4) & 32400 & 10 & {\bf0.9014$\pm$0.0066} &\\ [0.01cm]
9 & (24, 24) & (4, 4) & 11664 & 10 & $0.8939\pm0.0084$ &\\ [0.01cm]
9 & (32, 32) & (3, 3) & 3969 & 50 & $0.8764\pm0.0105$ &\\ [0.01cm]
5 & (16, 16) & (4, 4) & 18000 & 10 & $0.8945\pm0.0084$ &\\ [0.01cm]
7 & (16, 16) & (4, 4) & 25200 & 10 & $0.9003\pm0.0096$ &\\ [0.01cm]
5 & (8, 8) & (1, 1) & 6845 & 50 & $0.8456\pm0.0080$ &\\ [0.01cm]
5 & (16, 16) & (1, 1) & 1620 & 50 & $0.8381\pm0.0113$ &\\ [0.01cm]
5 & (24, 24) & (1, 1) & 720 & 50 & $0.8594\pm0.0094$ &\\ [0.01cm]
5 & (32, 32) & (1, 1) & 405 & 50 & $0.8581\pm0.0111$ &\\ [0.01cm]
5 & (16, 16) & (1, 1) & 1620 & 50 & $0.8381\pm0.0113$ &\\ [0.01cm]
7 & (16, 16) & (1, 1) & 2268 & 50 & $0.8488\pm0.0092$ &\\ [0.01cm]
9 & (8, 8) & (1, 1) & 12321 & 50 & $0.8539\pm0.0084$ &\\ [0.01cm] 
\hline \\ [-0.2cm]
\multicolumn{6}{c}{(b) LSST-like data (LSST10)} \\ [0.01cm]
\hline \\ [-0.2cm]
4 & (3, 3) & (3, 3) & 6084 & 50 & $0.6155\pm0.0049$ &\\ [0.01cm]
3 & (4, 4) & (3, 3) & 2187 & 100 & {\bf0.6680$\pm$0.0089} &\\ [0.01cm]
4 & (5, 5) & (3, 3) & 1764 & 50 & $0.6472\pm0.0039$ &\\ [0.01cm]
4 & (7, 7) & (3, 3) & 576 & 100 & $0.6512\pm0.0127$ &\\ [0.01cm]
4 & (4, 4) & (3, 3) & 2916 & 100 & $0.6583\pm0.0031$ &\\ [0.01cm]
6 & (4, 4) & (3, 3) & 4374 & 50 & $0.6506\pm0.0150$ &\\ [0.01cm]
9 & (7, 7) & (3, 3) & 1296 & 100 & $0.6405\pm0.0065$ &\\ [0.01cm]
9 & (3, 3) & (1, 1) & 2025 & 100 & $0.5400\pm0.0074$ &\\ [0.01cm]
4 & (4, 4) & (1, 1) & 484 & 100 & $0.5867\pm0.0175$ &\\ [0.01cm]
4 & (5, 5) & (1, 1) & 324 & 50 & $0.5800\pm0.0044$ &\\ [0.01cm]
6 & (7, 7) & (1, 1) & 216 & 100 & $0.5936\pm0.0097$ &\\ [0.01cm]
3 & (4, 4) & (2, 2) & 1200 & 500 & $0.6567\pm0.0082$ &\\ [0.01cm]
4 & (2, 2) & (1, 1) & 1936 & 50 & $0.5597\pm0.0045$ &\\ [0.01cm]
6 & (3, 3) & (1, 1) & 1350 & 50 & $0.5525\pm0.0032$ &\\ [0.01cm]
\hline
\end{tabular}
\end{center}
\label{tab:gridsearch}

\tablecomments{Panel (a) shows a subsample of the results of a grid
  search for HST across a range of HOG parameters, feature vector
  length, and regularization parameter for Logistic Regression,
  C$_{LogReg}$, from Equation~\ref{eqn:log_reg_regularized}. Panel (b)
  shows a subsample of the results of a grid search for LSST10. Each
  use a data set of size $2\times8000$ for cross validation to get the
  average scores and standard deviation.  We explore different HOG
  parameters in each dataset due to resolution and image size
  differences.  We highlight the best performance from the grid search
  in bold.}
\end{table}

We run a grid search across parameters that should reasonably sample
the arc edges in either the HST- or LSST-like mock observations, and
illustrate the results in Table~\ref{tab:gridsearch}. The grid search
procedure uses a 3-fold cross validation to help choose the best
parameters for the different simulated datasets.  Here, the 3-fold
cross validation consists of splitting the data into three parts; we
train on two of the three parts and test on the third to get a score
(accuracy), and rotate in which two are trained vs tested.  This is
how we derived an errorbar on the grid search results in Table 1, the
score (or the accuracy, which is the fraction of examples correctly
classified) with threshold for classification of 0.5.

Recall, the HST-like images are $300\times300$ pixels per image, while
the LSST-like mock observations are $45\times45$ pixels per image.

We first estimate the size of a cell that will contain a coherent arc
feature.  To first order approximation, subdivisions of cells that are
1/100th the area of the entire image should contain coherent arc edges
that span an elongated shape within arc-containing cells.  Therefore,
we sample the pixels per cell parameter from (8, 8) to (32, 32) for
the HST-like images, and (3, 3) to (7, 7) for the LSST-like
images in our grid search.

Next, the cells per block parameter determines the normalization of
each cell with respect to the neighboring cell.  In general, this will
down-weight arc-like edges in cells that neighbor very bright cells,
such as cells that cover the central lens galaxy.  We therefore vary
the cells per block parameter between (2, 2) and (4, 4) for the
LSST-like images and between (1, 1) and (4, 4) for the HST-like
images.

\begin{figure*}[t]
\begin{center}
\includegraphics[width=2\columnwidth]{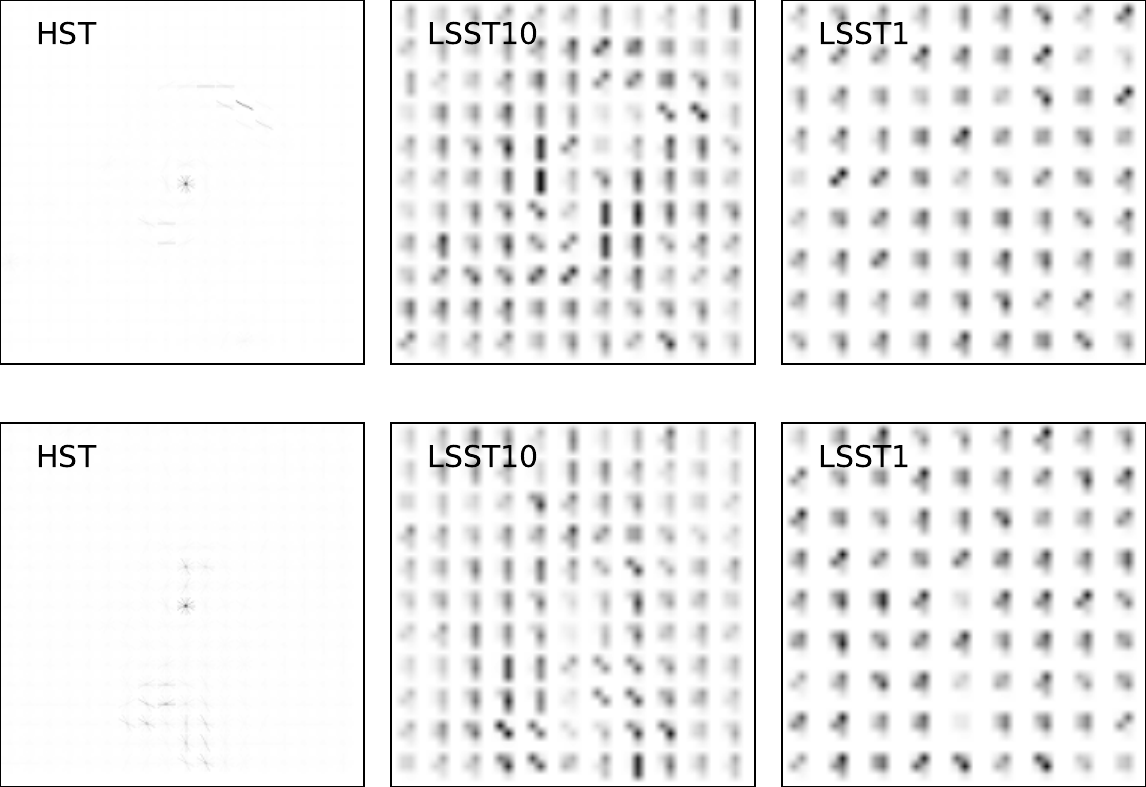}
\caption{Left to right show example image transforms of mock images
  from Figure~\ref{fig:mockimages} with a visualized histogram of
  oriented gradients.  The image transform picks up edge features,
  with arc features showing up as edges across radial orientations.
  Each of the oriented gradients within a cell is color-coded by
  magnitude, and represented as a line in the direction perpendicular
  to that gradient.  The actual extracted features fill a
  one-dimensional feature vector comprised of the magnitudes of each
  of the oriented edges within the visualized
  cells.}\label{fig:hogimages}
\end{center}
\end{figure*}

The number of orientations will determine the sampling of rounded
edges.  For example, if we only have two orientations, an arc-like
feature in a cell directly north of the lensing galaxy will appear in
our HOG visualization as a strong horizontal line (e.g. see top left
in Figure~\ref{fig:hogimages}), and an arc-like feature north-east of
the lensing galaxy will appear as an L-shape.  However, contributions
from a cluster or line-of-sight galaxy in the same cell will tend to
contribute edges in all orientations of the histogram (e.g. bottom
right in Figure~\ref{fig:hogimages}).  

Finally, the resolution of the overall image will also limit the
additional information that an increase in $N_\text{orient}$ will
provide.  From the grid search, the best case number of orientations
for each dataset is $N_\text{orient,HST}=9$,
$N_\text{orient,LSST10}=3$, and $N_\text{orient,LSST-best}=5$.

The image resolution affects the length of the HOG feature vector,
which has a monotonically increasing relationship with the time
required to train the model.  Additionally, for fixed memory
restrictions, there is a tradeoff between the length of the feature
vector and the size of the training set.  We will discuss how the
training set size affects the train time for each data set in
Section~\ref{sec:trainsetsize}.

\subsubsection{Machine Learning Algorithm: Logistic Regression}\label{sec:LR}
The problem of detecting gravitational lenses in images falls under
the general category of {\em{classification}} in machine learning. In
general, the task is to find a function that assigns data points $x$ to
one of two or more classes, denoted by the class label $y$.  This is
equivalent to specifying a decision boundary, or decision boundaries
between the classes in the space of the data points.  (Compare this to
{\em{regression}} in which the task is to find a function $y=f(x)$,
where $y$ is a continuous, rather than discrete, variable.) In our
case, we have two classes: lens and non-lens containing images, and
the data points $x$ are the HOG feature vectors extracted from the
images.  In this paper we use the {\em Logisitic Regression}
(LR) algorithm, for which the decision boundary is a hyperplane.  (The
equivalent in the regression setting would be linear regression.) In
LR we determine the optimal hyperplane by minimizing the objective
function
\begin{equation}
\label{eqn:log_reg_objective}
L(A,b) = \sum_i \log \left[1 + \exp \left(-y_i (A \cdot x_i + b) \right) \right]
\end{equation}
where $x_i$ is a data point (HOG feature vector), $y_i$ is the
known label for that data point ($1$ for a lens containing
image, $-1$ for a non-lens containing), and $A$ and $b$ are
the parameters of the hyperplane. Eq. \ref{eqn:log_reg_objective}
is to be minimized with respect to $A$ and $b$.
Other more complicated machine learning algorithms exist which
do not necessarily produce a linear decision boundary, such as
{\em Support Vector Machines} (SVM), {\em Random Forests}, and
{\em Neural Networks} \citep{hastie09}.

The HOG feature vectors in this paper can be very high-dimensional.
When dealing with high-dimensional data, where the number of
dimensions becomes comparable to the number of data points,
overfitting can become an issue. (An example of an extreme case
of overfitting is fitting a degree $n$ polynomial to $n$
points.  The polynomial would simply wiggle so that it goes
through every point, and would have no predictive power if you
tried to interpolate or extrapolate.) In machine learning,
overfitting is avoided using different {\em{regularization}}
techniques. A common choice for logistic regression is to
add a penalty term to Equation~\ref{eqn:log_reg_objective}
\begin{equation}\label{eqn:log_reg_regularized}
L_\text{Reg}(A,b,C_\text{LogReg}) = L(A,b) + \frac{1}{2C_\text{LogReg}} \|A\|
\end{equation}
where the norm $\|A\|$ is typically taken as either the
sum of squares of the coefficients ($L_2$ norm) or the sum
of absolute values of the coefficients ($L_1$ norm).  In this
paper we use the former.

The amount of regularization is controlled by the parameter
$C_\text{LogReg}$: larger values of $C_\text{LogReg}$ correspond to
increasing model complexity.  If $C_\text{LogReg}$ is too large then
the model will overfit, and if it is too small the model will
underfit.  To determine whether a model is overfit or underfit, the
model is trained, (i.e. Eq. \ref{eqn:log_reg_regularized} is
minimized), on a subset of the data called the {\em{training set}} and
its performance (goodness of fit) is evaluated both on the
training set and a separate {\em{test set}}
that was not used in constructing the model.
Figure~\ref{fig:regularization} shows the performance of a model with
selected HOG parameters as a function of the regularization parameter,
$C_\text{LogReg}$, for both the training and the test set.  (The
performance can be measured by the accuracy, i.e. percent of images
correctly classified, or by some other metric, such as the area under
the Receiver Operating Characteristic curve described in
Section~\ref{sec:ROC}.)

When $C_\text{LogReg}$ is small, the performance of the model improves
with increasing $C_\text{LogReg}$ on both the training and test set,
meaning that $C_\text{LogReg}$ is
still in the underfitting regime. Eventually the performance on the
test set reaches a maximum and starts to decrease, even while the
performance on the training set continues to increase.  This means
that the model is no longer generalizing well and is starting to
overfit. The optimal $C_\text{LogReg}$ occurs when the performance
of the test set is at its maximum; this is the value of $C_\text{LogReg}$
that should be used in the final model.

In practice, there is something of a trade-off between accuracy and
computational resources because a larger value of $C_\text{LogReg}$
will also increase the training time, since a larger $C_\text{LogReg}$
corresponds to a less constrained parameter space being searched.
We discuss the performance and training time dependence on $C_\text{LogReg}$
in Section\ref{sec:regularization}.

\section{Results}
\label{sec:results}

We show results of the HOG and LR generated models using our mock HST
and LSST data sets. For each of these, we also explore the performance
of our models trained on mock data in absence of the lens galaxy as an
idealized test of perfectly modeling out the lens.  The data with
removed lens are labeled as nHST, nLSST-best, and nLSST-10.  In the
final subsection of our resuls, we also examine the performance of
models trained on mock HST data, and tested on real observed data from
the The Sloan Lens ACS Survey (SLACS) \citep{boltonetal08}.

\subsection{Receiver Operating Characteristic}\label{sec:ROC}

\begin{figure}[t]
\begin{center}
\includegraphics[width=\columnwidth]{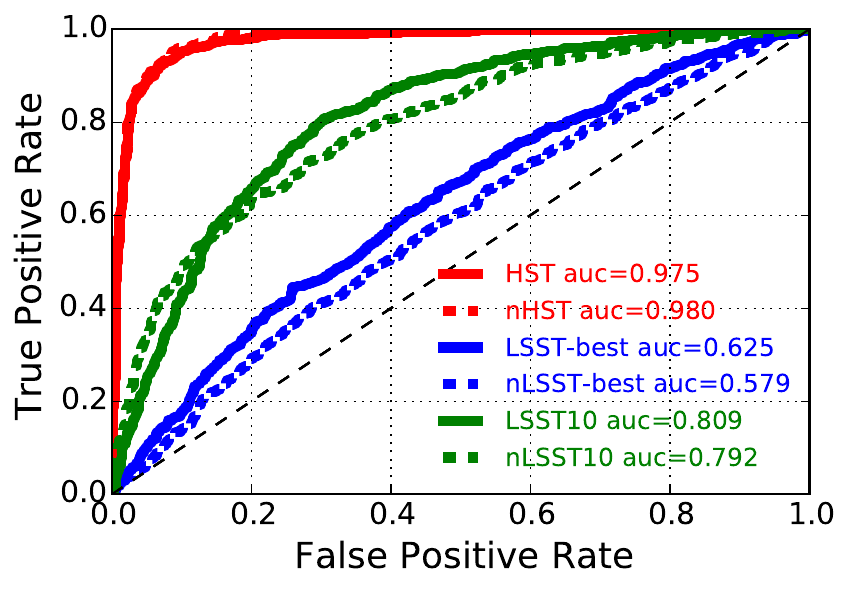}
\caption{Red, blue, green: ROC curves for models trained on our whole
  10,000 training set and tested on our holdout set of 1,000.  These
  respectively correspond to the HST, LSST 1 year, and LSST 10 year
  data.  The solid lines are for data that include the lensing central
  galaxy, and the dashed lines for the data where there is no lensing
  galaxy, mimicking an ideal removal of the lens.  Model performance
  can be summarized by the area under the curve (AUC), labeled in the
  legend. $AUC=1$ is a perfect model, and $AUC=0.5$ is a useless
  model.}\label{fig:ROCcompilation}
\end{center}
\end{figure}

In this section, we discuss the Receiver Operating Characteristic
(ROC) curve (see Figure~\ref{fig:ROCcompilation}), which shows the
true positive rate (tpr) as a function of false positive rate (fpr)
for a given model and test set.  The true positive rate is defined as
the number of lenses correctly identified as positive divided by the
total number of real lenses.  The false positive rate is defined as
the number of non-lenses incorrectly identified as positive divided by
the total number of non-lenses.  The ROC curve illustrates the
performance of our trained model as we vary the discrimination
threshold.

The classifier model assigns a score to each test image, which is a
probability that the image is a strong lensing system.  To construct
the ROC curve, we rank the test images by probability, and calculate
the tpr and fpr for decreasing discrimination threshold.

Higher discrimination thresholds correspond to higher true positive
rates, but will have more false negatives (bottom left region of the
ROC).  For a very low discrimination threshold, we have fewer false
negatives but more false positives (top right region of the ROC).  The
ideal model would have an ROC curve with data points that go from $(x,
y) = (0, 0)$ to $(x, y) = (0, 1)$ to $(x, y) = (1, 1)$.

In the context of strong lensing systems, we wish to maximize the true
positive rate so we have a representative count of the fraction of
strong lensing systems in an observed volume of the universe.  We also
want to minimize the false positive rate.  Positively identified
strong lensing systems will require expensive spectroscopic follow-up
for validation.  The steepness of the ROC curve indicates how well the
model will optimize the two. One way to characterize the performance
of a model is with the area under the curve (AUC).  The ideal model
would have an AUC of 1.  We show the ROC curves of our best performing
models in each dataset.

Figure~\ref{fig:ROCcompilation} shows the ROC curves for models
trained using the entire 10,000 training sample, with best-case HOG
and regularization parameters.  The models have been evaluated on a
hold-out set of 1,000 images that were not used in the parameter
search.  We show the mock HST, LSST-best, and LSST10 results
respectively in red, blue, and green.  Solid lines correspond to a
model trained and tested on images with the lensing galaxy.  Dashed
lines correspond to a model trained and tested on images where the
lensing galaxy is excluded from the mock observation, simulating an
ideal modeling and subtraction of the lensing galaxy, which has been
one proposed method to improve the identification of strong lensing
systems.  The corresponding AUC is listed in the legend.

The model performance for the mock HST data is AUC=0.975 for images
with the lens galaxy, and AUC=0.98 for images without the lens galaxy
(red solid and dashed).  On the other hand, the model for our
LSST-like dataset for one year has an AUC=0.625 with the lens galaxy
and AUC=0.579 without the lens galaxy (blue solid and dashed), and the
model for our LSST-like dataset for 10 years has an AUC=0.809 with the
lens galaxy and AUC=0.792 without the lens galaxy (green solid and
dashed).  Removal of the lens galaxy does not systematically perform
better, and is actually dependent on the size of the training set.  We
discuss relative model performance and complexity for images with and
without the lens galaxy in Section~\ref{sec:trainsetsize}.  

While the ROC curve is a standard metric for supervised classification
in the machine learning community, we note that it does not fully
capture the practicality of an algorithm since it is measured for a
dataset with equal lenses and non-lenses.  This is not the true ratio
between the two classes in the classification.  To complement the ROC
curve, we also discuss the Precision-Recall (PR) curve. The recall
axis is the same as the true positive rate axis in the ROC curve
(number of positively identified lenses divided by the number of real
lenses), also called the completeness of a sample. Precision is the
number of lenses correctly identified as positive divided by the total
number of positive identifications, also known as the purity of a
sample.

\begin{figure}[t]
\begin{center}
\includegraphics[width=\columnwidth]{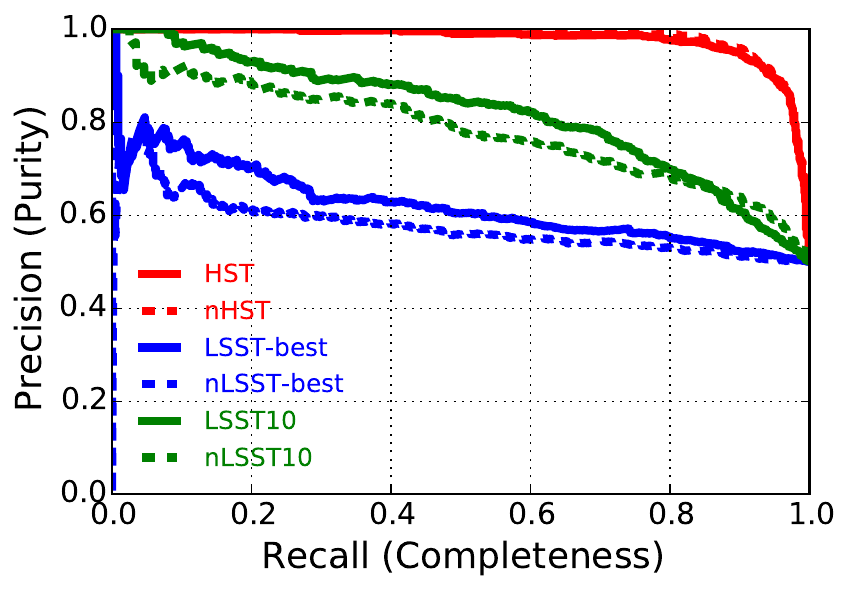}
\caption{Red, blue, green: Precision-recall (PR) curves for models
  trained on our whole 10,000 training set and tested on our holdout
  set of 1,000.  These respectively correspond to the same models and
  data shown in Figure~\ref{fig:ROCcompilation}.  An ideal model would
  reach both a precision (purity) and recall (completeness) that equal
  1.  Note, that this performance describes a data set with a 50-50
  split between lens and non-lens containing
  images.}\label{fig:PRcompilation}
\end{center}
\end{figure}

Figure~\ref{fig:PRcompilation} shows the PR curve for our models.
Each point in the figure is calculated with a varying threshold for
identification.  Since our sample has a class balance of 50-50 between
lens and non-lens, the most lenient threshold that classifies all
objects as positive would yield a precision of 0.5 at a recall of 1.0.
It is important to note that this figure changes as the class balance
changes; if we had 90\% non-lenses and 10\% lenses, the most lenient
threshold would yield a precision of 1/9 at a recall of 1.

In full application to real data, the precision quantity is what
determines the efficiency of follow-up by spectroscopic measurements
or human inspection, and we expect the number of lenses to non-lenses
to be 1-1000.  For approximate comparison, a 30\% recall
(completeness) for our mock HST data set has a 0.9967 precision in the
50-50 class balance, which then corresponds to a $23\%$ precision in
the realistic class balance of 1-1000. However, for LSST10, a 30\%
recall corresponds has a 0.89 precision in the 50-50 balance, which
then corresponds to 1\% in precision in 1-1000.  The precision in our
Our HST data is relatively idealized, so we expect the purity to be an
absolute upper limit estimate for how well the HOG/LR methods might be
able to do in real observations.  For comparison, the precision-recall
values quoted in other simulation-based tests are 94\%-100\% in
precision with 96\%-100\% recall in \citet{jacobsetal17} using
convolutional neural networks, where the lens to non-lens ratio is
approximately 1:1.  As another example, in \citet{gavazzietal14}, the
values are 29\% in precision and 42\% in recall using {\rm RingFinder}
in their sample. But, these values are only directly comparable to
tests where the ratio of lens to non-lens is the same.


\subsection{Effects of Regularization on Model Performance}\label{sec:regularization}

As described in Section~\ref{sec:LR}, LR trains a model with
complexity determined by the regularization parameter coefficient,
$C_\text{LogReg}$.  Larger values of $C_\text{LogReg}$ are less
regularized and allow for increased model complexity.  The highest
values of $C_\text{LogReg}$ will better describe features in the
training set.  However, an overly complex model will overfit the
training set at the expense of its performance on any independent test
set.  The regularization parameter ultimately defines the model
performance, and we must perform a parameter search to identify the
optimal value for $C_\text{LogReg}$.

\begin{figure*}[t]
\begin{center}
\includegraphics[width=.66\columnwidth]{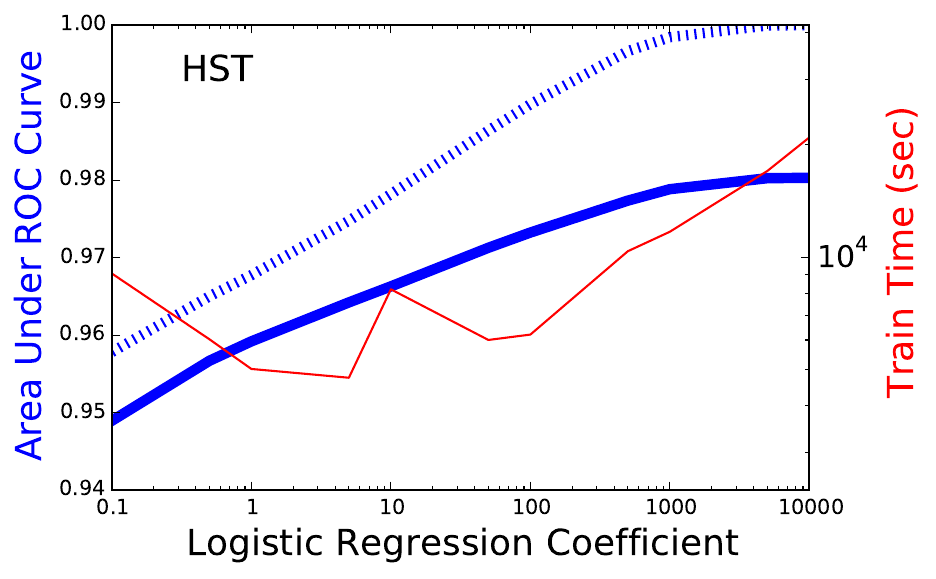}
\includegraphics[width=.66\columnwidth]{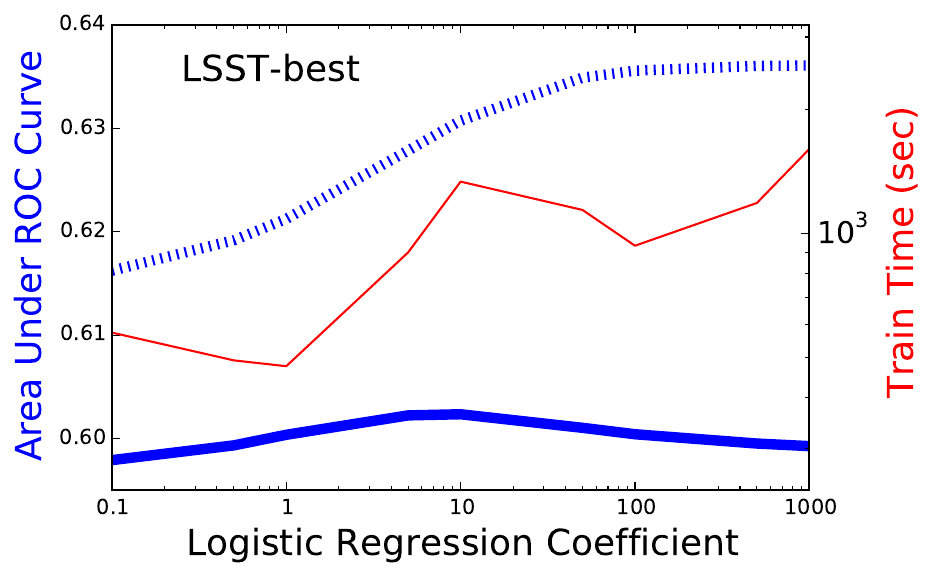}
\includegraphics[width=.66\columnwidth]{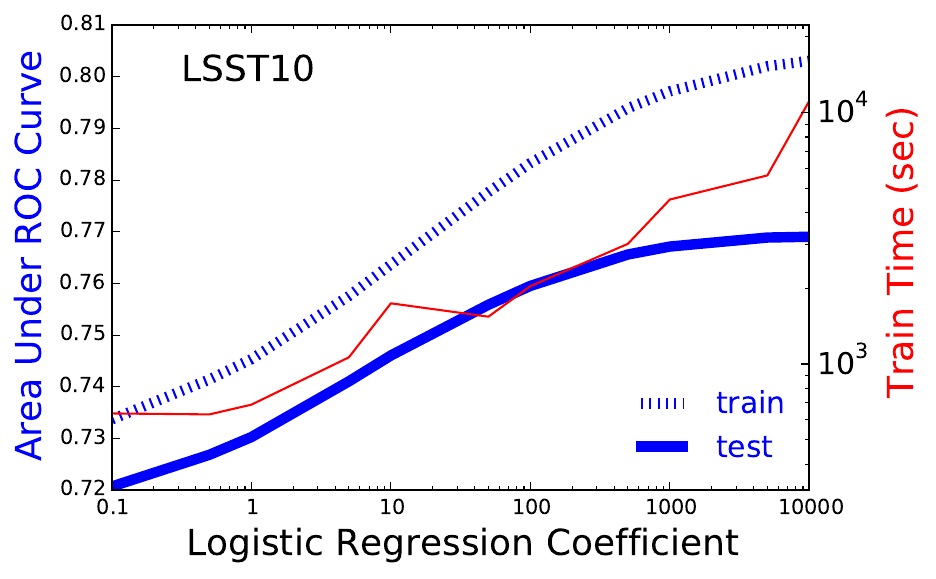}
\caption{AUC of the model with varying LR regularization coefficient
  parameter, $C_{LogReg}$, used when training the model classifier.
  We use a subset of the 10,000 training images to search over the LR
  $C_{LogReg}$ parameter, training on 8,000 and testing on 1,000.
  Each panel corresponds to a different mock observation.  From left
  to right: HST, LSST for one year, and LSST for 10 years.  The solid
  blue lines correspond to the AUC of the test set, and the dotted
  blue lines to the AUC of the training set.  To avoid overfitting, we
  choose the smallest parameter for which the AUC of the test set is
  maximal: 5000, 10, and 5000, respectively.  In thin red solid lines,
  we show the train time of the model, which roughly increases in a
  log-log scaling with logistic regression coefficient parameter.  The
  train time tick marks are on the right side of each figure in units
  of seconds.}\label{fig:regularization}
\end{center}
\end{figure*}

Figure~\ref{fig:regularization} shows the model performance as a
function of regularization parameter for each data set HST, LSST-best,
and LSST10.  The solid and dotted blue lines respectively correspond
to the model performance on the test and training set, with the AUC as
a metric for performance.  In red, we show the train time as a
function of $C_{LogReg}$.

As expected, the training set AUC increases and asymptotes with
$C_\text{LogReg}$.  With increasing model complexity, the model better
fits the training data set.  This is analogous to fitting a seventh
order polynomial to seven data points, where the fitting function will
go through every point but will not likely predict additional points.
With increasing model complexity, we are able to better able to
capture features that are generally characteristic of strong lensing
systems with arcs. However, past a certain $C_\text{LogReg}$, the
model performance on the test set decreases or asymptotes, as it has
overfit the training set.  We use the scaling of AUC with
$C_\text{LogReg}$ when training 8,000 out of our full 10,000 training
data set to determine the best value for $C_\text{LogReg}$.  However,
the optimal parameter is also dependent on the size of the training
set (see Section~\ref{sec:trainsetsize}), so this choice is not
generalizable.

For fixed training set size, the log of the train time roughly scales
linearly with the log of $C_\text{LogReg}$.  Since lower values of
$C_\text{LogReg}$ correspond to a more regularized model, there is a
smaller volume in hyperparameter space to search for the best fit
coefficients.  The solution, on average, will converge more quickly,
for more regularized models.  The scaling is not purely monotonic
because the fitting still has some randomness associated with the path
it takes to convergence.

\subsection{Dataset Size Dependence}\label{sec:datasize}
\subsubsection{Effects of Training Set Size on Performance and Train Time}\label{sec:trainsetsize}

In this section, we show the effects of training set size on model
performance on the hold-out test set of 1,000 images.  We also compare
the improvement between images that include the lens galaxy and images
with no lens galaxy.

\begin{figure}[t]
\begin{center}
\includegraphics[width=\columnwidth]{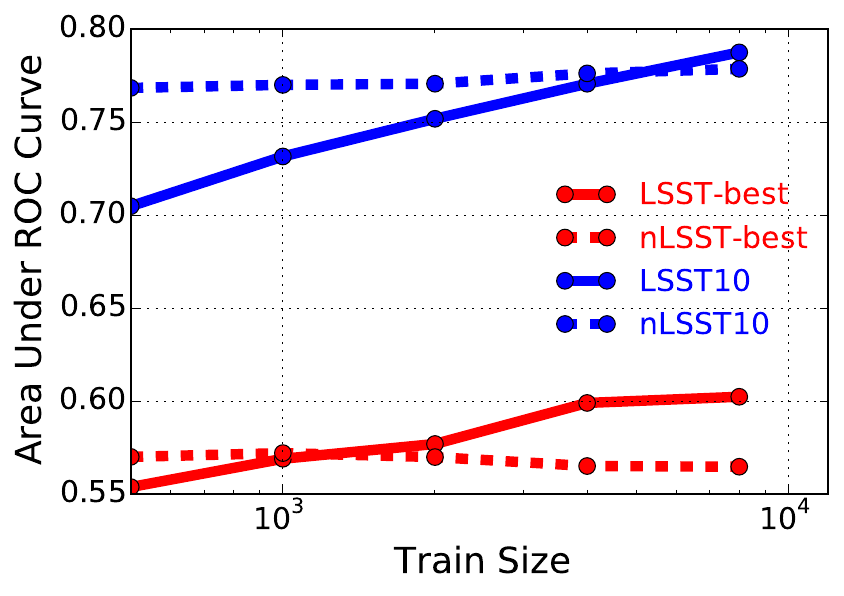}
\caption{Solid (dotted) blue line: AUC for models with varying size of
  the training set for LSST10 (nLSST10).  Solid (dotted) red line:
  Same for LSST-best. The improvement of AUC scales roughly linearly
  with the log of the training set size.  However, LSST10 and
  LSST-best have a steeper improvement with data training size.  The
  performance of nLSST10 and nLSST-best models trained on smaller
  training data sets is better than the respective LSST10 and
  LSST-best models trained on the same size data, but the LSST10 and
  LSST-best models outperform with larger size training data
  set.}\label{fig:trainsizeLSST}
\end{center}
\end{figure}

Figure~\ref{fig:trainsizeLSST} shows how the AUC depends on the log of
the size of the training set for both the LSST10 (LSST-best) data in
the solid blue (red) line, and the nLSST10 (nLSST-best) data in dashed
blue (red) line.  The AUC for models trained on the LSST10 data
improves almost linearly with the log of the training set size,
increasing from AUC=0.705 to AUC=0.788 when the train size is
increased from $2\times500$ lens/non-lens images to $2\times8000$
lens/non-lens images.  However, for the nLSST10 data, where the lens
galaxy has been removed from the images, the improvement is less
dramatic.  With the same increase in training size, the AUC for
nLSST10 changes from just below 0.77 to just below 0.78.

In the LSST10 case, the trained model can incorporate the additional
information of the edges from the lens galaxy, which is correlated
with the lensing cross-section and likelihood of the image being a
lensing system.  The nLSST10 images do not contain this information,
but provide cleaner signals of the lensed image for lensed images that
occur close to the lens galaxy.  The cleaner signal in nLSST10 allows
for better model performance for smaller training data set sizes
($N_\text{train}\sim5,000$).  However, models trained on LSST10
improve more rapidly with train size, since the additional information
from the lens galaxy better describes the lens containing images.  For
$N_\text{train}\gsim 5,000$, the models trained on LSST10 outperform
those trained on nLSST10.

In red, we see an analogous improvement for LSST-best and nLSST-best.
Here, the crossover happens at train size of $2x1000$, where an
increase in larger training data set size does not help improve the
AUC for nLSST-best.

Since the models that contain information from the lens galaxy edges
in LSST10 are more complex, the models require a larger training set
size for a better fit.  While model performance for LSST10 appears to
steadily increase, this comes at the cost of increased train time,
which is two-fold.  The train time will increase due to both an
increase in data to fit, and also an increase in optimal
$C_\text{LogReg}$ where the volume of hyperparameter space for allowed
solutions is larger (see red lines in
Figure~\ref{fig:regularization}).  

\begin{figure}[t]
\begin{center}
  \includegraphics[width=\columnwidth]{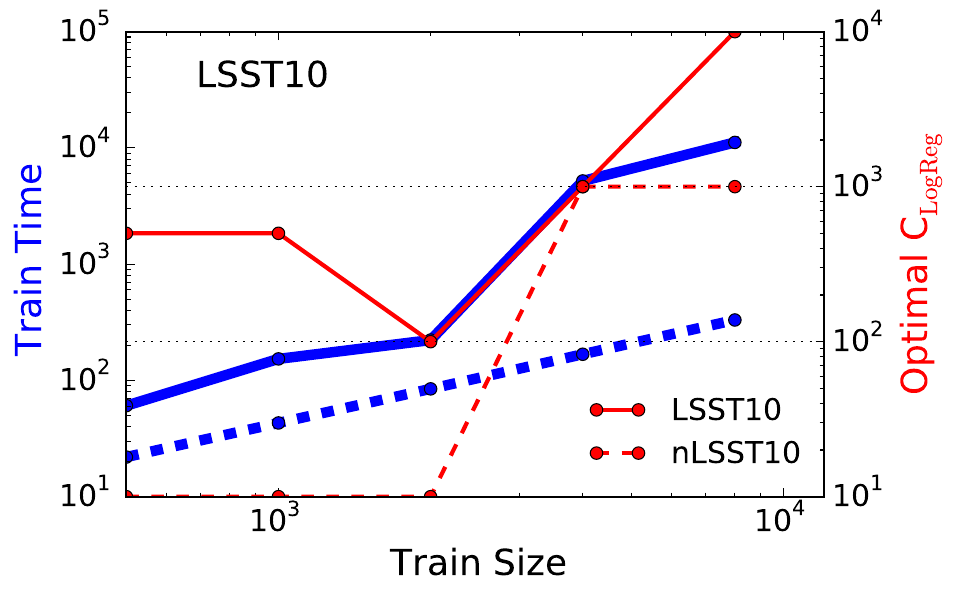}
\caption{Solid (dotted) blue line: Train time for models with varying
  size of the training set for LSST10 (nLSST10).  Train time roughly
  scales logarithmically with the train size, but the train time is
  also affected by model complexity.  Solid (dotted) red line: Best
  regularization parameter as a function of train size.  Note, LSST10
  requires more model complexity to exceed the performance of nLSST10
  (see blue solid and dotted lines in
  Figure~\ref{fig:trainsizeLSST}), and therefore requires more
  training time for continual increase in
  performance. }\label{fig:traintimeLSST10}
\end{center}
\end{figure}

\begin{figure}[t]
\begin{center}
  \includegraphics[width=\columnwidth]{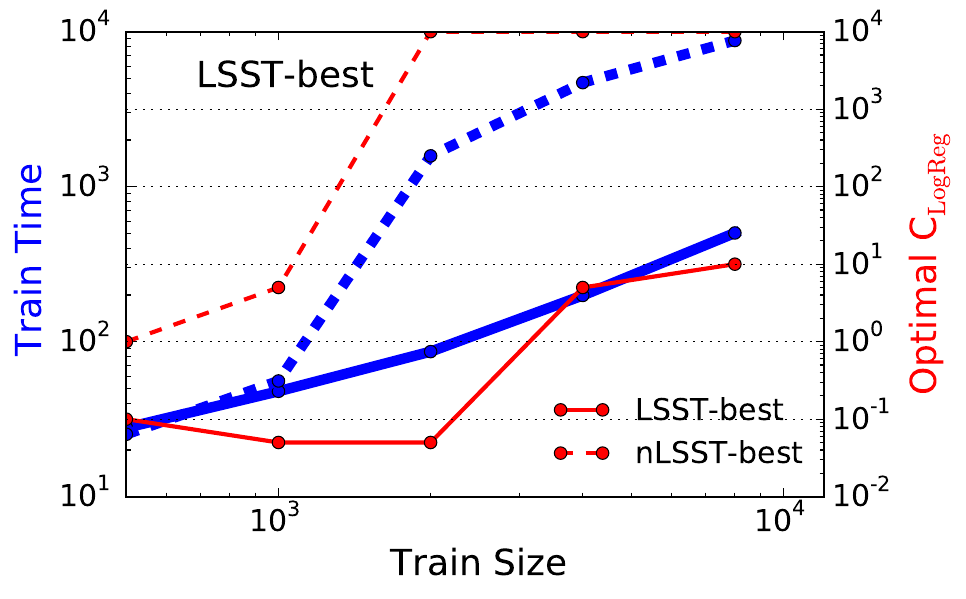}
\caption{Same as Figure~\ref{fig:traintimeLSST10}, but for LSST-best.
  Here, nLSST-best requires more model complexity than
  LSST-best.  Note, that LSST-best and nLSST-best
  have better seeing but reduced overall signal to noise than LSST10
  and nLSST10.  This corresponds to better resolution, and therefore
  sharper edge features that prominently corespond to
  arcs.}\label{fig:traintimeLSST1}
\end{center}
\end{figure}

We illustrate the dependence of train time on both the size of the
training set and model complexity in Figures~\ref{fig:traintimeLSST10}
and \ref{fig:traintimeLSST1}.  In red,
Figure~\ref{fig:traintimeLSST10} shows that the optimal values for
$C_\text{LogReg}$.  For LSST10, $C_\text{LogReg}$ roughly scales
logarithmically with the log of the train size, with an exception of
the data point corresponding to train size of 2000.  Generally, a
larger training set allows for an increase in model complexity without
reducing its ability to generalize.  This is also true for the optimal
$C_\text{LogReg}$ dependence on the number of training images in the
nLSST10 data.  But, the required complexity is systematically less
than for the LSST10 images.

In blue, Figure~\ref{fig:traintimeLSST10} also shows the train time
for LSST10 and nLSST10 as a function of the size of the training set
for the optimal regularization parameter for that subset of the
training data.  Each model uses features extracted with the same HOG
parameterization from the grid search and the optimal regularization
parameter for that subset of the training data.  The train time of a
given model generally increases for increasing regularization
parameter.  For the subset of train size $N_\text{train}=2000$ in
LSST10, the optimal regularization parameter happened to be
$C_{LogReg}=100$, whereas it was $C_{LogReg}=500$ for the subset of
trainsize 1000, and $C_{LogReg}=1000$ for the subset of trainsize
4000.  This makes the train time increase at train size
$N_\text{train}=2000$ for LSST10 less dramatic than the average
log-log slope of approximately 2.

Since nLSST10 does not contain the lens galaxies, fewer of the
extracted features describe the lens system, requiring decreased model
complexity.  The increase in train time for nLSST10 as a function of
train size is mostly due to only having more data to fit in the
regression, leading to a steady and slow increase of train time with
number of training images with log-log slope of approximately 1.

For LSST-best, Figure~\ref{fig:traintimeLSST1} shows the same
relationships between train time and train size in blue, and the
optimal regulariztion parameter (or model complexity) in red.
Contrary to what we found when comparing LSST10 to nLSST10, nLSST-best
requires more model complexity than LSST-best.  Recall that LSST-best
and nLSST-best correspond to single epoch simulated exposures with the
best seeing; these images exhibit better resolution but worse signal
to noise.  The switch in required model complexity corresponds to a
trade-off between features from the lens galaxy providing additional
information or swamping the signal from a strong lensing arc.


\subsubsection{Effects of Rotation on AUC}

To augment our training set, we rotated each image in the set by
multiples of $90^o$.  Since our feature extraction method of HOG is
not rotationally invariant, augmenting our data by a factor of four
naturally optimizes the use of available training data.  This has an
equivalent improvement to the study illustrated in
Figure~\ref{fig:trainsizeLSST}.

\begin{figure}[t]
\begin{center}
\includegraphics[width=\columnwidth]{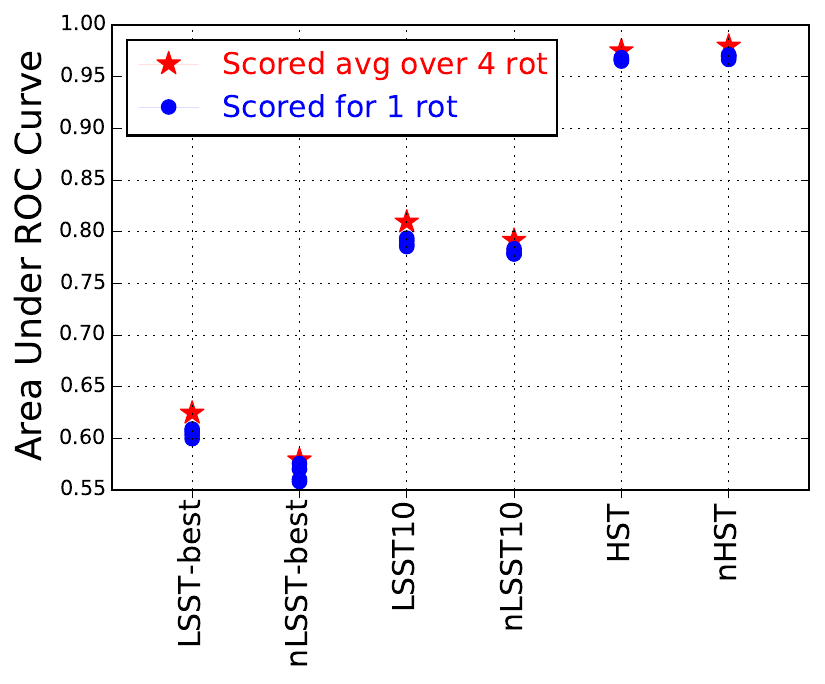}
\caption{Our summary figure: The AUCs of models trained on the full
  10,000 and tested on the holdout 1,000.  Blue circles: AUC
  calculated from scores of images at a given rotation (e.g. 0, 90,
  180, and 270 degrees).  Red stars: AUC calculated from the average
  score of all rotations of each image.  The average score produces an
  improved AUC in all data sets.  We expect the AUC to further improve
  with increased train size.}\label{fig:rotation_test}
\end{center}
\end{figure}

We also tested the effects of evaluating our model on all four
rotations of the test set, and using the average score of each test
image to calculate the AUC.  In Figure~\ref{fig:rotation_test}, we
show the AUC for different orientations of the datasets.  The x-axis
corresponds to each of our three datasets, with and without the lens
galaxy.  The y-axis shows the AUC.  The filled blue circles correspond
to the four AUCs calculated when the model evaluated images at each of
the four rotations.  The filled red stars correspond to the AUC
calculated from the average of all four test scores, which are
systematically higher than any one rotation.  The average score across
all rotations for each image is likely to be less noisy for the whole
test sample, giving an improved AUC.

Figure~\ref{fig:rotation_test} also summarizes the best-case results
of our models trained on our entire 10,000 training sets, and tested
on our holdout 1,000 test set.  Recall, however, that we expect model
performance on images containing the lens galaxy to improve further
with larger training sets (see Figure~\ref{fig:trainsizeLSST}).

\subsection{Image Classification Performance}\label{sec:performance}

\begin{figure*}[t]
\begin{center}
\includegraphics[width=.8\columnwidth]{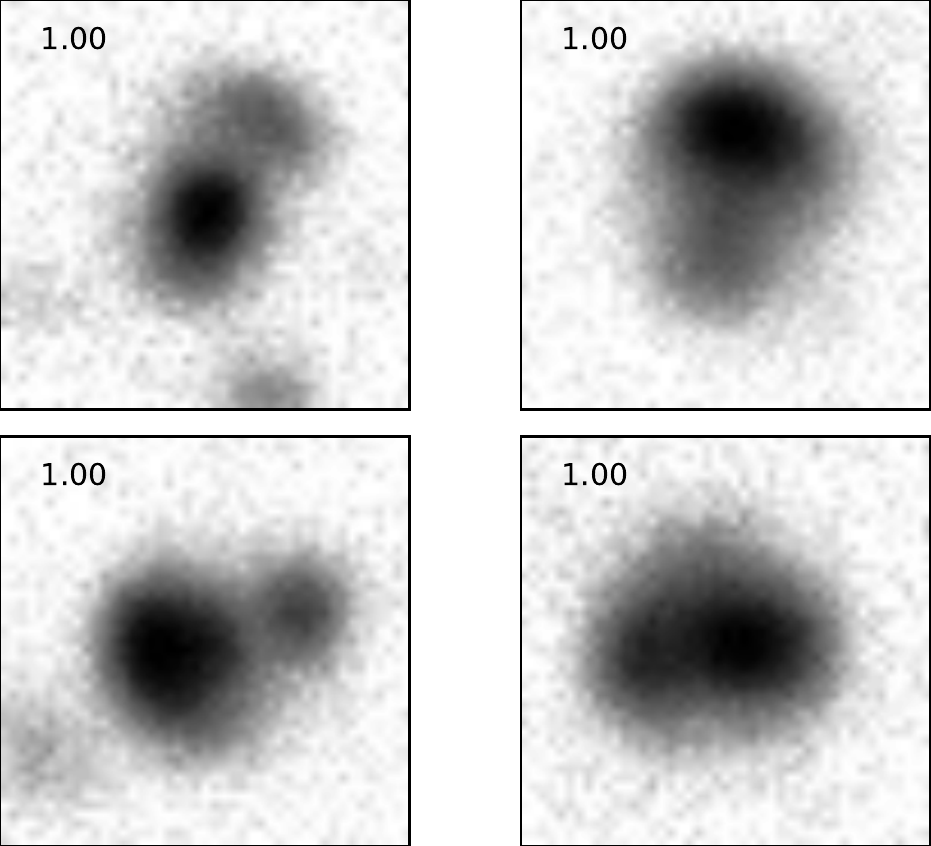}\hspace{15pt}
\includegraphics[width=.8\columnwidth]{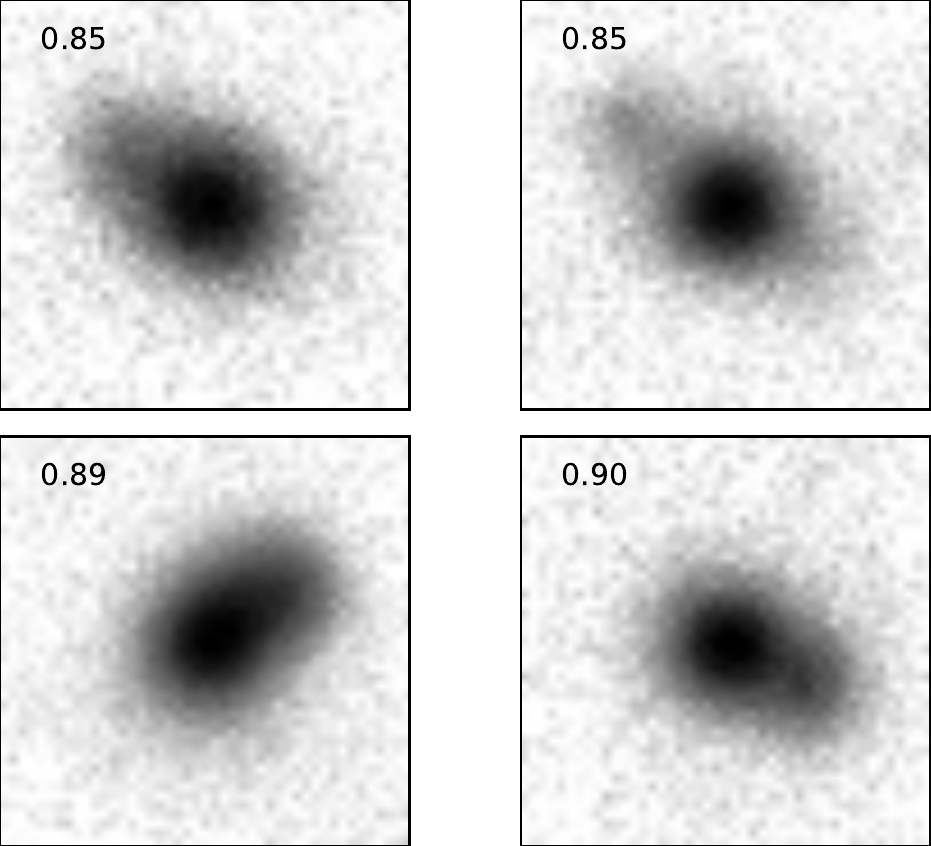}\\\vspace{5pt}
\includegraphics[width=.8\columnwidth]{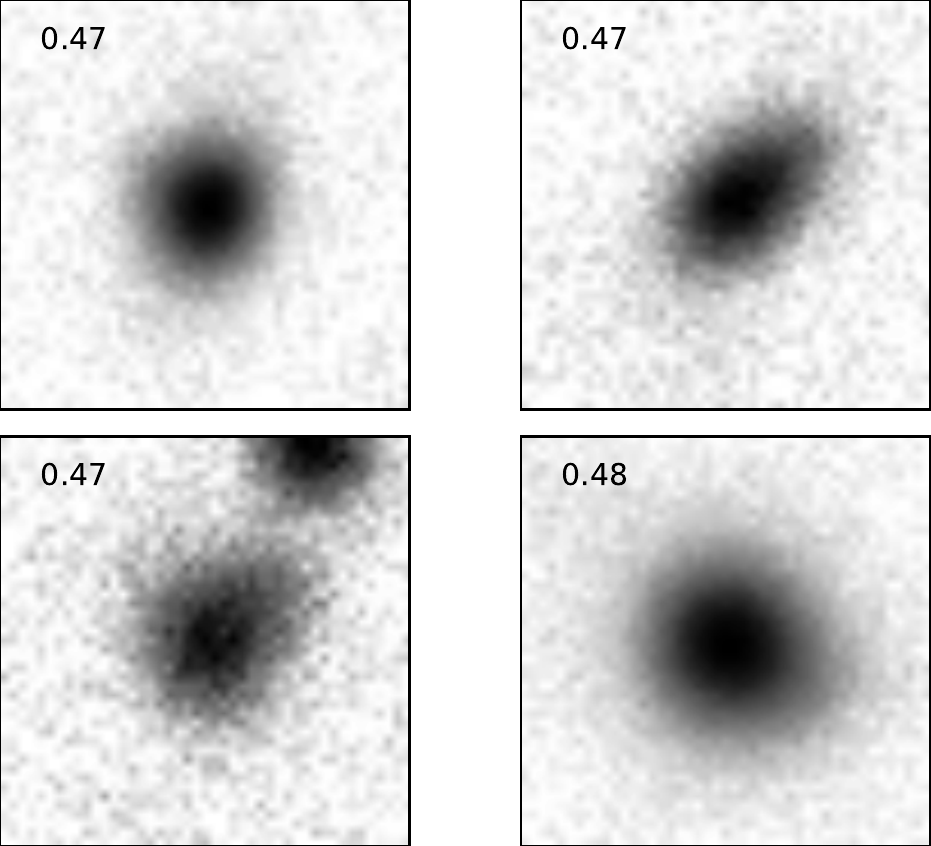}\hspace{15pt}
\includegraphics[width=.8\columnwidth]{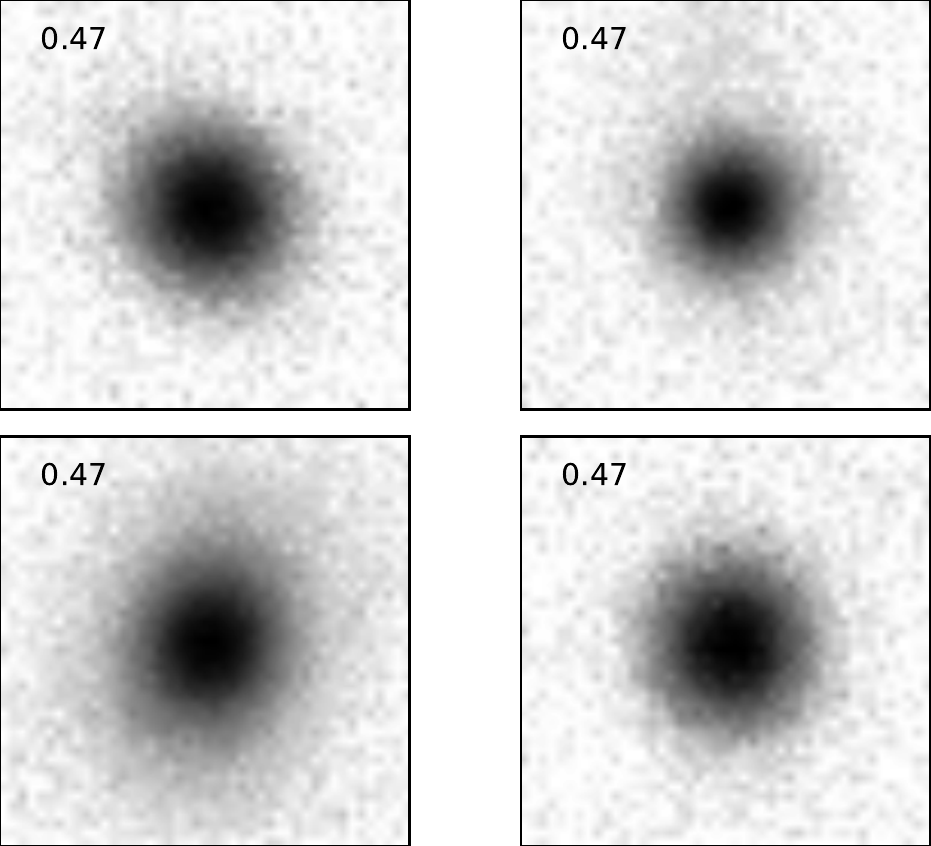}\\\vspace{5pt}
\includegraphics[width=.8\columnwidth]{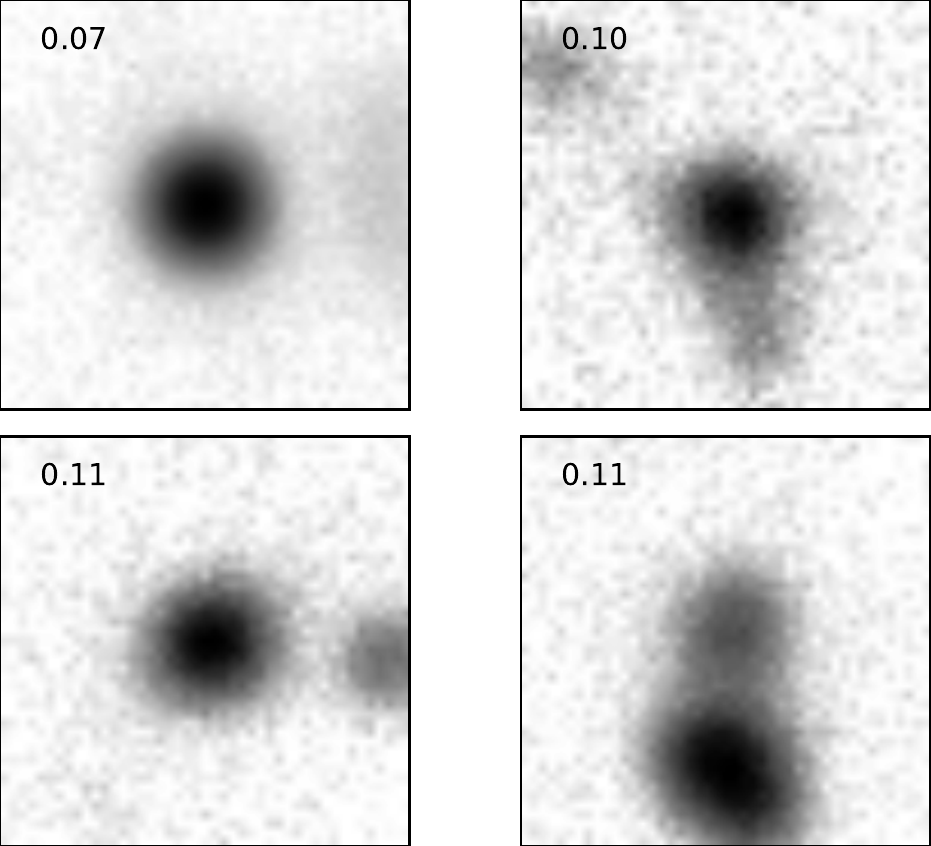}\hspace{15pt}
\includegraphics[width=.8\columnwidth]{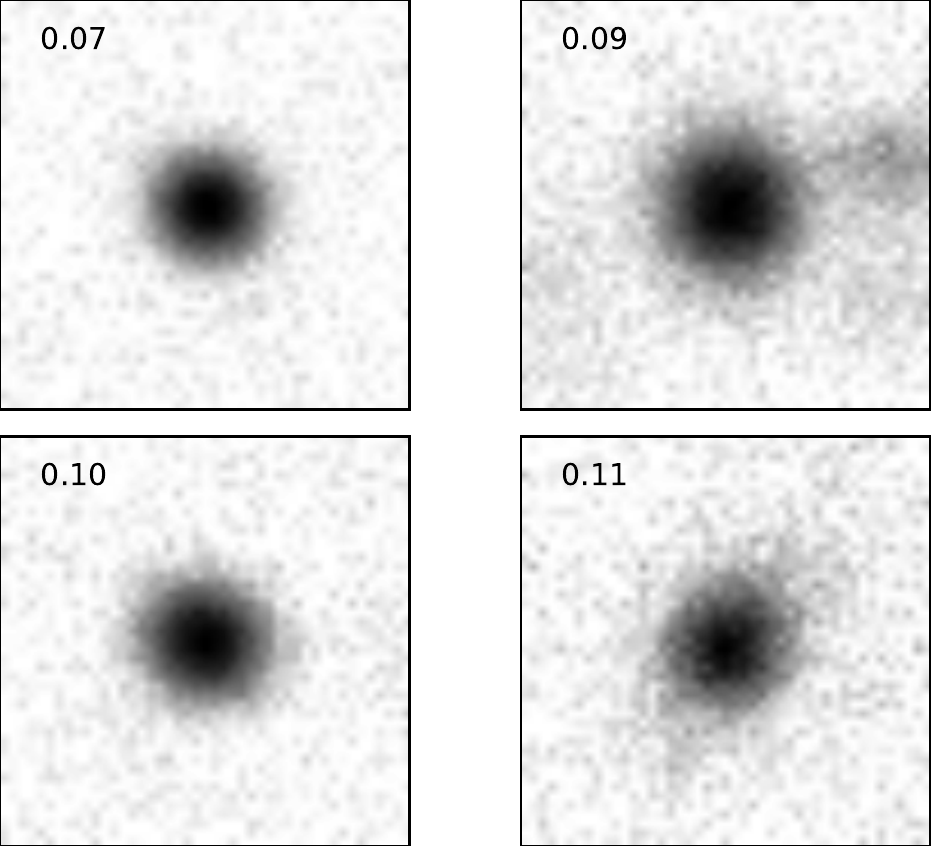}\\
\caption{LSST 10 year mock images. Left two columns: Lens containing
  images, annotated with the image score assigned by our trained
  classifier.  Right two columns: Non-lens containing images,
  annotated with the image score.  The top two rows show
  characteristic images that will be accepted with a high threshold
  for classification, contributing to the bottom left of the ROC curve
  in Figure~\ref{fig:ROCcompilation}.  The middle two rows show
  characteristic images that will be accepted with a moderate
  threshold, contributing to the knee of the ROC curve, with true
  positive rates and false positive rates of $tpr\approx0.8$ and
  $fpr\approx0.25$, respectively.  The bottom two rows show
  characteristic images that will only be accepted with an extremely
  lenient threshold, contributing to the top right area of the ROC
  curve.}\label{fig:ROCsamplesLSST10}
\end{center}
\end{figure*}

\subsubsection{Populating the ROC Curve}
In this section, we discuss the different image types that our model
is most and least able to successfully classify.  We have six
paradigms of model performance on the mock images based on the score
an image receives when evaluated by the trained model, and its true
label.  From highest scoring to lowest scoring: True Positives (tp),
False Positives (fp), Borderline Positives (bp), Borderline Negatives
(bn), False Negatives (fn), and True Negatives (tn).

Figure~\ref{fig:ROCsamplesLSST10} illustrates four images from each
paradigm for the hold-out test set of LSST10.  The trained model used
the entire 10,000 image training set.  The left two columns show lens
systems, and the right two columns show non-lens systems.  The
annotation in the top left corner of each images shows the score.

In general, the true positives (the highest scoring lens systems) have
lensed images with large magnification.  The the true negatives (the
lowest scoring non-lens systems) have small lens galaxies with
galaxies along the line of sight that are rounded.  The successful
classification of these two paradigms are least sensitive to the
threshold.  On the other hand the failed classification of the false
positives (the highest scoring non-lens systems) and the false
negatives (the lowest scoring lens systems) are also least sensitive
to the threshold.  False negatives are typically lens systems with
lensed images of smaller magnification and minor distortions that
mimic along the line of sight galaxies that the model has learned to
ignore.  False positives are often non-lens systems with elongated,
elliptical, or ``fuzzy'', galaxies along the line of sight whose signal
blends with the lens galaxy contributing to the false positive rate
even for conservative thresholds.  Visually, these false positives are
virtually indistinguishable from true arcs, and would require
spectroscopic follow-up.

The middle two rows of Figure~\ref{fig:ROCsamplesLSST10} illustrate
the borderline positives and borderline negatives. The successful
classification of the borderline positives and negatives are most
sensitive to the threshold, and would be the first candidates for
alternative classification methods, such as visual follow-up.
Thresholds set around these scores yield a true positive rate and
false positive rate of $tpr\approx0.8$ and $fpr\approx0.25$,
respectively.

\subsubsection{Dependence on Lens-Model Parameters}

Here, we examine lens-model parameters that affect how well our
pipeline can classify the system.  The lens-model parameters we
examined are the redshift, ellipticity, orientation angle, and
velocity dispersion of the lensing galaxy, and also the magnification
of the lensed image compared with its original size.  We found that
the magnification of the lensed image is the most correlated
lens-model parameter with our trained model performance, with a
secondary and related correlation with lens galaxy velocity dispersion
that is encapsulated in the Einstein radius.  The more strongly lensed
an image is, the larger its magnification, and the easier it is for
our trained model to classify.

\begin{figure*}[t]
  \centering 
  \mbox{
    \subfigure{\includegraphics[width=.66\columnwidth,trim=13 0 10 0,
        clip]{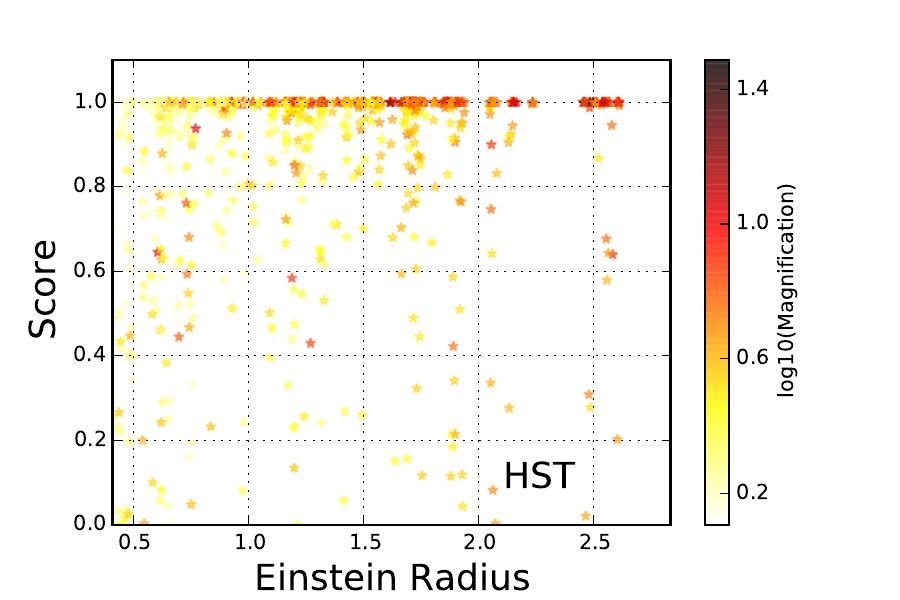}}\hfill
    \subfigure{\includegraphics[width=.66\columnwidth,trim=13 0 10 0,
        clip]{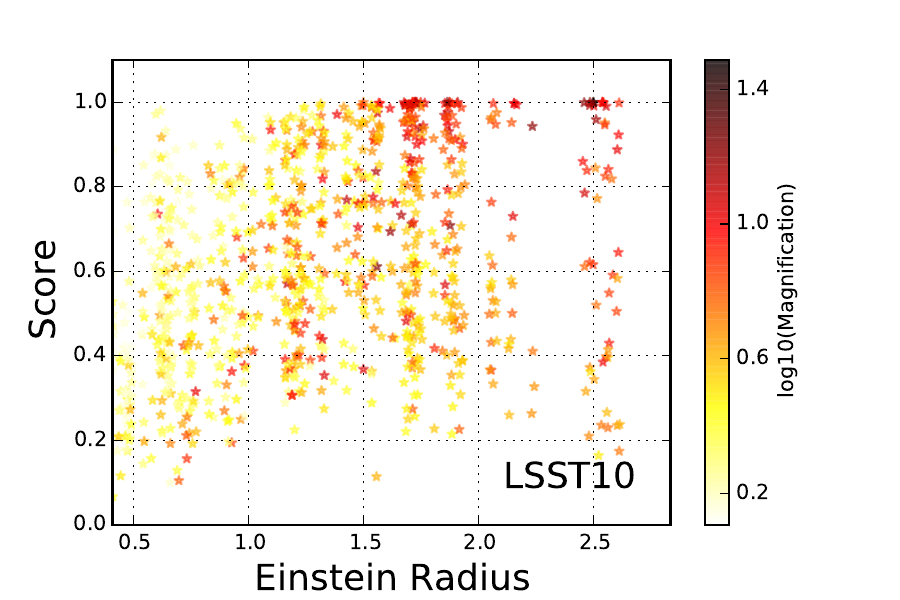}}\hfill
    \subfigure{\includegraphics[width=.66\columnwidth,trim=13 0 10 0,
        clip]{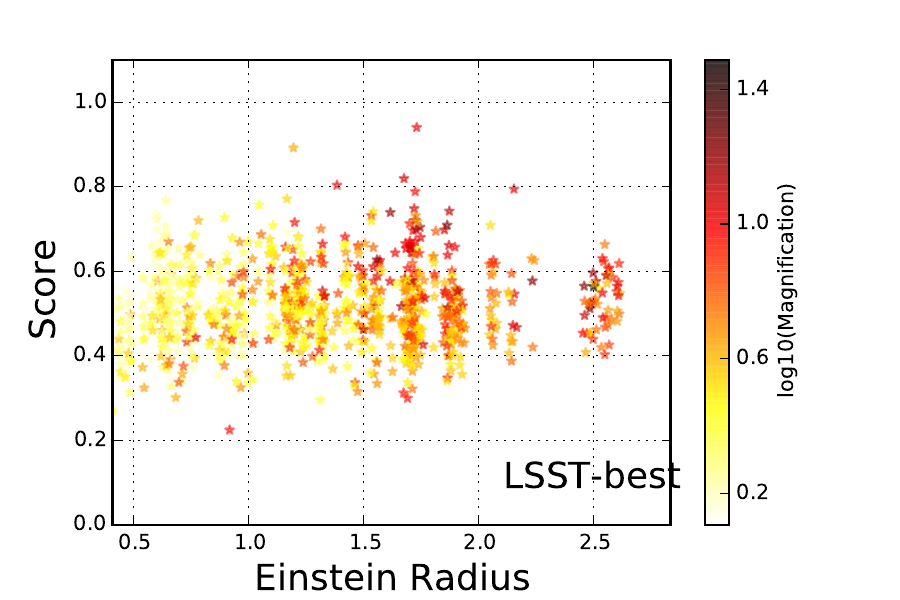}\hfill}
  }\\ 
  \mbox{
    \subfigure{\includegraphics[width=.66\columnwidth,trim=13 0 10 0,
        clip]{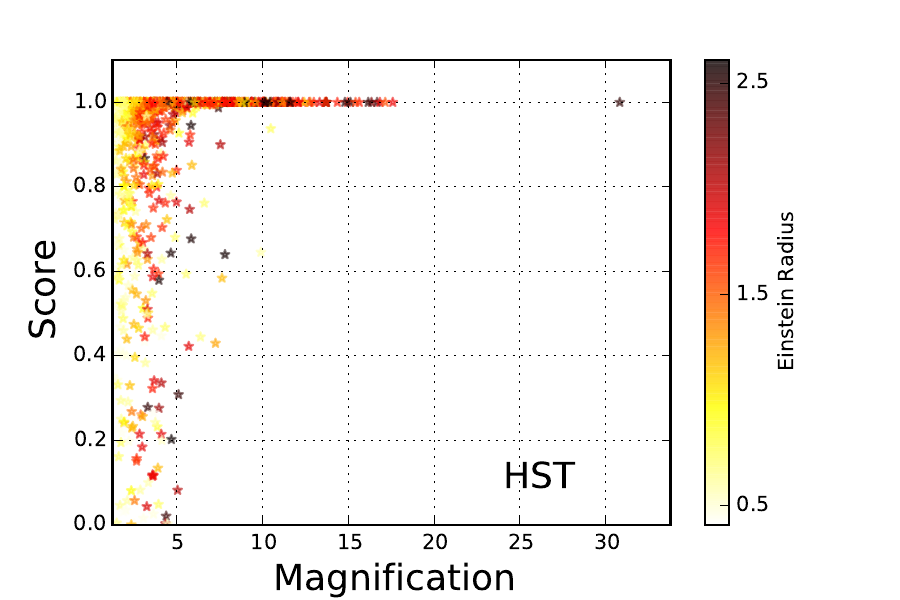}}\hfill
    \subfigure{\includegraphics[width=.66\columnwidth,trim=13 0 10 0,
        clip]{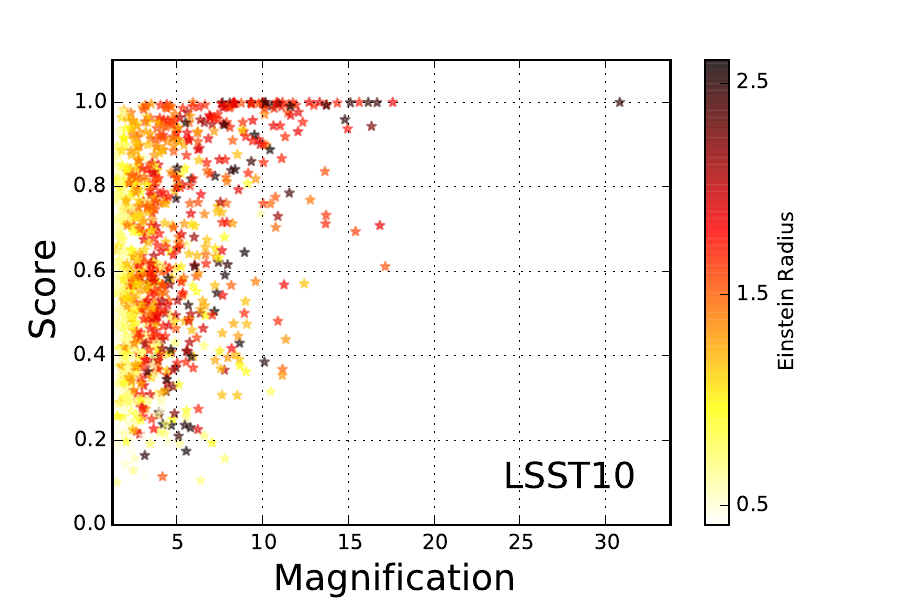}}\hfill
    \subfigure{\includegraphics[width=.66\columnwidth,trim=13 0 10 0,
        clip]{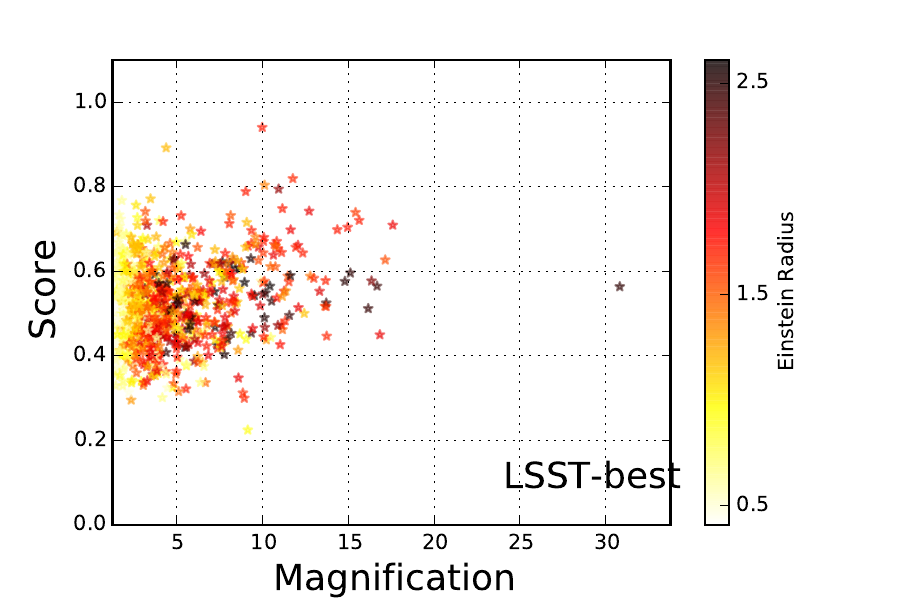}\hfill}
  }\\ 
\caption{Left to right: Image classification score on HST, LSST10,
  LSST-best mock observed lensing images as a function of Einstein
  radius (top) and magnification (bottom). The top panels are
  color-coded by magnification, and the bottom by Einstein radius to
  visualize the combined effects. Magnification is the strongest
  indicator of how likely a lensing system will be successfully
  classified.  The Einstein radius has a weaker correlation with how
  easily a lensing system might be classified, since it contains
  information on both the velocity dispersion of the lens and the
  redshifts of the lens and source galaxies.  High velocity dispersion
  lens galaxies are likely to produce high magnification images of the
  source galaxies.  }\vspace{10pt}
\label{fig:scorevslensparams}
\end{figure*}

In Figure~\ref{fig:scorevslensparams}, we show the classification
score as a function of Einstein radius and image magnification of the
source galaxy in each of our samples. In an HST- or LSST10-like
observation, lensing systems with images that have magnification
$\gsim7$ will likely be classified as positive with threshold scores
above $\gsim0.5$.  These systems are also those that are most easily
classified by eye.  However, our trained model has varying performance
for systems with lower magnifications.

Highly magnified systems typically have lenses that are at higher
redshifts of and/or lens galaxies with velocity dispersions larger
than $\sigma_v\gsim230.$~km/s.  Lensing galaxies with smaller velocity
dispersions are less massive and therefore have a smaller efficiency
of lensing cross section.  The smaller efficiency means that a
background galaxy is less likely to be strongly lensed with high
magnification.  Also, due to hierarchical structure formation,
galaxies are less massive at higher redshifts, so the trends of model
performance with these three parameters are somewhat degenerate with
one another.

The relationship between our model performance and lens parameters
indicates that the magnification of lensed images is the most
relevant, and the distribution of image magnification in a data set
will impact trained model performance.  We did not find strong
correlations in model performance with other lens parameters.  It is
also useful to keep in mind that lensed galaxy images with
magnification $\lsim 5$ are often visually indistinguishable from
edge-on disk galaxies along the line of sight, which can lead to false
positives.  Since the latter can lead to false positives, the model
has learned to downweight the related features.  Note that the model
would be more sensitive to systems with lower magnification without
galaxies along the line of sight.

In a forthcoming paper, we will discuss how class imbalance, or
differences between the lens-model parameter distributions in the
training and test sets affect model performance and will explore a
method to correct for this.

\subsubsection{Methods Applied to Real Data}\label{sec:SLACS}

We examine the performance of the HOG/LR methodology on data from The
Sloan Lens ACS Survey (SLACS), real space-based HST observations
\citet{boltonetal08}. The main conclusion from our tests is that HOG
is a feature extraction method where parameters {\it can} be varied to
compensate for imperfections and details that the mock training data
does not capture.  However, the HOG parameterization that best
captures the geometric features of an arc and lens galaxy will vary
depending on the quality of the image. This subsection also examines
how potential further steps in using HOG might mitigate differences
between simulated and real data with a focus on HST images.  For
shorthand, we provide the HOG parmeterization of $n_{pix}\times
n_{pix}$ pixels-per-cell and $m\times m$ cells-per-block as
ppc-$n_{pix}$-cpb-$m$.

We note that a test of our model on real ground-based data is plotted
in Figure~8 of \citet{metcalfetal18}.  Consistent with other methods
explored in \citet{metcalfetal18}, the performance of our method
decreases when evaluated only on the subset of real data.  We do not
explore how HOG parameterizations impact the performance on the real
data evaluated in \citet{metcalfetal18}.  Instead, we focus on SLACS
data, which is most similar to our mock HST data set that we have for
model training and testing.

The SLACS data set is comprised of images selected for high-redshift
emission lines and a lower redshift continuum in a single spectrum
from the Sloan Digital Sky Survey.  In the data set we examined, there
are 64 clear lensing systems, and 27 non-lensing systems as classified
by the authors through visual examination in the direct images and
model-subtracted residual images.  We do not use ambiguously
classified systems from their sample in our test.  Each system has an
image in at least one of three filters: F814W, F555W, and F435W.  For
34 of these systems, there is another exposure at the same filter.

\begin{figure*}[t]
  \centering 
  \mbox{
    \subfigure{\includegraphics[width=.4\columnwidth,trim=90 10 10 60,
        clip]{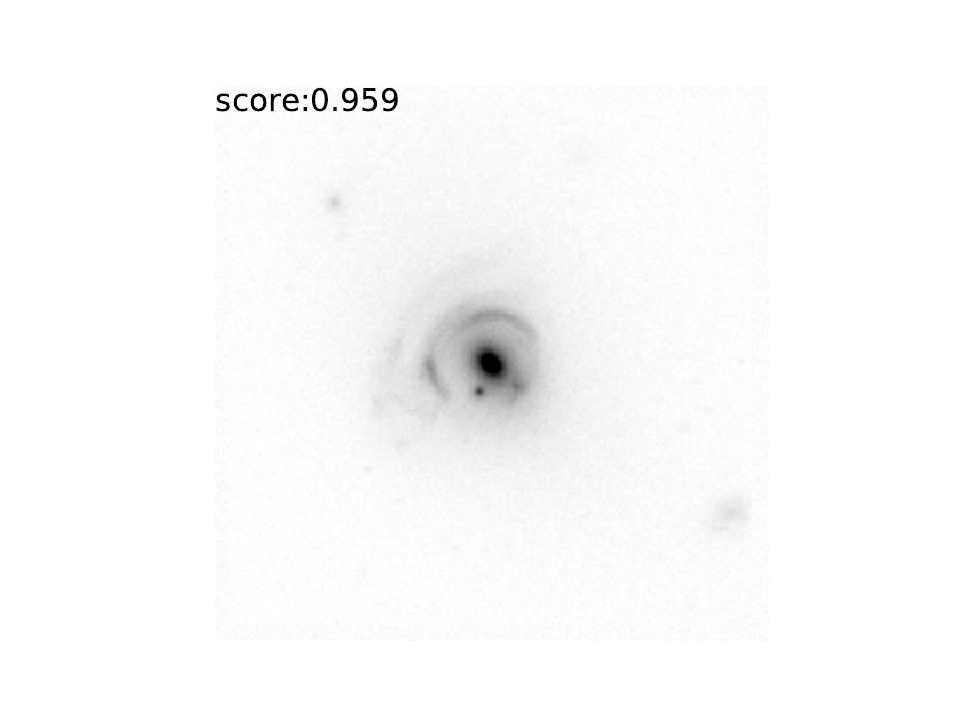}}\hfill
    \subfigure{\includegraphics[width=.4\columnwidth,trim=90 10 10 10,
        clip]{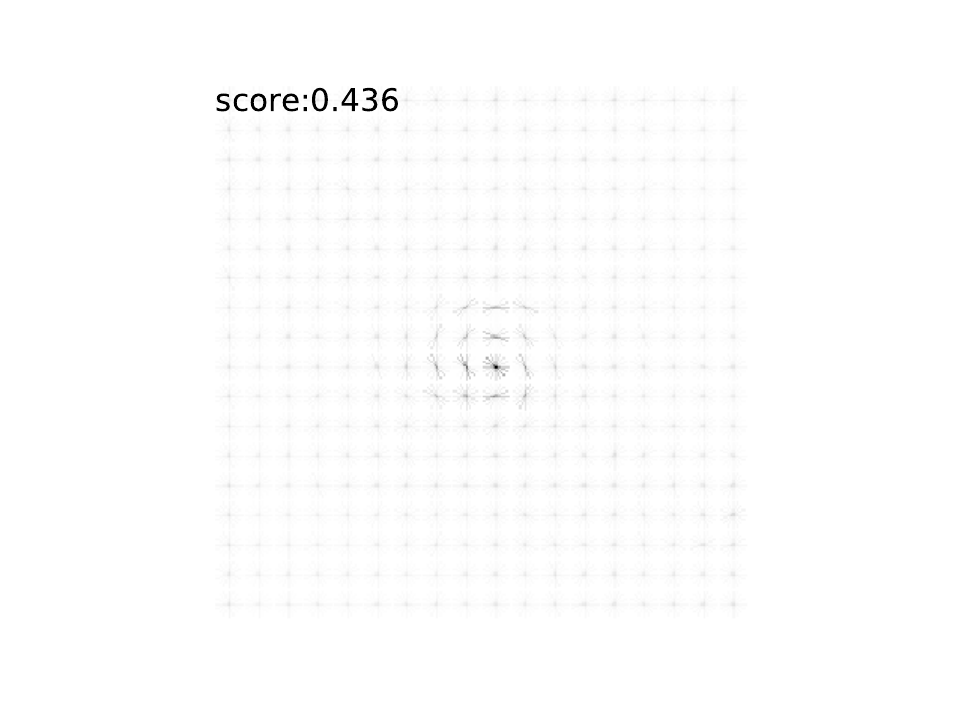}}\hfill
    \subfigure{\includegraphics[width=.4\columnwidth,trim=90 10 10 10,
        clip]{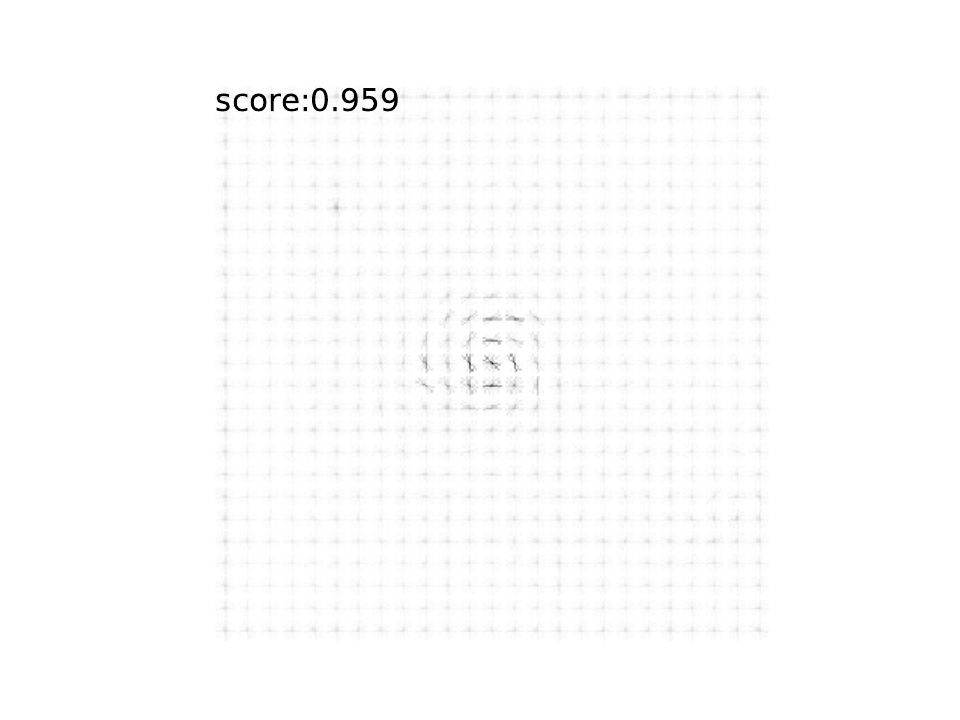}\hfill}
    \subfigure{\includegraphics[width=.4\columnwidth,trim=10 0 11 0,
        clip]{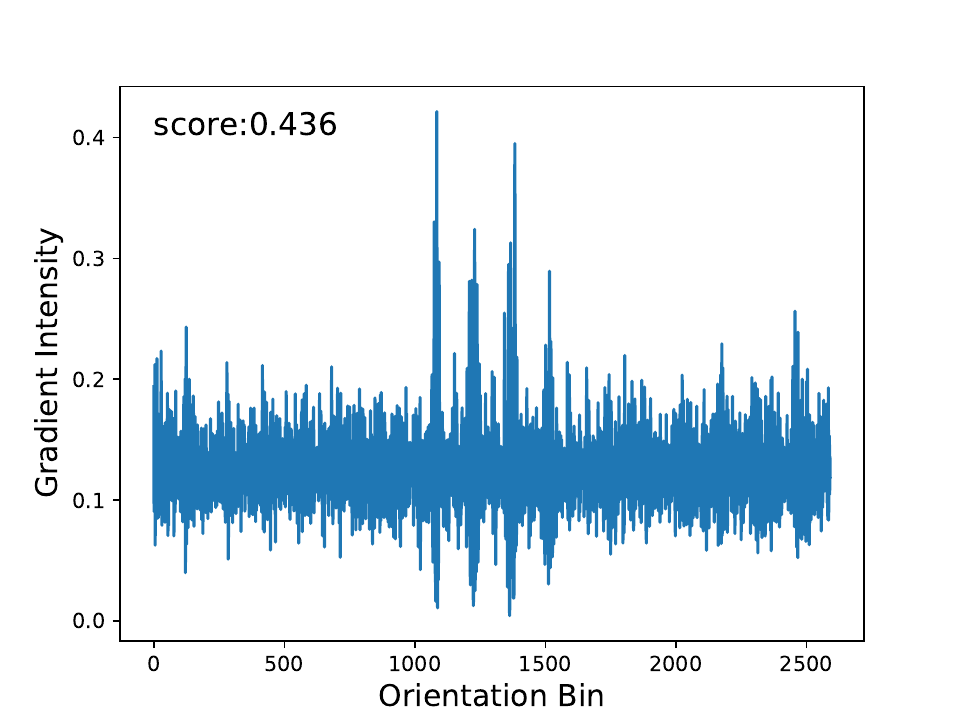}}\hfill
    \subfigure{\includegraphics[width=.4\columnwidth,trim=10 0 11 0,
        clip]{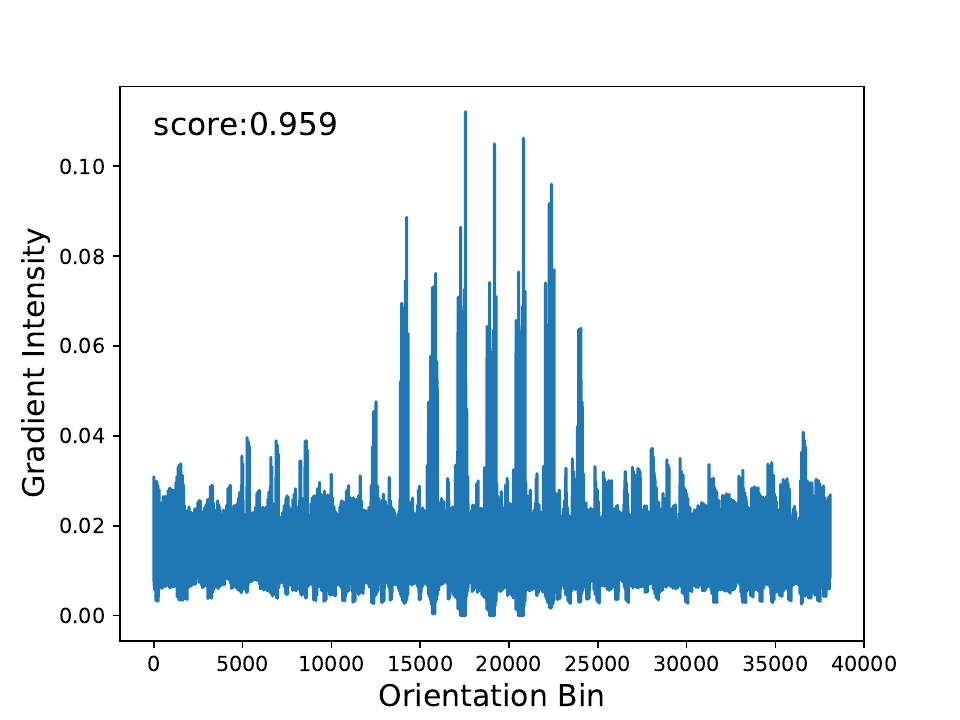}}\hfill
  }\\ 
  \mbox{
    \subfigure{\includegraphics[width=.4\columnwidth,trim=90 10 10 60,
        clip]{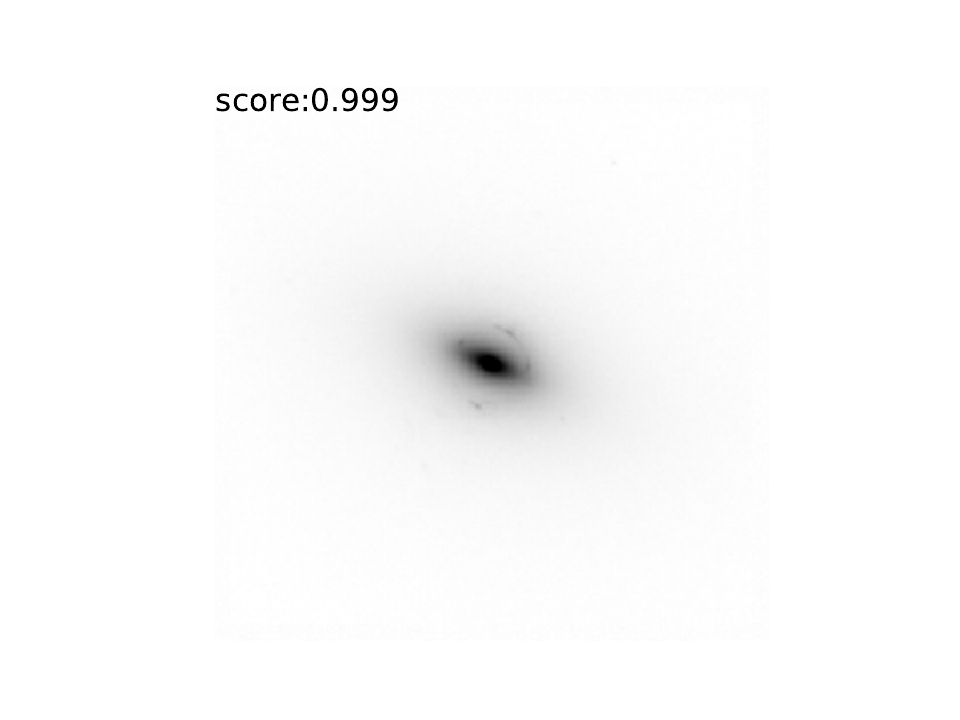}}\hfill
    \subfigure{\includegraphics[width=.4\columnwidth,trim=90 10 10 10,
        clip]{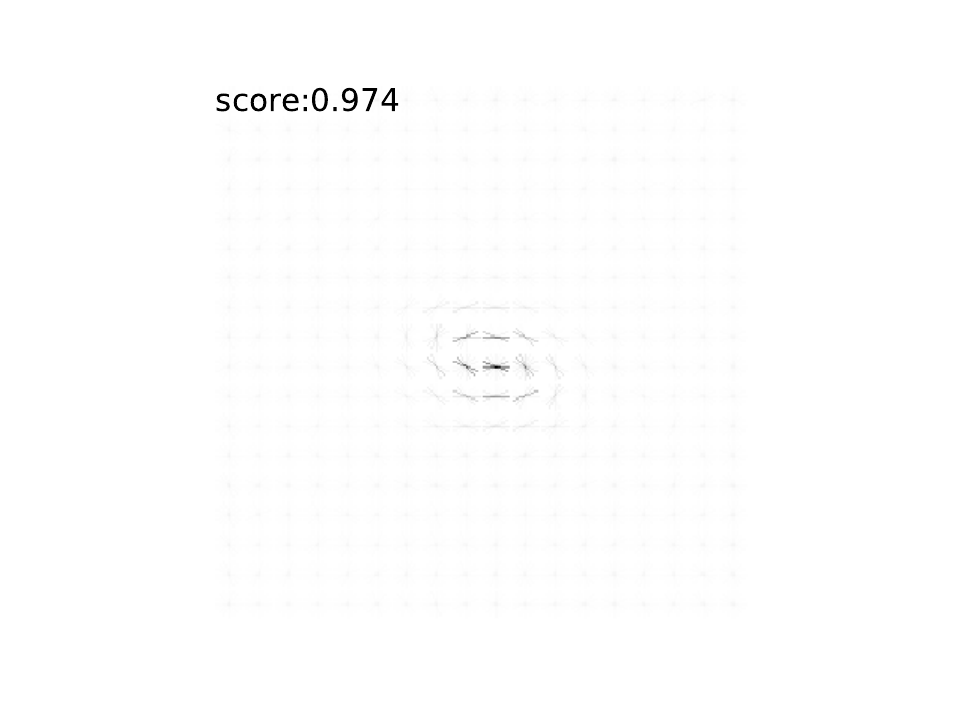}}\hfill
    \subfigure{\includegraphics[width=.4\columnwidth,trim=90 10 10 10,
        clip]{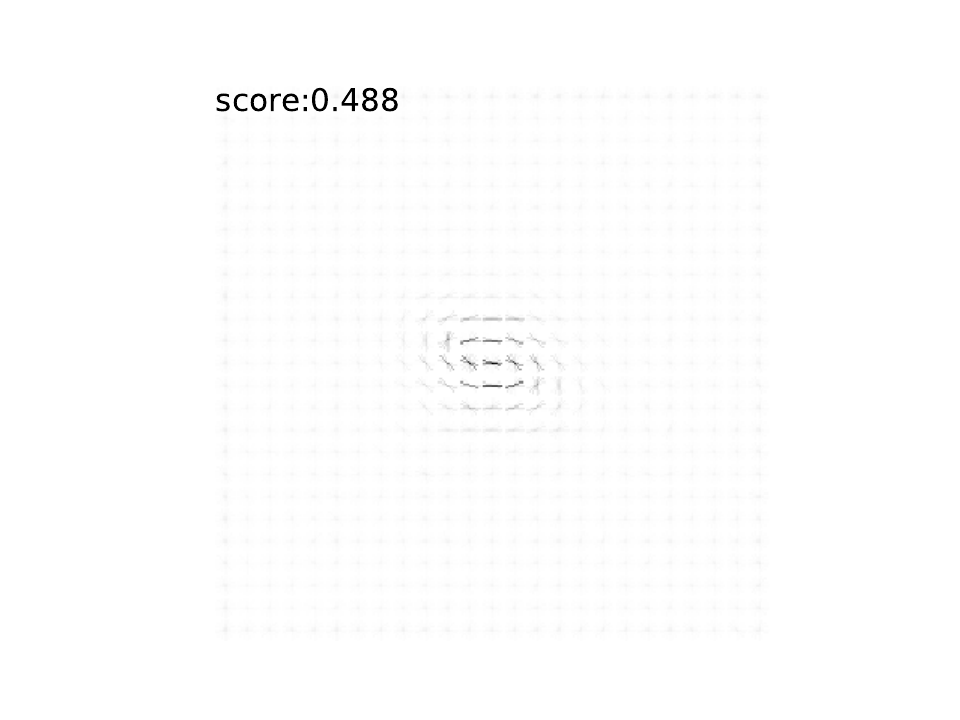}\hfill}
    \subfigure{\includegraphics[width=.4\columnwidth,trim=10 0 11 0,
        clip]{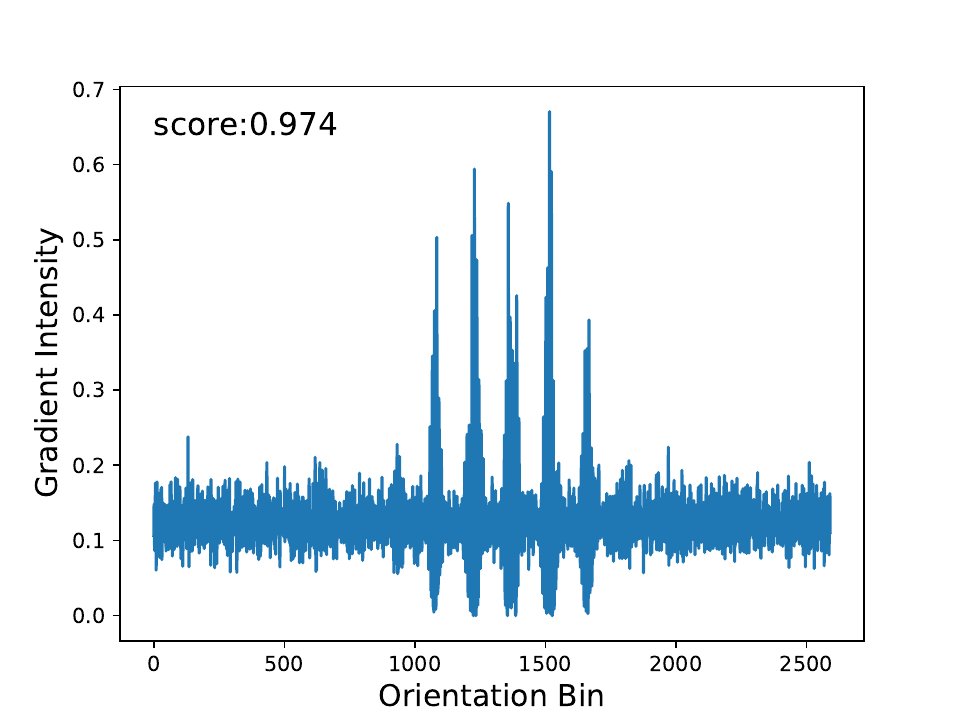}}\hfill
    \subfigure{\includegraphics[width=.4\columnwidth,trim=10 0 11 0,
        clip]{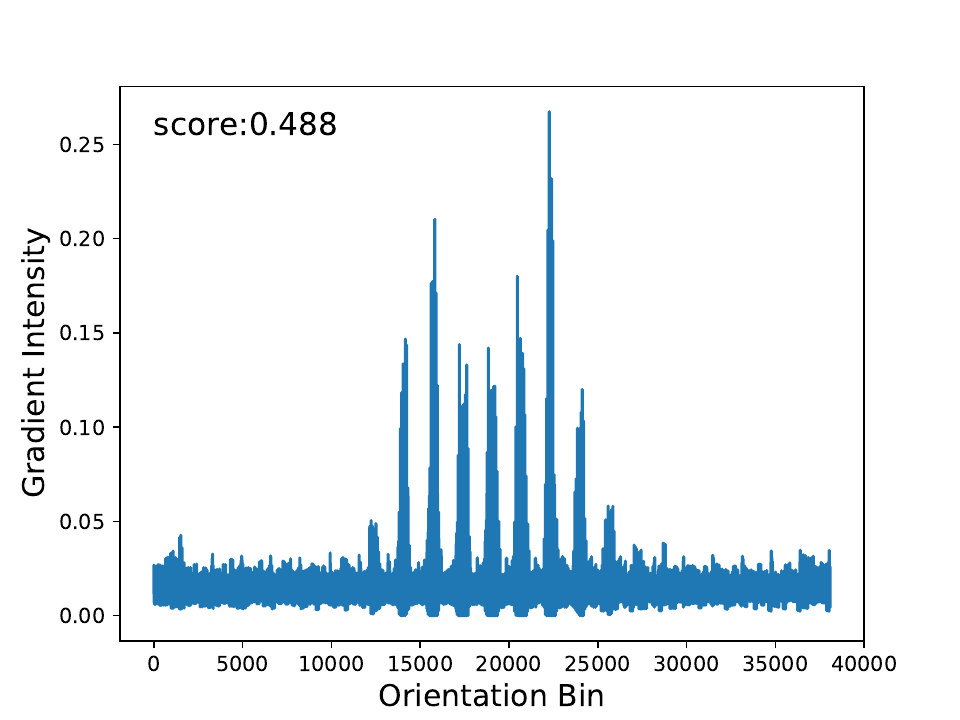}}\hfill
  }\\ 
\caption{Top row: From left to right, the image of SLACSJ0956+5100 in
  the F814W filter, its HOG visualization for ppc-16-cpb-1, for
  ppc-12-cpb-3, the histogram for ppc-16-cpb-1, and the histogram
  for ppc-12-cpb-3.  The asymmetric arc features are better captured
  in the latter HOG parameterization.  Bottom row: From left to right,
  the image of SLACSJ1420+6019 in the F555W filter and the analogous
  HOG visualizations and histograms.  For this image, the arc features
  are better captured in the former HOG parmeterization, where arc
  features contribute to the right side of the
  histogram.}\vspace{10pt}
\label{fig:hogvariations}
\end{figure*}

The model trained on mock HST images using the best grid search output
parameterization, ppc-16-cpb-1, does not uniformly perform well on all
images from the SLACS sample.  For example, the evaluated score for
SLACSJ0956+5100 imaged in the F814W filter increased from 0.436 to
0.959 with ppc-12-cpb-3. From left to right, the top row of
Figure~\ref{fig:hogvariations} shows the image, HOG visualization for
ppc-16-cpb-1, HOG visualization for ppc-12-cpb-3, histogram for
ppc-16-cpb-1, and histogram for ppc-12-cpb-3.  As a counter-example,
the bottom row of Figure~\ref{fig:hogvariations} shows SLACSJ1420+6019
imaged in F555W filter and its transition went from a score of 0.974
to 0.488 in ppc-12-cpb-3.

\begin{figure*}[t]
  \centering 
  \mbox{
    \subfigure{\includegraphics[width=.66\columnwidth,trim=13 0 10 0,
        clip]{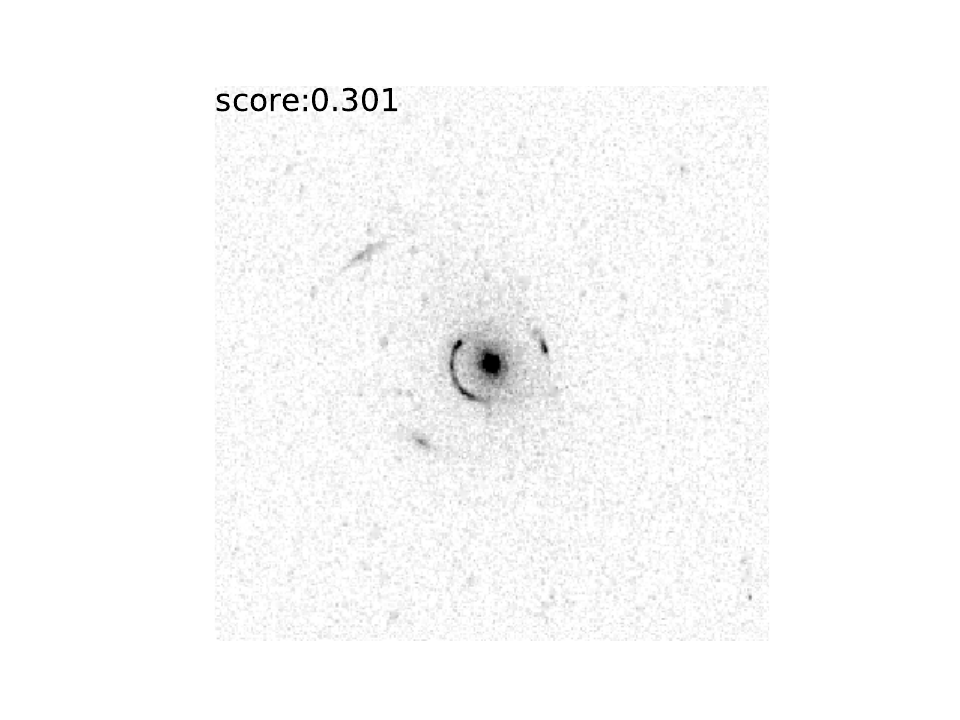}}\hfill
    \subfigure{\includegraphics[width=.66\columnwidth,trim=13 0 10 0,
        clip]{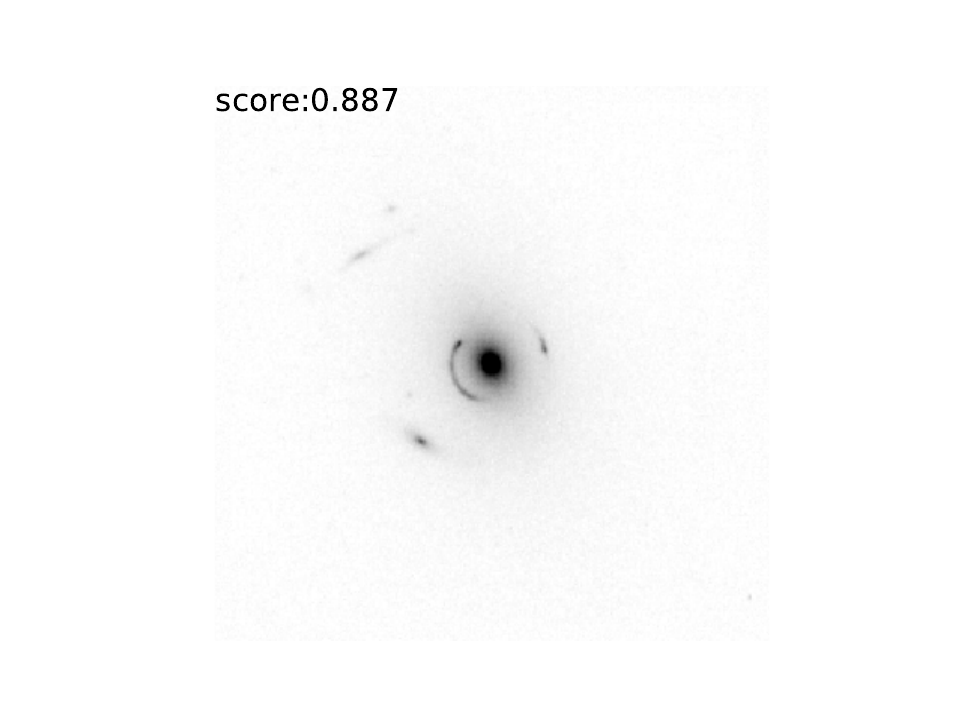}}\hfill
    \subfigure{\includegraphics[width=.66\columnwidth,trim=13 0 10 0,
        clip]{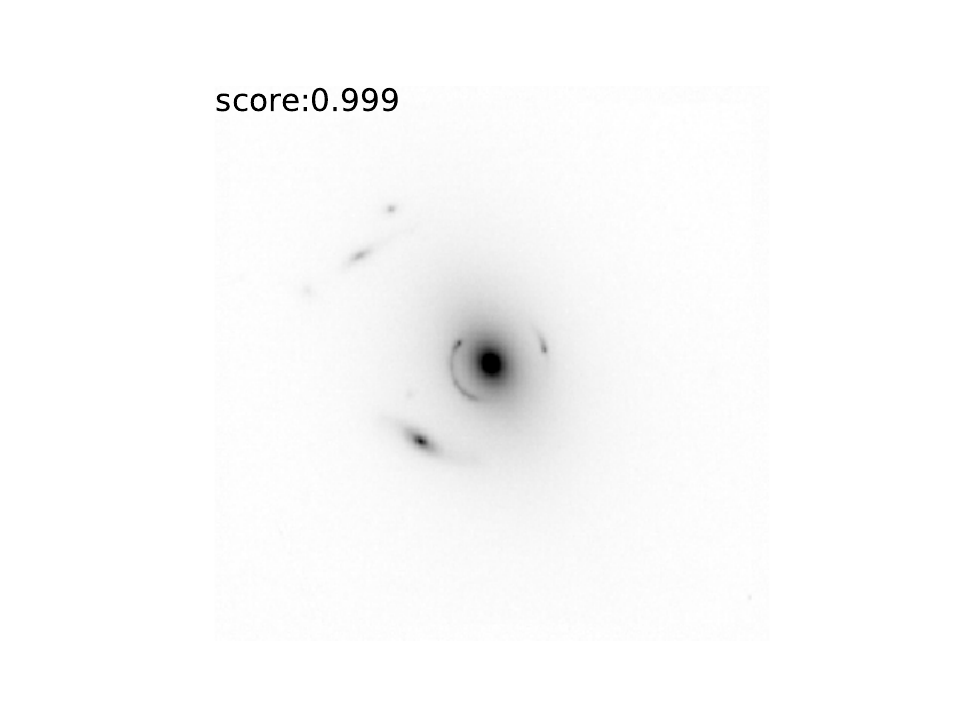}\hfill}
  }\\ 
  \mbox{
    \subfigure{\includegraphics[width=.66\columnwidth,trim=13 0 10 0,
        clip]{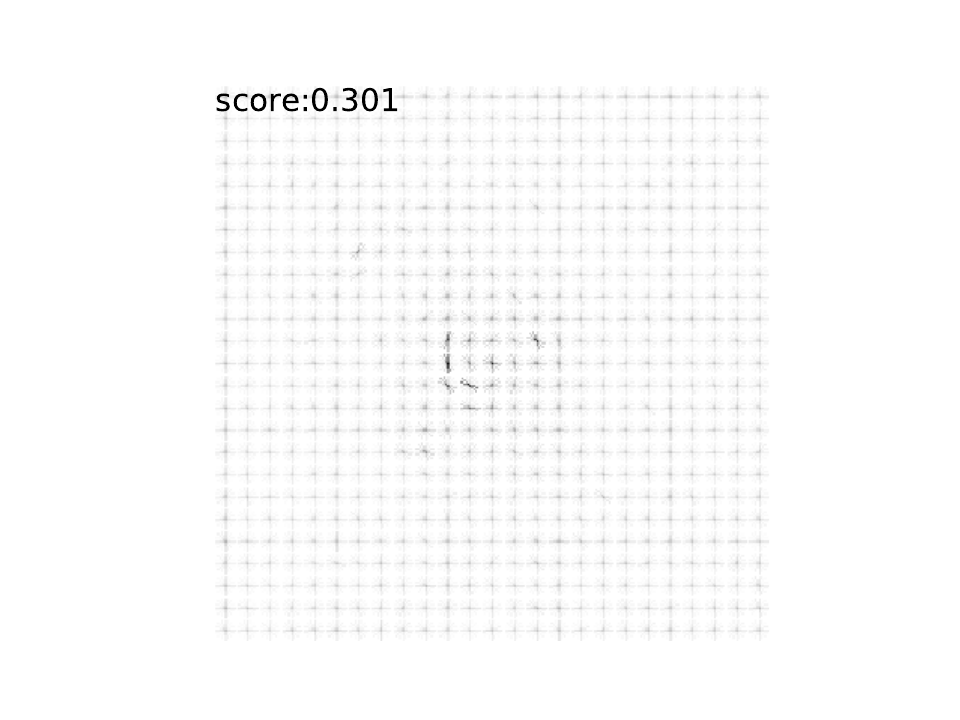}}\hfill
    \subfigure{\includegraphics[width=.66\columnwidth,trim=13 0 10 0,
        clip]{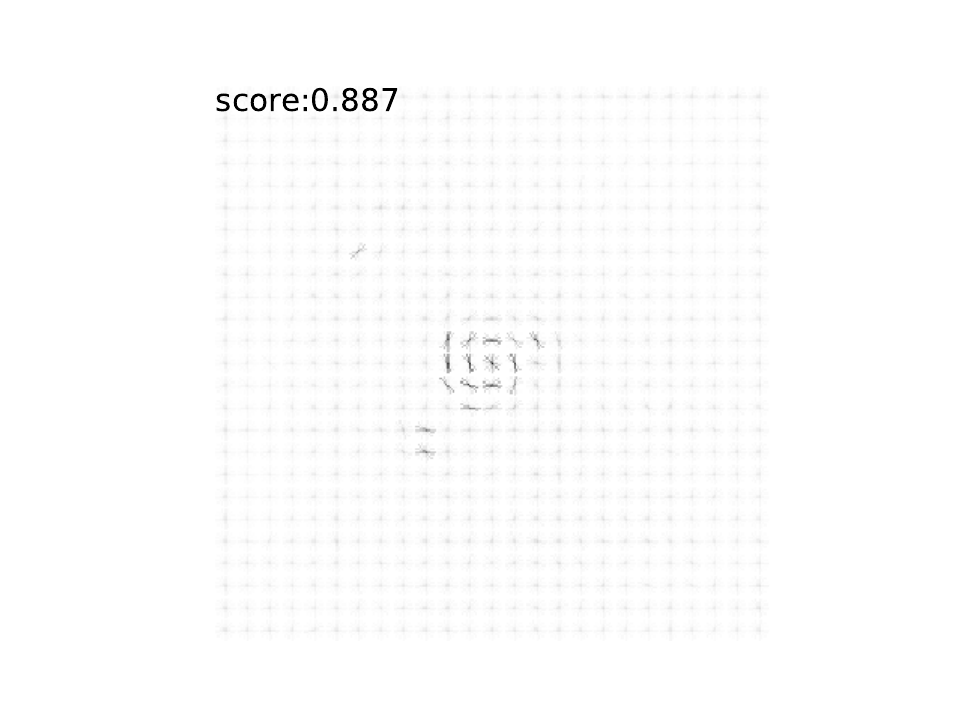}}\hfill
    \subfigure{\includegraphics[width=.66\columnwidth,trim=13 0 10 0,
        clip]{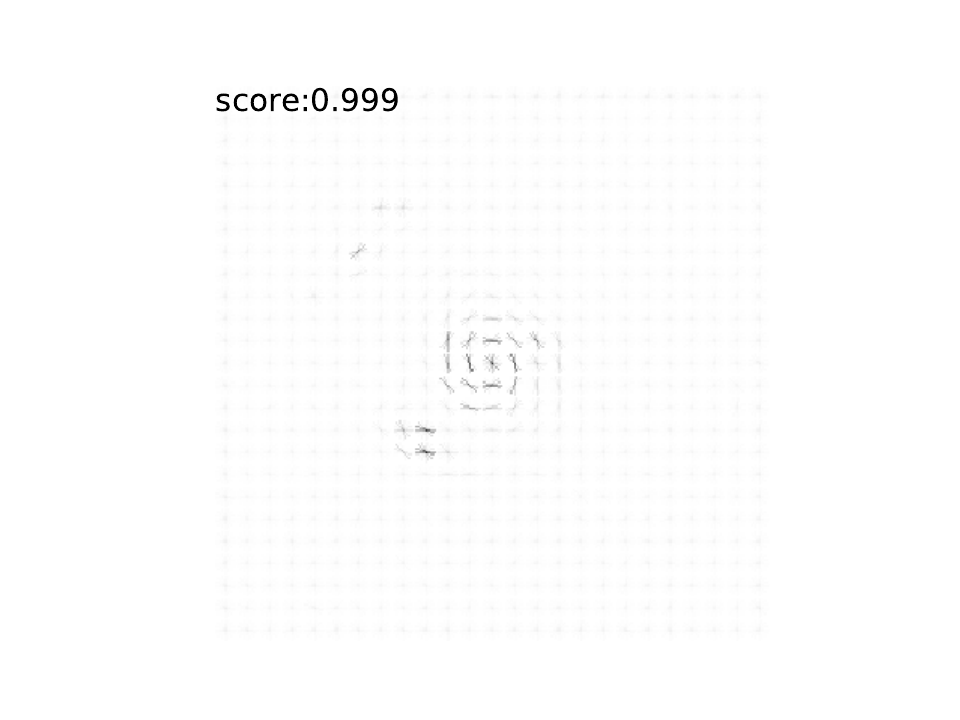}\hfill}
  }\\ 
  \mbox{
    \subfigure{\includegraphics[width=.65\columnwidth,trim=10 0 11 0,
        clip]{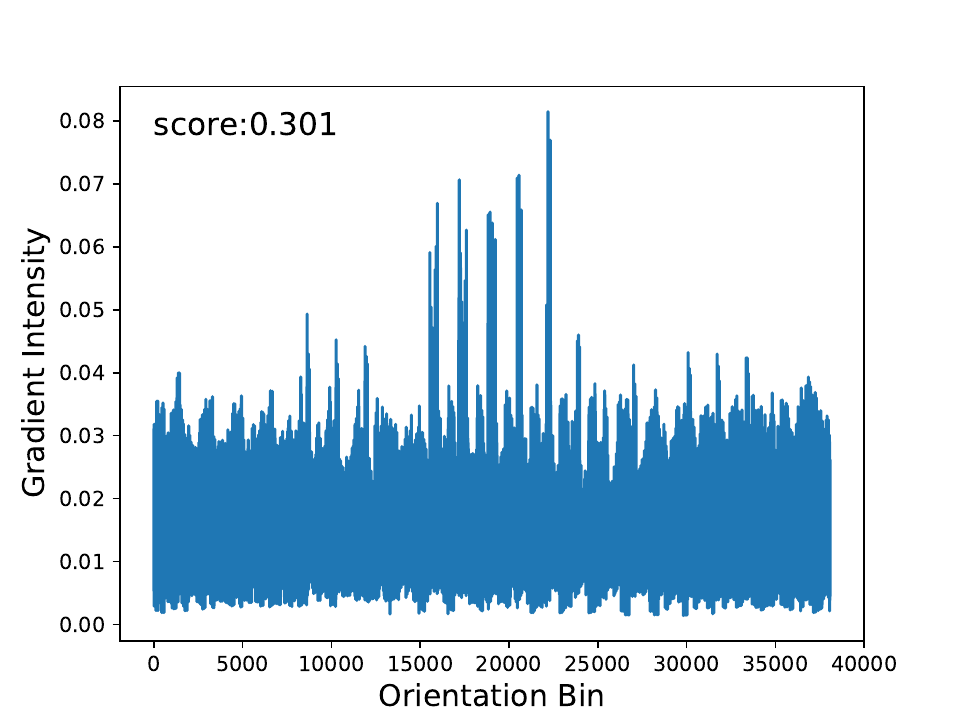}}\hfill
    \subfigure{\includegraphics[width=.65\columnwidth,trim=10 0 11 0,
        clip]{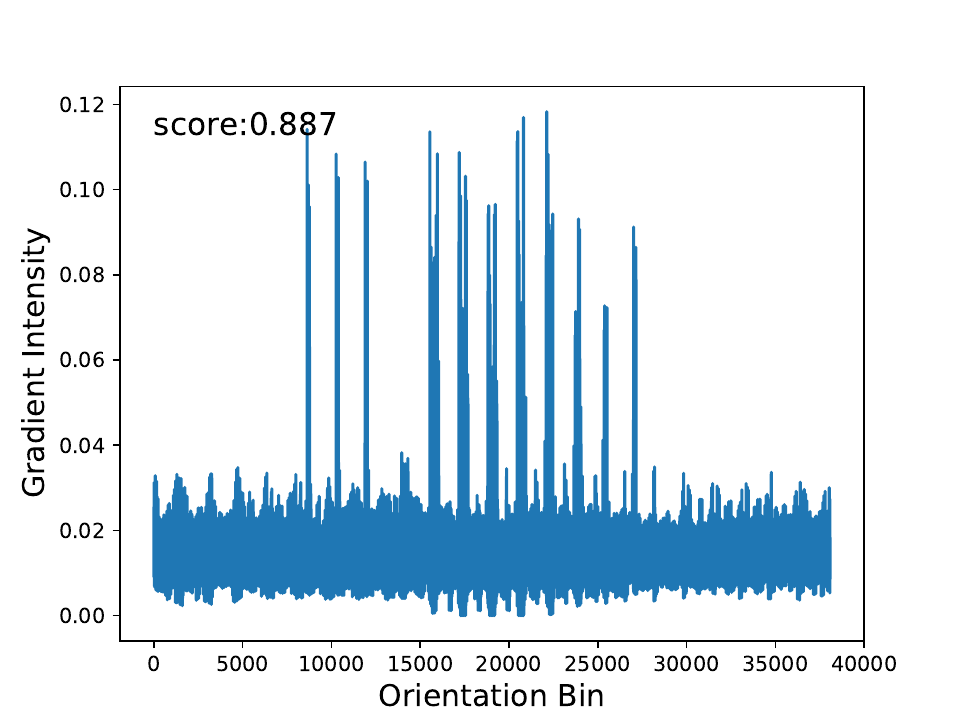}}\hfill
    \subfigure{\includegraphics[width=.65\columnwidth,trim=10 0 11 0,
        clip]{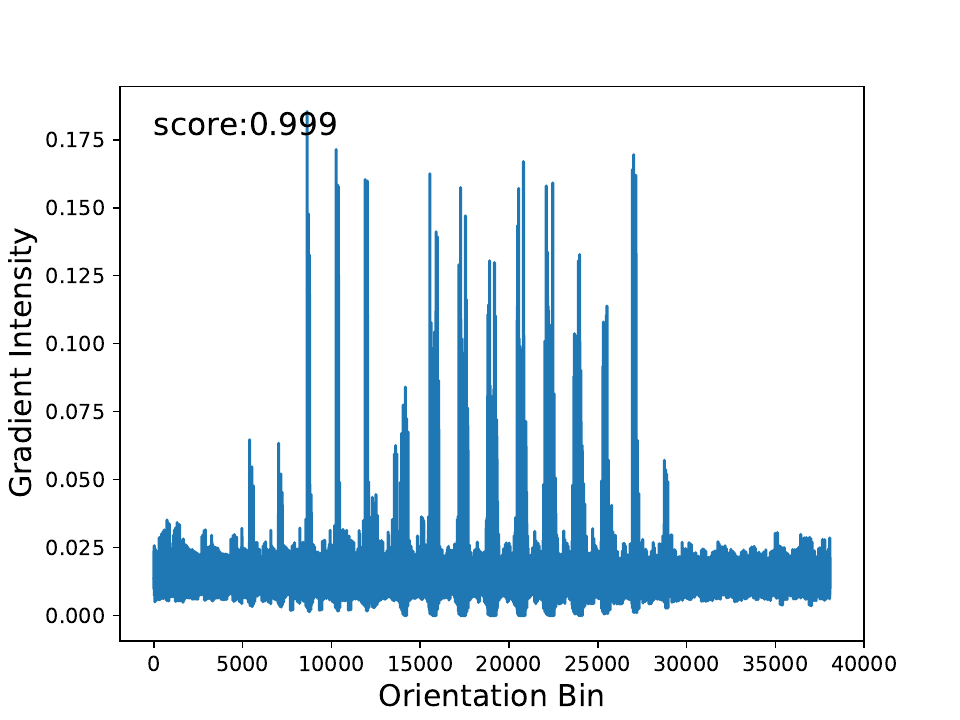}\hfill}
  }\\ 
\caption{Top row: Image of SLACSJ1205+4910 in the F435W, F555W, and
  the F814W filters.  Middle row: Corresponding visualization of the
  HOG of each image with ppc-12-cpb3.  Bottom row: The histogrammed
  edges.  Grainy features in F435W increases the histogram values in
  every direction, swamping the edge signatures in the left side of
  the histogram and lowering the score.}\vspace{10pt}
\label{fig:bestSLACS}
\end{figure*}

The image graininess present in real data can impact how well a given
HOG parameterization can capture the mophological features of an arc.
In Figure~\ref{fig:bestSLACS}, SLACSJ1205+4910 is an example of a
visibly clear lens that is highly scored for its image in the F814W
and F555W filters, but has a significantly lower score in the F435W
filter.  We also show the feature vector and the visualization of the
HOG in the middle and right panels.  The grainy features correspond to
a higher normalization in additional bins of oriented edges, swamping
the signal from the lens edge.

To assess the variations in how well a given HOG parameterization can
capture lens parameters, we trained models on only 1000 lensed and
non-lensed mock HST images in 8 different sets of HOG
parameterizations in the spirit of a grid search: ppc-6-cpb3,
ppc-8-cpb-2, ppc-8-cpb3, ppc-8-cpb-4, ppc-8-cpb-10, ppc-12-cpb-3,
ppc-16-cpb-1, ppc-16-cpb-3.  We tested each of these models on a
separate 1000 lensed and non-lensed mock images, deriving an AUC on
the mock dataset to assess the robustness of these models within the
simulated data.  We then measured the AUC of each model when evaluated
on the SLACS data divided into bands.  The results of our test are
summarized in Table~\ref{tab:mockandSLACS}.  Note, the AUCs of the
mock test data are not as high as the results quoted in
Figure~\ref{fig:rotation_test} because of the difference in training
set size.

\begin{table*}
\caption{Effects of HOG Parameterization on Model Evaluation}
\begin{center}
\begin{tabular}{ccccccccccccccc}
\hline \\ [-0.2cm]
PPC & CPB & AUC$_{\rm mock}$ & AUC$_{\rm SL}$ & AUC$_{\rm 814}$ & AUC$_{\rm 435}$ & P$_{\rm mock}$ & R$_{\rm mock}$ & P$_{\rm SL}$ & R$_{\rm SL}$ & P$_{\rm 814}$ & R$_{\rm 814}$ & P$_{\rm 435}$ & R$_{\rm 435}$ \\ 
\hline \\ 
(6, 6) & (3, 3) &  0.901 & 0.469 (0.470$\pm$0.052) & 0.470 (0.471$\pm$0.066) & 0.311 (0.312$\pm$0.099) & 0.713 & 0.916 & 0.741 & 0.496 & 0.758 & 0.588 & 0.333 & 0.125 \\ [0.01cm]
(8, 8) & (2, 2) & 0.889 & 0.540 (0.541$\pm$0.053) & 0.543 (0.543$\pm$0.066) & 0.417 (0.417$\pm$0.108) & 0.670 & 0.936 & 0.781 & 0.678 & 0.789 & 0.750 & 0.563 & 0.375 & \\ [0.01cm]
(8, 8) & (3, 3) & 0.920 & 0.566 (0.566$\pm$0.053) & 0.573 (0.573$\pm$0.066) & 0.426 (0.426$\pm$0.113) & 0.712 & 0.954 & 0.777 & 0.603 & 0.779 & 0.663 & 0.538 & 0.292 & \\ [0.01cm]
(8, 8) & (4, 4) & 0.934 & 0.574 (0.575$\pm$0.052) & 0.578 (0.579$\pm$0.064) & 0.423 (0.422$\pm$0.117) & 0.731 & 0.955 & 0.795 & 0.545 & 0.800 & 0.650 & 0.333 & 0.083 & \\ [0.01cm]
(10, 10) & (3, 3) & 0.925 & 0.567 (0.566$\pm$0.053) & 0.548 (0.548$\pm$0.068) & 0.433 (0.432$\pm$0.111) & 0.729 & 0.924 & 0.787 & 0.612 & 0.500 & 0.208 & 0.787 & 0.738 & \\ [0.01cm]
(12, 12) & (3, 3) & 0.932 & 0.612 (0.611$\pm$0.052) & 0.607 (0.607$\pm$0.065) & 0.503 (0.505$\pm$0.107) & 0.724 & 0.943 & 0.817 & 0.554 & 0.809 & 0.688 & 0.333 & 0.042 & \\ [0.01cm]
(16, 16) & (1, 1) & 0.820 & 0.509 (0.509$\pm$0.056) & 0.481 (0.481$\pm$0.067) & 0.474 (0.475$\pm$0.108) & 0.581 & 0.933 & 0.782 & 0.711 & 0.769 & 0.750 & 0.667 & 0.500 & \\ [0.01cm]
(16, 16) & (3, 3) & 0.942 & 0.580 (0.580$\pm$0.052) & 0.576 (0.574$\pm$0.063) & 0.439 (0.437$\pm$0.116) & 0.698 & 0.963 & 0.797 & 0.620 & 0.787 & 0.738 & 0.400 & 0.083 & \\ [0.01cm]
\hline \\
\end{tabular}
\end{center}
\label{tab:mockandSLACS}

\tablecomments{A table summary of a grid-search-like test to check if
  a given HOG parameterization might best capture arc morphology in
  the varying levels of graininess in observed data.  The columns
  correspond to the pixels-per-cell HOG parameterization,
  cells-per-block, the AUC of the ROC curve of mock test data, the AUC
  of the ROC curve of all of the SLACS data (SL), the subselection of
  F814W filter images (814), and the subselection of F435W filter
  images 435 in addition to the respective precision and recall
values at a threshold of 0.5.  We do not provide AUC values in the
F555W filter because there are no non-lens images in the filter.  The
AUC columns for the SLACS data include the average and standard
deviation of the AUC from bootstrapping.  We also include precision
and recall values at the 0.5 threshold.}
\end{table*}

First, we should note that our overall sample to test on is relatively
small, so the uncertainty of the AUC values for the SLACS dataset is
rather large.  To quantify this, we compute the bootstrap average and
standard deviation, which are shown in the parenthesis of the table
column for the AUCs.  Note, when subselecting on individual bands, the
standard deviation can be almost 25\%.

Next, of the HOG parameterizations we tested, ppc-12-cpb3 performed
the best on all of the SLACS data, and in particular on images from
the F814W filter.  Finally, the main point we would like to emphasize
is that the models were trained on mock HST data do not contain the
same artifiacts or variations in graininess as the real observations
in the SLACS sample.  But, for a given image, a HOG parameterization
can be selected that ignores grainy features and keeps features that
correspond to arcs.  The current problem is that the SLACS sample is
too small to run a robust systematic search.  In absence of a single
ideal HOG parameterization for a SLACS-like sample, we could identify
a handful of HOG parameterizations that describe subsets of the data
and concatenate the vectors from each HOG parameterization for the
feature vector. Feature vector concatenation is how we leveraged
multi-band data from the ground-based sample in
\citet{metcalfetal18}.  Another alternative would be to quantify the
quality of an image by a metric that correlates best with HOG
parameterization.

\section{Summary and Discussions}
\label{sec:conclusions}

We have presented a supervised classification pipeline to
automatically identify galaxy-galaxy strong lensing systems using a
histogram of oriented gradients (HOG) as a feature extractor, and
logistic regression (LR) as a machine learning algorithm.  Our
pipeline can easily be extended to identify other strong lensing
features, such as multiply imaged quasars, and to test alternative
features and/or machine learning algorithms.

We have also made use of a new sophisticated set of mock observations,
which will be made publicly available.  The lensing systems have
lens galaxies generated with a realistic redshift distribution and
along the line of sight galaxies drawn from Hubble Ultra-Deep field
observations.  We have explored results from mock HST, one year LSST,
and ten year LSST observations.

We summarize key points below:

\begin{itemize}

\item We have designed our pipeline to easily select and add image
  pre-processing and feature extraction methods, and to select a
  machine learning algorithm for classification.  Additionally, the
  user can easily perform parameter searches to train a model with the
  best parameters for a given problem.

\item We have tested and run parameter tests for a histogram of
  oriented gradients as an efficient and effective feature extractor
  for galaxy-galaxy strong lensing systems in both a space based
  (HST-like) and ground based (LSST-like) observation.  We have also
  tested and run parameter tests for logistic regression as a
  scalable, cheap, and effective machine learning algorithm.

\item We find AUC values of ROC curves of optimized classifier models
  to yield $AUC=0.975$ for the HST-like data, $AUC=0.809$ for the
  stacked LSST-like data.  Model performance exhibits continual
  increase with the training size.

\item While removal of the lens galaxy improves model performance for
  smaller size training samples, features from the lens galaxy improve
  model performance for larger training data sets.

\item Images that were easiest for our model to classify typically
  were lens systems that had high lensed image magnification and a
  lens galaxy with large velocity dispersion or non-lens systems with
  lens galaxies with smaller velocity dispersion and non-elongated
  along the line of sight galaxies.

\item We have explored the potential of HOG/LR in mitigating the
  problem where simulations are not able to capture imperfections and
  details in real data from the SLACS sample.  The results indicate
  that different HOG parameterizations can be robust to different
  amounts of noise/defects that are not captured by our simulations.
  However, no single HOG parameterization is able to maximally perform
  on the ensemble of SLACS images.  With a larger data set, a
  systematic study that couples image quality to the best HOG
  parameterizations would be possible.
\end{itemize}

We emphasize that simple linear classifiers, such as logistic
regression, are scalable and relatively easy to parallelize with open
source tools such as {\em Apache
  Spark}\footnote{http://spark.apache.org/}.  Our work indicates that
HOG feature extraction plus a linear classifier captures much of the
morphological complexity in the arc finding problem.  We have also
tested how HOG feature extraction can be parameterized to be less
sensitive to real image quality variations that are not captured by
simulations, but further tests on a larger sample of data will be
necessary.  The methods also scale to large datasets on a computing
cluster, if needed.

One major caveat to our results is the fact that our mock data does
not describe the full distribution of lens and non-lens images that
will be observed, a shortcoming to be addressed in future mock data
work.  For example, the along the line of sight galaxies in our
training and test images all come from the Hubble Ultra Deep field,
sampling a smaller range of potential contaminants that are not
associated with a lensed image.  A limitation of the CANDELS sources
is that the sources are observed with HST PSF, and would not resolve
the clumpiness within a true HST arc.  Also, the redshift of the
source galaxies have been fixed to $z_s=2$.  Varying the source
redshifts affects the relative brightness between the lens and the
source.  Note, in our lens-classification method, the HOG image
processing step normalizes the contrast of local histograms within
blocks of the image, providing an option to enable results that are
more invariant to changes in brightness across the image.

Finally, we must take the class balance, or relative number of lens to
non-lens systems, into account when assessing a metric.  The
Precision-Recall (purity-completeness) metric is sensitive to the
ratio of lens and non-lens systems in the data.  Both of our training
and test data sets have $50\%$ lensed and $50\%$ unlensed images,
which is not expected in observations.  Again, the precision of a
method is what would enable efficient spectroscopic or human-based
follow-up.

For images selected for a massive elliptical, the number of non-lenses
will outnumber lenses by at least 1000 to 1.  Therefore, a sample that
is $50\%$ pure requires a classifier with a false positive rate of
0.001.  Looking at the solid green line in
Figure~\ref{fig:PRcompilation}, we can set a high classification
threshold and obtain a sample with close to 100\% purity and up to
$\sim$10\% recall of all of our lenses, before contamination from
non-lenses leaks into the selection. While information from other
bands will certainly improve the model performance, a maximally large
and pure sample from our method would likely require further
filtering, e.g. by citizen science, or modification to how the HOG
features are used by machine learning classifiers.  However, neural
networks have the current best performance in application to the
strong lens finding problem and are the best single approach to the
pure lens-finding problem \citep{metcalfetal18}.

The ROC curve metric is insensitive to the ratio, but is sensitive to
the sampling.  Given alternative lens and non-lens sample splittings,
our true positive and false positive rates in the ROC curves would
stay the same, making the ROC curve a more standard metric in the
literature.  On the other hand, the ROC curves show a representative
rate for lens and source distributions that are evenly sampled.  We do
{\it not} expect this sampling to be representative of what we might
expect from a an observational survey.  We leave these additional
challenges to future work.

\acknowledgments CA acknowledges support from the Enrico Fermi
Institute at the University of Chicago, and the University of Chicago
Provost's Office. NL would like to thank the funding support from
NSFC, grant no.11503064, {as well as a UK Science and Technology
  Facilities Council research grant}. WL thanks the funding support
from Shanghai Natural Science Foundation, grant no. 15ZR1446700. This
work was also supported in part by the Kavli Institute for
Cosmological Physics at the University of Chicago through grant NSF
PHY-1125897 and an endowment from the Kavli Foundation and its founder
Fred Kavli. The authors thank the referee for their careful reading of
the manuscript and constructive comments that have significantly
improved the paper, Adam Bolton for providing the SLACS data, and Mike
Gladders for useful discussions.
\lastpagefootnotes

\bibliography{ms}

\begin{thebibliography}{}
\expandafter\ifx\csname natexlab\endcsname\relax\def\natexlab#1{#1}\fi

\bibitem[{{Agnello} {et~al.}(2015){Agnello}, {Kelly}, {Treu}, \&
  {Marshall}}]{agnelloetal15}
{Agnello}, A., {Kelly}, B.~C., {Treu}, T., \& {Marshall}, P.~J. 2015,
  \href{http://dx.doi.org/10.1093/mnras/stv037}{\mnras},
  \href{http://adsabs.harvard.edu/abs/2015MNRAS.448.1446A}{448},
  \href{http://adsabs.harvard.edu/abs/2015MNRAS.448.1446A}{1446}

\bibitem[{{Alard}(2006)}]{alard06}
{Alard}, C. 2006, ArXiv Astrophysics e-prints,
  \href{http://adsabs.harvard.edu/abs/2006astro.ph..6757A}{astro-ph/0606757}

\bibitem[{{Allam} {et~al.}(2007){Allam}, {Tucker}, {Lin}, {Diehl}, {Annis},
  {Buckley-Geer}, \& {Frieman}}]{allametal07}
{Allam}, S.~S., {Tucker}, D.~L., {Lin}, H., {et~al.} 2007,
  \href{http://dx.doi.org/10.1086/519520}{\apjl},
  \href{http://adsabs.harvard.edu/abs/2007ApJ...662L..51A}{662},
  \href{http://adsabs.harvard.edu/abs/2007ApJ...662L..51A}{L51}

\bibitem[{{Bezecourt} {et~al.}(1998){Bezecourt}, {Pello}, \&
  {Soucail}}]{bezecourtetal98}
{Bezecourt}, J., {Pello}, R., \& {Soucail}, G. 1998, \aap,
  \href{http://adsabs.harvard.edu/abs/1998A%26A...330..399B}{330},
  \href{http://adsabs.harvard.edu/abs/1998A%26A...330..399B}{399}

\bibitem[{{Bolton} {et~al.}(2008){Bolton}, {Burles}, {Koopmans}, {Treu},
  {Gavazzi}, {Moustakas}, {Wayth}, \& {Schlegel}}]{boltonetal08}
{Bolton}, A.~S., {Burles}, S., {Koopmans}, L.~V.~E., {et~al.} 2008,
  \href{http://dx.doi.org/10.1086/589327}{\apj},
  \href{http://adsabs.harvard.edu/abs/2008ApJ...682..964B}{682},
  \href{http://adsabs.harvard.edu/abs/2008ApJ...682..964B}{964}

\bibitem[{{Bom} {et~al.}(2016){Bom}, {Makler}, {Albuquerque}, \&
  {Brandt}}]{bometal16}
{Bom}, C.~R., {Makler}, M., {Albuquerque}, M.~P., \& {Brandt}, C.~H. 2016,
  ArXiv e-prints,
  arXiv:\href{http://adsabs.harvard.edu/abs/2016arXiv160704644B}{1607.04644}

\bibitem[{{Bonvin} {et~al.}(2017){Bonvin}, {Courbin}, {Suyu}, {Marshall},
  {Rusu}, {Sluse}, {Tewes}, {Wong}, {Collett}, {Fassnacht}, {Treu}, {Auger},
  {Hilbert}, {Koopmans}, {Meylan}, {Rumbaugh}, {Sonnenfeld}, \&
  {Spiniello}}]{bonvinetal17}
{Bonvin}, V., {Courbin}, F., {Suyu}, S.~H., {et~al.} 2017,
  \href{http://dx.doi.org/10.1093/mnras/stw3006}{\mnras},
  \href{http://adsabs.harvard.edu/abs/2017MNRAS.465.4914B}{465},
  \href{http://adsabs.harvard.edu/abs/2017MNRAS.465.4914B}{4914}

\bibitem[{{Brault} \& {Gavazzi}(2015)}]{braultandgavazzi15}
{Brault}, F., \& {Gavazzi}, R. 2015,
  \href{http://dx.doi.org/10.1051/0004-6361/201425275}{\aap},
  \href{http://adsabs.harvard.edu/abs/2015A%26A...577A..85B}{577},
  \href{http://adsabs.harvard.edu/abs/2015A%26A...577A..85B}{A85}

\bibitem[{{Chae}(2003)}]{chae03}
{Chae}, K.-H. 2003,
  \href{http://dx.doi.org/10.1111/j.1365-2966.2003.07092.x}{\mnras},
  \href{http://adsabs.harvard.edu/abs/2003MNRAS.346..746C}{346},
  \href{http://adsabs.harvard.edu/abs/2003MNRAS.346..746C}{746}

\bibitem[{{Collett}(2015)}]{collett15}
{Collett}, T.~E. 2015,
  \href{http://dx.doi.org/10.1088/0004-637X/811/1/20}{\apj},
  \href{http://adsabs.harvard.edu/abs/2015ApJ...811...20C}{811},
  \href{http://adsabs.harvard.edu/abs/2015ApJ...811...20C}{20}

\bibitem[{{Collett} \& {Auger}(2014)}]{collettandauger14}
{Collett}, T.~E., \& {Auger}, M.~W. 2014,
  \href{http://dx.doi.org/10.1093/mnras/stu1190}{\mnras},
  \href{http://adsabs.harvard.edu/abs/2014MNRAS.443..969C}{443},
  \href{http://adsabs.harvard.edu/abs/2014MNRAS.443..969C}{969}

\bibitem[{{Collett} {et~al.}(2012){Collett}, {Auger}, {Belokurov}, {Marshall},
  \& {Hall}}]{collettetal12}
{Collett}, T.~E., {Auger}, M.~W., {Belokurov}, V., {Marshall}, P.~J., \&
  {Hall}, A.~C. 2012,
  \href{http://dx.doi.org/10.1111/j.1365-2966.2012.21424.x}{\mnras},
  \href{http://adsabs.harvard.edu/abs/2012MNRAS.424.2864C}{424},
  \href{http://adsabs.harvard.edu/abs/2012MNRAS.424.2864C}{2864}

\bibitem[{{Connolly} {et~al.}(2010){Connolly}, {Peterson}, {Jernigan}, {Abel},
  {Bankert}, {Chang}, {Claver}, {Gibson}, {Gilmore}, {Grace}, {Jones},
  {Ivezic}, {Jee}, {Juric}, {Kahn}, {Krabbendam}, {Krughoff}, {Lorenz},
  {Pizagno}, {Rasmussen}, {Todd}, {Tyson}, \& {Young}}]{connollyetal10}
{Connolly}, A.~J., {Peterson}, J., {Jernigan}, J.~G., {et~al.} 2010, in
  \procspie, Vol. 7738, Modeling, Systems Engineering, and Project Management
  for Astronomy IV, 77381O, doi:10.1117/12.857819

\bibitem[{{Dalal} \& {Triggs}(2005)}]{dalalandtriggs05}
{Dalal}, N., \& {Triggs}, B. 2005,
  \href{http://dx.doi.org/10.1109/CVPR.2005.177}{IEEE Computer Society
  Conference on Computer Vision and Pattern Recognition},
  doi:10.1109/CVPR.2005.177

\bibitem[{{Dieleman} {et~al.}(2015){Dieleman}, {Willett}, \&
  {Dambre}}]{dielemanetal15}
{Dieleman}, S., {Willett}, K.~W., \& {Dambre}, J. 2015,
  \href{http://dx.doi.org/10.1093/mnras/stv632}{\mnras},
  \href{http://adsabs.harvard.edu/abs/2015MNRAS.450.1441D}{450},
  \href{http://adsabs.harvard.edu/abs/2015MNRAS.450.1441D}{1441}

\bibitem[{{Dye} {et~al.}(2008){Dye}, {Evans}, {Belokurov}, {Warren}, \&
  {Hewett}}]{dyeetal08}
{Dye}, S., {Evans}, N.~W., {Belokurov}, V., {Warren}, S.~J., \& {Hewett}, P.
  2008, \href{http://dx.doi.org/10.1111/j.1365-2966.2008.13401.x}{\mnras},
  \href{http://adsabs.harvard.edu/abs/2008MNRAS.388..384D}{388},
  \href{http://adsabs.harvard.edu/abs/2008MNRAS.388..384D}{384}

\bibitem[{{Estrada} {et~al.}(2007){Estrada}, {Annis}, {Diehl}, {Hall}, {Las},
  {Lin}, {Makler}, {Merritt}, {Scarpine}, {Allam}, \& {Tucker}}]{estradaetal07}
{Estrada}, J., {Annis}, J., {Diehl}, H.~T., {et~al.} 2007,
  \href{http://dx.doi.org/10.1086/512599}{\apj},
  \href{http://adsabs.harvard.edu/abs/2007ApJ...660.1176E}{660},
  \href{http://adsabs.harvard.edu/abs/2007ApJ...660.1176E}{1176}

\bibitem[{{Galametz} {et~al.}(2013){Galametz}, {Grazian}, {Fontana},
  {Ferguson}, {Ashby}, {Barro}, {Castellano}, {Dahlen}, {Donley}, {Faber},
  {Grogin}, {Guo}, {Huang}, {Kocevski}, {Koekemoer}, {Lee}, {McGrath}, {Peth},
  {Willner}, {Almaini}, {Cooper}, {Cooray}, {Conselice}, {Dickinson}, {Dunlop},
  {Fazio}, {Foucaud}, {Gardner}, {Giavalisco}, {Hathi}, {Hartley}, {Koo},
  {Lai}, {de Mello}, {McLure}, {Lucas}, {Paris}, {Pentericci}, {Santini},
  {Simpson}, {Sommariva}, {Targett}, {Weiner}, {Wuyts}, \& {the CANDELS
  Team}}]{galametzetal13}
{Galametz}, A., {Grazian}, A., {Fontana}, A., {et~al.} 2013,
  \href{http://dx.doi.org/10.1088/0067-0049/206/2/10}{\apjs},
  \href{http://adsabs.harvard.edu/abs/2013ApJS..206...10G}{206},
  \href{http://adsabs.harvard.edu/abs/2013ApJS..206...10G}{10}

\bibitem[{{Gavazzi} {et~al.}(2014){Gavazzi}, {Marshall}, {Treu}, \&
  {Sonnenfeld}}]{gavazzietal14}
{Gavazzi}, R., {Marshall}, P.~J., {Treu}, T., \& {Sonnenfeld}, A. 2014,
  \href{http://dx.doi.org/10.1088/0004-637X/785/2/144}{\apj},
  \href{http://adsabs.harvard.edu/abs/2014ApJ...785..144G}{785},
  \href{http://adsabs.harvard.edu/abs/2014ApJ...785..144G}{144}

\bibitem[{{Gavazzi} {et~al.}(2007){Gavazzi}, {Treu}, {Rhodes}, {Koopmans},
  {Bolton}, {Burles}, {Massey}, \& {Moustakas}}]{gavazzietal07}
{Gavazzi}, R., {Treu}, T., {Rhodes}, J.~D., {et~al.} 2007,
  \href{http://dx.doi.org/10.1086/519237}{\apj},
  \href{http://adsabs.harvard.edu/abs/2007ApJ...667..176G}{667},
  \href{http://adsabs.harvard.edu/abs/2007ApJ...667..176G}{176}

\bibitem[{{Gladders} {et~al.}(2003){Gladders}, {Hoekstra}, {Yee}, {Hall}, \&
  {Barrientos}}]{gladdersetal03}
{Gladders}, M.~D., {Hoekstra}, H., {Yee}, H.~K.~C., {Hall}, P.~B., \&
  {Barrientos}, L.~F. 2003, \href{http://dx.doi.org/10.1086/376518}{\apj},
  \href{http://adsabs.harvard.edu/abs/2003ApJ...593...48G}{593},
  \href{http://adsabs.harvard.edu/abs/2003ApJ...593...48G}{48}

\bibitem[{{Grogin} {et~al.}(2011){Grogin}, {Kocevski}, {Faber}, {Ferguson},
  {Koekemoer}, {Riess}, {Acquaviva}, {Alexander}, {Almaini}, {Ashby}, {Barden},
  {Bell}, {Bournaud}, {Brown}, {Caputi}, {Casertano}, {Cassata}, {Castellano},
  {Challis}, {Chary}, {Cheung}, {Cirasuolo}, {Conselice}, {Roshan Cooray},
  {Croton}, {Daddi}, {Dahlen}, {Dav{\'e}}, {de Mello}, {Dekel}, {Dickinson},
  {Dolch}, {Donley}, {Dunlop}, {Dutton}, {Elbaz}, {Fazio}, {Filippenko},
  {Finkelstein}, {Fontana}, {Gardner}, {Garnavich}, {Gawiser}, {Giavalisco},
  {Grazian}, {Guo}, {Hathi}, {H{\"a}ussler}, {Hopkins}, {Huang}, {Huang},
  {Jha}, {Kartaltepe}, {Kirshner}, {Koo}, {Lai}, {Lee}, {Li}, {Lotz}, {Lucas},
  {Madau}, {McCarthy}, {McGrath}, {McIntosh}, {McLure}, {Mobasher},
  {Moustakas}, {Mozena}, {Nandra}, {Newman}, {Niemi}, {Noeske}, {Papovich},
  {Pentericci}, {Pope}, {Primack}, {Rajan}, {Ravindranath}, {Reddy}, {Renzini},
  {Rix}, {Robaina}, {Rodney}, {Rosario}, {Rosati}, {Salimbeni}, {Scarlata},
  {Siana}, {Simard}, {Smidt}, {Somerville}, {Spinrad}, {Straughn}, {Strolger},
  {Telford}, {Teplitz}, {Trump}, {van der Wel}, {Villforth}, {Wechsler},
  {Weiner}, {Wiklind}, {Wild}, {Wilson}, {Wuyts}, {Yan}, \&
  {Yun}}]{groginetal11}
{Grogin}, N.~A., {Kocevski}, D.~D., {Faber}, S.~M., {et~al.} 2011,
  \href{http://dx.doi.org/10.1088/0067-0049/197/2/35}{\apjs},
  \href{http://adsabs.harvard.edu/abs/2011ApJS..197...35G}{197},
  \href{http://adsabs.harvard.edu/abs/2011ApJS..197...35G}{35}

\bibitem[{{Hastie, Trevor J. and Tibshirani, Robert John and Friedman, Jerome
  H.}(2009)}]{hastie09}
{Hastie, Trevor J. and Tibshirani, Robert John and Friedman, Jerome H.} 2009,
  {The elements of statistical learning : data mining, inference, and
  prediction}, {Springer series in statistics} (New York: Springer),
  doi:10.1007/978-0-387-84858-7

\bibitem[{{Jacobs} {et~al.}(2017){Jacobs}, {Glazebrook}, {Collett}, {More}, \&
  {McCarthy}}]{jacobsetal17}
{Jacobs}, C., {Glazebrook}, K., {Collett}, T., {More}, A., \& {McCarthy}, C.
  2017, \href{http://dx.doi.org/10.1093/mnras/stx1492}{\mnras},
  \href{http://adsabs.harvard.edu/abs/2017MNRAS.471..167J}{471},
  \href{http://adsabs.harvard.edu/abs/2017MNRAS.471..167J}{167}

\bibitem[{{Joseph} {et~al.}(2014){Joseph}, {Courbin}, {Metcalf}, {Giocoli},
  {Hartley}, {Jackson}, {Bellagamba}, {Kneib}, {Koopmans}, {Lemson},
  {Meneghetti}, {Meylan}, {Petkova}, \& {Pires}}]{josephetal14}
{Joseph}, R., {Courbin}, F., {Metcalf}, R.~B., {et~al.} 2014,
  \href{http://dx.doi.org/10.1051/0004-6361/201423365}{\aap},
  \href{http://adsabs.harvard.edu/abs/2014A%26A...566A..63J}{566},
  \href{http://adsabs.harvard.edu/abs/2014A%26A...566A..63J}{A63}

\bibitem[{{Jullo} {et~al.}(2010){Jullo}, {Natarajan}, {Kneib}, {D'Aloisio},
  {Limousin}, {Richard}, \& {Schimd}}]{julloetal10}
{Jullo}, E., {Natarajan}, P., {Kneib}, J.-P., {et~al.} 2010,
  \href{http://dx.doi.org/10.1126/science.1185759}{Science},
  \href{http://adsabs.harvard.edu/abs/2010Sci...329..924J}{329},
  \href{http://adsabs.harvard.edu/abs/2010Sci...329..924J}{924}

\bibitem[{{Kneib} \& {Natarajan}(2011)}]{kneibandnatarajan11}
{Kneib}, J.-P., \& {Natarajan}, P. 2011,
  \href{http://dx.doi.org/10.1007/s00159-011-0047-3}{\aapr},
  \href{http://adsabs.harvard.edu/abs/2011A%26ARv..19...47K}{19},
  \href{http://adsabs.harvard.edu/abs/2011A%26ARv..19...47K}{47}

\bibitem[{{Kochanek}(1996)}]{kochanek96}
{Kochanek}, C.~S. 1996, \href{http://dx.doi.org/10.1086/178175}{\apj},
  \href{http://adsabs.harvard.edu/abs/1996ApJ...473..595K}{473},
  \href{http://adsabs.harvard.edu/abs/1996ApJ...473..595K}{595}

\bibitem[{{Koekemoer} {et~al.}(2011){Koekemoer}, {Faber}, {Ferguson}, {Grogin},
  {Kocevski}, {Koo}, {Lai}, {Lotz}, {Lucas}, {McGrath}, {Ogaz}, {Rajan},
  {Riess}, {Rodney}, {Strolger}, {Casertano}, {Castellano}, {Dahlen},
  {Dickinson}, {Dolch}, {Fontana}, {Giavalisco}, {Grazian}, {Guo}, {Hathi},
  {Huang}, {van der Wel}, {Yan}, {Acquaviva}, {Alexander}, {Almaini}, {Ashby},
  {Barden}, {Bell}, {Bournaud}, {Brown}, {Caputi}, {Cassata}, {Challis},
  {Chary}, {Cheung}, {Cirasuolo}, {Conselice}, {Roshan Cooray}, {Croton},
  {Daddi}, {Dav{\'e}}, {de Mello}, {de Ravel}, {Dekel}, {Donley}, {Dunlop},
  {Dutton}, {Elbaz}, {Fazio}, {Filippenko}, {Finkelstein}, {Frazer}, {Gardner},
  {Garnavich}, {Gawiser}, {Gruetzbauch}, {Hartley}, {H{\"a}ussler},
  {Herrington}, {Hopkins}, {Huang}, {Jha}, {Johnson}, {Kartaltepe},
  {Khostovan}, {Kirshner}, {Lani}, {Lee}, {Li}, {Madau}, {McCarthy},
  {McIntosh}, {McLure}, {McPartland}, {Mobasher}, {Moreira}, {Mortlock},
  {Moustakas}, {Mozena}, {Nandra}, {Newman}, {Nielsen}, {Niemi}, {Noeske},
  {Papovich}, {Pentericci}, {Pope}, {Primack}, {Ravindranath}, {Reddy},
  {Renzini}, {Rix}, {Robaina}, {Rosario}, {Rosati}, {Salimbeni}, {Scarlata},
  {Siana}, {Simard}, {Smidt}, {Snyder}, {Somerville}, {Spinrad}, {Straughn},
  {Telford}, {Teplitz}, {Trump}, {Vargas}, {Villforth}, {Wagner}, {Wandro},
  {Wechsler}, {Weiner}, {Wiklind}, {Wild}, {Wilson}, {Wuyts}, \&
  {Yun}}]{koekemoeretal11}
{Koekemoer}, A.~M., {Faber}, S.~M., {Ferguson}, H.~C., {et~al.} 2011,
  \href{http://dx.doi.org/10.1088/0067-0049/197/2/36}{\apjs},
  \href{http://adsabs.harvard.edu/abs/2011ApJS..197...36K}{197},
  \href{http://adsabs.harvard.edu/abs/2011ApJS..197...36K}{36}

\bibitem[{{Koopmans} {et~al.}(2006){Koopmans}, {Treu}, {Bolton}, {Burles}, \&
  {Moustakas}}]{koopmansetal06}
{Koopmans}, L.~V.~E., {Treu}, T., {Bolton}, A.~S., {Burles}, S., \&
  {Moustakas}, L.~A. 2006, \href{http://dx.doi.org/10.1086/505696}{\apj},
  \href{http://adsabs.harvard.edu/abs/2006ApJ...649..599K}{649},
  \href{http://adsabs.harvard.edu/abs/2006ApJ...649..599K}{599}

\bibitem[{{Kubo} \& {Dell'Antonio}(2008)}]{kuboanddellantonio08}
{Kubo}, J.~M., \& {Dell'Antonio}, I.~P. 2008,
  \href{http://dx.doi.org/10.1111/j.1365-2966.2008.12880.x}{\mnras},
  \href{http://adsabs.harvard.edu/abs/2008MNRAS.385..918K}{385},
  \href{http://adsabs.harvard.edu/abs/2008MNRAS.385..918K}{918}

\bibitem[{{Lanusse} {et~al.}(2018){Lanusse}, {Ma}, {Li}, {Collett}, {Li},
  {Ravanbakhsh}, {Mandelbaum}, \& {P{\'o}czos}}]{lanusseetal17}
{Lanusse}, F., {Ma}, Q., {Li}, N., {et~al.} 2018,
  \href{http://dx.doi.org/10.1093/mnras/stx1665}{\mnras},
  \href{http://adsabs.harvard.edu/abs/2018MNRAS.473.3895L}{473},
  \href{http://adsabs.harvard.edu/abs/2018MNRAS.473.3895L}{3895}

\bibitem[{{Lee}(2017)}]{lee17}
{Lee}, C.-H. 2017, ArXiv e-prints,
  arXiv:\href{http://adsabs.harvard.edu/abs/2017arXiv170207557L}{1702.07557}

\bibitem[{{Lenzen} {et~al.}(2004){Lenzen}, {Schindler}, \&
  {Scherzer}}]{lenzenetal04}
{Lenzen}, F., {Schindler}, S., \& {Scherzer}, O. 2004,
  \href{http://dx.doi.org/10.1051/0004-6361:20034619}{\aap},
  \href{http://adsabs.harvard.edu/abs/2004A%26A...416..391L}{416},
  \href{http://adsabs.harvard.edu/abs/2004A%26A...416..391L}{391}

\bibitem[{{Li} \& {Chen}(2009)}]{liandchen09}
{Li}, N., \& {Chen}, D.-M. 2009,
  \href{http://dx.doi.org/10.1088/1674-4527/9/11/001}{Research in Astronomy and
  Astrophysics}, \href{http://adsabs.harvard.edu/abs/2009RAA.....9.1173L}{9},
  \href{http://adsabs.harvard.edu/abs/2009RAA.....9.1173L}{1173}

\bibitem[{{Li} {et~al.}(2016){Li}, {Gladders}, {Rangel}, {Florian}, {Bleem},
  {Heitmann}, {Habib}, \& {Fasel}}]{lietal16}
{Li}, N., {Gladders}, M.~D., {Rangel}, E.~M., {et~al.} 2016,
  \href{http://dx.doi.org/10.3847/0004-637X/828/1/54}{\apj},
  \href{http://adsabs.harvard.edu/abs/2016ApJ...828...54L}{828},
  \href{http://adsabs.harvard.edu/abs/2016ApJ...828...54L}{54}

\bibitem[{{Linder}(2004)}]{linder04}
{Linder}, E.~V. 2004,
  \href{http://dx.doi.org/10.1103/PhysRevD.70.043534}{\prd},
  \href{http://adsabs.harvard.edu/abs/2004PhRvD..70d3534L}{70},
  \href{http://adsabs.harvard.edu/abs/2004PhRvD..70d3534L}{043534}

\bibitem[{{Lintott} {et~al.}(2008){Lintott}, {Schawinski}, {Slosar}, {Land},
  {Bamford}, {Thomas}, {Raddick}, {Nichol}, {Szalay}, {Andreescu}, {Murray}, \&
  {Vandenberg}}]{lintottetal08}
{Lintott}, C.~J., {Schawinski}, K., {Slosar}, A., {et~al.} 2008,
  \href{http://dx.doi.org/10.1111/j.1365-2966.2008.13689.x}{\mnras},
  \href{http://adsabs.harvard.edu/abs/2008MNRAS.389.1179L}{389},
  \href{http://adsabs.harvard.edu/abs/2008MNRAS.389.1179L}{1179}

\bibitem[{{Lynds} \& {Petrosian}(1986)}]{lyndsandpetrosian86}
{Lynds}, R., \& {Petrosian}, V. 1986, in \baas, Vol.~18, Bulletin of the
  American Astronomical Society, 1014

\bibitem[{{Marshall} {et~al.}(2009){Marshall}, {Hogg}, {Moustakas},
  {Fassnacht}, {Brada{\v c}}, {Schrabback}, \& {Blandford}}]{marshalletal09}
{Marshall}, P.~J., {Hogg}, D.~W., {Moustakas}, L.~A., {et~al.} 2009,
  \href{http://dx.doi.org/10.1088/0004-637X/694/2/924}{\apj},
  \href{http://adsabs.harvard.edu/abs/2009ApJ...694..924M}{694},
  \href{http://adsabs.harvard.edu/abs/2009ApJ...694..924M}{924}

\bibitem[{{Marshall} {et~al.}(2016){Marshall}, {Verma}, {More}, {Davis},
  {More}, {Kapadia}, {Parrish}, {Snyder}, {Wilcox}, {Baeten}, {Macmillan},
  {Cornen}, {Baumer}, {Simpson}, {Lintott}, {Miller}, {Paget}, {Simpson},
  {Smith}, {K{\"u}ng}, {Saha}, \& {Collett}}]{marshalletal16}
{Marshall}, P.~J., {Verma}, A., {More}, A., {et~al.} 2016,
  \href{http://dx.doi.org/10.1093/mnras/stv2009}{\mnras},
  \href{http://adsabs.harvard.edu/abs/2016MNRAS.455.1171M}{455},
  \href{http://adsabs.harvard.edu/abs/2016MNRAS.455.1171M}{1171}

\bibitem[{{Maturi} {et~al.}(2014){Maturi}, {Mizera}, \&
  {Seidel}}]{maturietal14}
{Maturi}, M., {Mizera}, S., \& {Seidel}, G. 2014,
  \href{http://dx.doi.org/10.1051/0004-6361/201321634}{\aap},
  \href{http://adsabs.harvard.edu/abs/2014A%26A...567A.111M}{567},
  \href{http://adsabs.harvard.edu/abs/2014A%26A...567A.111M}{A111}

\bibitem[{{Metcalf} {et~al.}(2018){Metcalf}, {Meneghetti}, {Avestruz},
  {Bellagamba}, {Bom}, {Bertin}, {Cabanac}, {Davies}, {Decenci{\`e}re},
  {Flamary}, {Gavazzi}, {Geiger}, {Hartley}, {Huertas-Company}, {Jackson},
  {Jullo}, {Kneib}, {Koopmans}, {Lanusse}, {Li}, {Ma}, {Makler}, {Li},
  {Lightman}, {Enrico Petrillo}, {Serjeant}, {Sch{\"a}fer}, {Sonnenfeld},
  {Tagore}, {Tortora}, {Tuccillo}, {Valent{\'{\i}}n}, {Velasco-Forero},
  {Verdoes Kleijn}, \& {Vernardos}}]{metcalfetal18}
{Metcalf}, R.~B., {Meneghetti}, M., {Avestruz}, C., {et~al.} 2018, ArXiv
  e-prints,
  arXiv:\href{http://adsabs.harvard.edu/abs/2018arXiv180203609M}{1802.03609}

\bibitem[{{Miralda-Escude} \& {Lehar}(1992)}]{miraldaescudeandlehar92}
{Miralda-Escude}, J., \& {Lehar}, J. 1992,
  \href{http://dx.doi.org/10.1093/mnras/259.1.31P}{\mnras},
  \href{http://adsabs.harvard.edu/abs/1992MNRAS.259P..31M}{259},
  \href{http://adsabs.harvard.edu/abs/1992MNRAS.259P..31M}{31P}

\bibitem[{{More} {et~al.}(2016){More}, {Verma}, {Marshall}, {More}, {Baeten},
  {Wilcox}, {Macmillan}, {Cornen}, {Kapadia}, {Parrish}, {Snyder}, {Davis},
  {Gavazzi}, {Lintott}, {Simpson}, {Miller}, {Smith}, {Paget}, {Saha},
  {K{\"u}ng}, \& {Collett}}]{moreetal16}
{More}, A., {Verma}, A., {Marshall}, P.~J., {et~al.} 2016,
  \href{http://dx.doi.org/10.1093/mnras/stv1965}{\mnras},
  \href{http://adsabs.harvard.edu/abs/2016MNRAS.455.1191M}{455},
  \href{http://adsabs.harvard.edu/abs/2016MNRAS.455.1191M}{1191}

\bibitem[{{Oguri} \& {Marshall}(2010)}]{oguriandmarshall10}
{Oguri}, M., \& {Marshall}, P.~J. 2010,
  \href{http://dx.doi.org/10.1111/j.1365-2966.2010.16639.x}{\mnras},
  \href{http://adsabs.harvard.edu/abs/2010MNRAS.405.2579O}{405},
  \href{http://adsabs.harvard.edu/abs/2010MNRAS.405.2579O}{2579}

\bibitem[{{Paraficz} {et~al.}(2016){Paraficz}, {Courbin}, {Tramacere},
  {Joseph}, {Metcalf}, {Kneib}, {Dubath}, {Droz}, {Filleul}, {Ringeisen}, \&
  {Sch{\"a}fer}}]{paraficzetal16}
{Paraficz}, D., {Courbin}, F., {Tramacere}, A., {et~al.} 2016,
  \href{http://dx.doi.org/10.1051/0004-6361/201527971}{\aap},
  \href{http://adsabs.harvard.edu/abs/2016A%26A...592A..75P}{592},
  \href{http://adsabs.harvard.edu/abs/2016A%26A...592A..75P}{A75}

\bibitem[{{Pedregosa} {et~al.}(2012){Pedregosa}, {Varoquaux}, {Gramfort},
  {Michel}, {Thirion}, {Grisel}, {Blondel}, {Prettenhofer}, {Weiss}, {Dubourg},
  {Vanderplas}, {Passos}, {Cournapeau}, {Brucher}, {Perrot}, \&
  {Duchesnay}}]{pedregosaetal12}
{Pedregosa}, F., {Varoquaux}, G., {Gramfort}, A., {et~al.} 2012, ArXiv
  e-prints,
  arXiv:\href{http://adsabs.harvard.edu/abs/2012arXiv1201.0490P}{1201.0490}

\bibitem[{{Petrillo} {et~al.}(2017){Petrillo}, {Tortora}, {Chatterjee},
  {Vernardos}, {Koopmans}, {Verdoes Kleijn}, {Napolitano}, {Covone},
  {Schneider}, {Grado}, \& {McFarland}}]{petrilloetal17}
{Petrillo}, C.~E., {Tortora}, C., {Chatterjee}, S., {et~al.} 2017,
  \href{http://dx.doi.org/10.1093/mnras/stx2052}{\mnras},
  \href{http://adsabs.harvard.edu/abs/2017MNRAS.472.1129P}{472},
  \href{http://adsabs.harvard.edu/abs/2017MNRAS.472.1129P}{1129}

\bibitem[{{Seidel} \& {Bartelmann}(2007)}]{seidelandbartelmann07}
{Seidel}, G., \& {Bartelmann}, M. 2007,
  \href{http://dx.doi.org/10.1051/0004-6361:20066097}{\aap},
  \href{http://adsabs.harvard.edu/abs/2007A%26A...472..341S}{472},
  \href{http://adsabs.harvard.edu/abs/2007A%26A...472..341S}{341}

\bibitem[{{Sharon} {et~al.}(2005){Sharon}, {Ofek}, {Smith}, {Broadhurst},
  {Maoz}, {Kochanek}, {Oguri}, {Suto}, {Inada}, \& {Falco}}]{sharonetal05}
{Sharon}, K., {Ofek}, E.~O., {Smith}, G.~P., {et~al.} 2005,
  \href{http://dx.doi.org/10.1086/452633}{\apjl},
  \href{http://adsabs.harvard.edu/abs/2005ApJ...629L..73S}{629},
  \href{http://adsabs.harvard.edu/abs/2005ApJ...629L..73S}{L73}

\bibitem[{{Soler} {et~al.}(2019){Soler}, {Beuther}, {Rugel}, {Wang}, {Clark},
  {Glover}, {Goldsmith}, {Heyer}, {Anderson}, {Goodman}, {Henning},
  {Kainulainen}, {Klessen}, {Longmore}, {McClure-Griffiths}, {Menten},
  {Mottram}, {Ott}, {Ragan}, {Smith}, {Urquhart}, {Bigiel}, {Hennebelle},
  {Roy}, \& {Schilke}}]{soleretal18}
{Soler}, J.~D., {Beuther}, H., {Rugel}, M., {et~al.} 2019,
  \href{http://dx.doi.org/10.1051/0004-6361/201834300}{\aap},
  \href{http://adsabs.harvard.edu/abs/2019A%26A...622A.166S}{622},
  \href{http://adsabs.harvard.edu/abs/2019A%26A...622A.166S}{A166}

\bibitem[{{Suyu} {et~al.}(2014){Suyu}, {Treu}, {Hilbert}, {Sonnenfeld},
  {Auger}, {Blandford}, {Collett}, {Courbin}, {Fassnacht}, {Koopmans},
  {Marshall}, {Meylan}, {Spiniello}, \& {Tewes}}]{suyuetal14}
{Suyu}, S.~H., {Treu}, T., {Hilbert}, S., {et~al.} 2014,
  \href{http://dx.doi.org/10.1088/2041-8205/788/2/L35}{\apjl},
  \href{http://adsabs.harvard.edu/abs/2014ApJ...788L..35S}{788},
  \href{http://adsabs.harvard.edu/abs/2014ApJ...788L..35S}{L35}

\bibitem[{{Suyu} {et~al.}(2017){Suyu}, {Bonvin}, {Courbin}, {Fassnacht},
  {Rusu}, {Sluse}, {Treu}, {Wong}, {Auger}, {Ding}, {Hilbert}, {Marshall},
  {Rumbaugh}, {Sonnenfeld}, {Tewes}, {Tihhonova}, {Agnello}, {Blandford},
  {Chen}, {Collett}, {Koopmans}, {Liao}, {Meylan}, \& {Spiniello}}]{suyuetal16}
{Suyu}, S.~H., {Bonvin}, V., {Courbin}, F., {et~al.} 2017,
  \href{http://dx.doi.org/10.1093/mnras/stx483}{\mnras},
  \href{http://adsabs.harvard.edu/abs/2017MNRAS.468.2590S}{468},
  \href{http://adsabs.harvard.edu/abs/2017MNRAS.468.2590S}{2590}

\bibitem[{{Walsh} {et~al.}(1979){Walsh}, {Carswell}, \&
  {Weymann}}]{walshetal79}
{Walsh}, D., {Carswell}, R.~F., \& {Weymann}, R.~J. 1979,
  \href{http://dx.doi.org/10.1038/279381a0}{\nat},
  \href{http://adsabs.harvard.edu/abs/1979Natur.279..381W}{279},
  \href{http://adsabs.harvard.edu/abs/1979Natur.279..381W}{381}

\bibitem[{{Warren} \& {Dye}(2003)}]{warrenanddye03}
{Warren}, S.~J., \& {Dye}, S. 2003,
  \href{http://dx.doi.org/10.1086/375132}{\apj},
  \href{http://adsabs.harvard.edu/abs/2003ApJ...590..673W}{590},
  \href{http://adsabs.harvard.edu/abs/2003ApJ...590..673W}{673}

\bibitem[{{Xu} {et~al.}(2016){Xu}, {Postman}, {Meneghetti}, {Seitz}, {Zitrin},
  {Merten}, {Maoz}, {Frye}, {Umetsu}, {Zheng}, {Bradley}, {Vega}, \&
  {Koekemoer}}]{xuetal16}
{Xu}, B., {Postman}, M., {Meneghetti}, M., {et~al.} 2016,
  \href{http://dx.doi.org/10.3847/0004-637X/817/2/85}{\apj},
  \href{http://adsabs.harvard.edu/abs/2016ApJ...817...85X}{817},
  \href{http://adsabs.harvard.edu/abs/2016ApJ...817...85X}{85}

\bibitem[{{Zahid} {et~al.}(2015){Zahid}, {Damjanov}, {Geller}, \&
  {Chilingarian}}]{zahidetal15}
{Zahid}, H.~J., {Damjanov}, I., {Geller}, M.~J., \& {Chilingarian}, I. 2015,
  \href{http://dx.doi.org/10.1088/0004-637X/806/1/122}{\apj},
  \href{http://adsabs.harvard.edu/abs/2015ApJ...806..122Z}{806},
  \href{http://adsabs.harvard.edu/abs/2015ApJ...806..122Z}{122}

\end{thebibliography}
\end{CJK*}
\end{document}